\documentclass[a4paper,fleqn]{cas-dc}

\usepackage{amsmath,amssymb,amsfonts}
\usepackage{algorithmic}
\usepackage{graphicx}
\usepackage{float}
\usepackage{enumitem}
\usepackage{array}
\usepackage{multirow}
\usepackage{booktabs}
\usepackage{longtable}
\usepackage{tabularx}
\usepackage{ltxtable}  
\usepackage{threeparttable}
\usepackage{pgfplots}
\pgfplotsset{compat=1.17}
\usepackage{microtype}    
\usepackage{hyphenat}
\usepackage{lipsum}
\usepackage{soul}         
\setuldepth{Berlin}

\usepackage[dvipsnames]{xcolor}
\definecolor{main}{HTML}{CFCFCF}
\definecolor{sub}{HTML}{CFCFCF}
\definecolor{darkgray}{rgb}{0.2, 0.2, 0.2}
\definecolor{pblue}{rgb}{0.13,0.13,1}
\definecolor{pgreen}{rgb}{0,0.5,0}
\definecolor{pred}{rgb}{0.9,0,0}
\definecolor{pgrey}{rgb}{0.46,0.45,0.48}
\definecolor{codebackground}{rgb}{0.95, 0.95, 0.92}
\definecolor{gray50}{gray}{.5}
\definecolor{gray40}{gray}{.6}
\definecolor{gray30}{gray}{.7}
\definecolor{gray20}{gray}{.8}
\definecolor{gray10}{gray}{.9}
\definecolor{gray05}{gray}{.95}
\definecolor{arsenic}{rgb}{0.23, 0.27, 0.29}
\definecolor{darkmagenta}{rgb}{0.55, 0.0, 0.55}
\definecolor{darkgreen}{RGB}{6, 46, 3}
\definecolor{amber}{rgb}{1.0, 0.75, 0.0}
\definecolor{ao(english)}{rgb}{0.0, 0.5, 0.0}
\definecolor{googleblue}{HTML}{4285F4}

\usepackage{colortbl}

\usepackage{flafter}
\usepackage{float}
\usepackage{flushend}
\usepackage[breakable,many]{tcolorbox}
\tcbset{
    sharp corners,
    colback = white,
    before skip = 0.5cm,
    after skip = 0.7cm
}

\newtcolorbox{boxCNoTitle}{
    top=10pt,
    rounded corners,
    coltitle=black,
    colframe=gray,
    colback=sub,
    enhanced,
    center,
    boxrule=0.5pt
}

\newtcolorbox{boxC}[2][]{aibox,title=#2,#1}
\tcbset{
  aibox/.style={
    top=10pt,
    boxrule=0.5pt,
    rounded corners,
    coltitle=black,
    colframe=gray,
    colbacktitle=sub,
    colback=sub,
    enhanced,
    center,
    attach boxed title to top left={yshift=-0.1in,xshift=0.15in},
    boxed title style={boxrule=0.5pt,colframe=gray,},
  }
}

\newcounter{keyTakeAwaysCounter}
\newenvironment{keyTakeAways}[1][Key Take Away]{
    \addtocounter{keyTakeAwaysCounter}{1}
    \begin{boxC}{\faLightbulbO ~ \thekeyTakeAwaysCounter. #1}
}{\end{boxC}}

\newcounter{keyRQAnswerCounter}
\newenvironment{keyRQAnswer}[1][RQAnswer]{
    \addtocounter{keyRQAnswerCounter}{1}
    \begin{boxC}{\faKey ~ \thekeyRQAnswerCounter. #1}
}{\end{boxC}}

\newcounter{keyLimitationsCounter}
\newenvironment{implications}[1][Key Limitations]{
    \addtocounter{keyLimitationsCounter}{1}
    \begin{boxC}{\faBook ~ \thekeyLimitationsCounter. #1}
}{\end{boxC}}

\usepackage{listings}
\usepackage{caption}
\usepackage{subcaption}
\usepackage{adjustbox}
\usepackage{tikz}
\usepackage{arydshln}
\usepackage{verbatim}
\usepackage{todonotes}

\usepackage{natbib}
\usepackage{url}
\usepackage{hyperref}
\hypersetup{hidelinks}
\usepackage{tablefootnote}

\usepackage{footmisc}
\usepackage{tablefootnote}
\usepackage{fontawesome}
\usepackage{pifont}

\begin{document}

\title[mode=title]{Emerging Trends in Software Architecture from the Practitioner’s Perspective: A Five-Year Review}
\shorttitle{Emerging Trends in Software Architecture from the Practitioner’s Perspective: A Five-Year Review}
\shortauthors{Su et al.}
\author[1]{Ruoyu Su}[orcid=0009-0008-6206-8787]
\ead{ruoyu.su@oulu.fi}

\author[1]{Noman Ahmad}[orcid=0009-0005-4228-2493]
\ead{noman.ahmad@oulu.fi}

\author[1]{Matteo Esposito}[orcid=0000-0002-8451-3668]
\ead{matteo.esposito@oulu.fi}

\author[2]{Andrea Janes}[orcid=0000-0002-1423-6773]
\ead{andrea.janes@unibz.it}

\author[1]{Davide Taibi}[orcid=0000-0002-3210-3990]
\ead{davide.taibi@oulu.fi}

\author[1]{Valentina Lenarduzzi}[orcid=0000-0003-0511-5133]
\ead{valentina.lenarduzzi@oulu.fi}

\affiliation[1]{
    organization={University of Oulu},
    country={Finland}
}

\affiliation[2]{
    organization={Free University of Bozen-Bolzano},
    country={Italy}
}

\begin{abstract}
\noindent\textbf{CONTEXT.} Software architecture plays a central role in the design, development, and maintenance of software systems. With the rise of cloud computing, microservices, and containers, architectural practices have diversified. Understanding these shifts is vital. This study analyzes software architecture trends across eight leading industry conferences over five years.

\noindent\textbf{AIM.} We investigate the evolution of software architecture by analyzing talks from top practitioner conferences, focusing on the motivations and contexts driving technology adoption.

\noindent\textbf{METHODS.} We analyzed 5,677 talks from eight major industry conferences (2020–2024), using large language models and expert validation to extract technologies, their purposes, and usage contexts. We also explored how technologies interrelate and fit within DevOps and deployment pipelines.

\noindent\textbf{RESULTS.} Among 450 technologies, Kubernetes, Cloud Native, Serverless, and Containers dominate by frequency and centrality. Practitioners present technology mainly related to deployment, communication, AI, and observability. We identify five technology communities covering automation, coordination, cloud AI, monitoring, and cloud-edge. Most technologies span multiple DevOps stages and support hybrid deployment. 

\noindent\textbf{CONCLUSIONS.} Our study reveals that a few core technologies, like Kubernetes and Serverless, dominate the contemporary software architecture practice. These are mainly applied in later DevOps stages, with limited focus on early phases like planning and coding. We also show how practitioners frame technologies by purpose and context, reflecting evolving industry priorities. Finally, we observe how only research can provide a more holistic lens on architectural design, quality, and evolution.
\end{abstract}


\maketitle

\section{Introduction}
\label{Introduction}

Software architecture serves as the backbone of software development. It defines the structure and therefore organization of systems, playing a vital role in shaping both their scalability and maintainability~\citep{nivedhaa2024software}. Beyond that, software architecture acts as a guiding framework throughout the development process, often driving key activities such as integration testing~\citep{mens2008software}.

Over the past decade, architectural paradigms have continuously evolved to keep pace with shifting technological landscapes and emerging requirements. This evolution often signals shifts in focus, for example, the widespread transition from monolithic systems to microservices~\citep{taibi2017processes}, the subsequent reconsideration of that shift with a return to monoliths~\citep{su2024microservice}.

With the rise of enabling technologies like Cloud Computing and DevOps, software architecture has grown more diverse, both in patterns and in supporting tools~\citep{bass2021software}. These transformations have introduced fresh challenges, but they have also opened new avenues for innovation. As a result, researchers and practitioners are actively exploring novel solutions to tackle these complexities and harness emerging opportunities.

Industry professionals tend to be at the forefront of researching such evolution and trends because they need to solve practical problems using innovative approaches in their work. 

Following this reasoning, industry conferences serve as an excellent source of data, often showcasing new practices, tools, and the challenges faced by practitioners~\citep{garousi2016challenges}. Unlike traditional academic conferences, industry conferences typically focus more on applied research and case studies, providing a unique perspective on the factors driving the field. However, despite their practical relevance, few insights from these conferences have been systematically analyzed to reveal broader patterns and trends~\citep{barroca2018bridging}.

Our goal is to explore software architecture trends by analyzing practitioner talks from leading conferences over the past five years. Specifically, our investigation (i) examines which technologies have emerged as prominent in the field, how different technologies interact with each other, and (ii) seeks to understand the purposes and contexts in which these technologies are adopted within software architecture practice. 

Our study provides the following contributions:
\begin{itemize}
    \item \textbf{Practitioner-oriented alignment}: Hints at how academic research interests should align with and focus on practitioners' current software architecture implementations.
    \item \textbf{Topic classification}: Develops a structured classification of software architecture topics.
    \item \textbf{Trend mapping}: Identifies key software architecture trends from five years of practitioner talks.
    \item \textbf{Temporal and co-occurrence analysis}: Tracks how topics evolved and appeared together in practice.
    \item \textbf{Industry reflection}: Offers a basis for practitioners to compare their practices with observed trends.
\end{itemize}

We identified 450 technologies from five years of top industry conferences highlights that Kubernetes, Cloud Native, Serverless, and Containers dominate software architecture talks, acting as key connectors across deployment, AI, and observability. Five major technology clusters emerge, with most tools supporting multiple DevOps phases and hybrid setups. Talks are shifting from basic introductions toward more strategic and innovation-driven uses, especially in cloud and AI. However, early DevOps stages like planning and coding remain less emphasized, pointing to evolving priorities in practice. Finally, we observe how only research can provide a more holistic lens on architectural design, quality, and evolution.

Section~\ref{sec:Methodology} describes our research method applied to our research, including the research objective and question, research strategy, data processing, and data analysis. In Section~\ref{sec:Results}, we analyze the processed data and present the results addressing the goals of the study. Section~\ref{sec:Discussion}, we state the discussion points, implications, limitations, and future work from our research analysis. Section~\ref{sec:rw} presents an overview of the related work. Section~\ref{sec:Threats} identifies the threats to validity, and finally, Section~\ref{sec:Conclusion} provides the conclusion.

\section{Methodology}
\label{sec:Methodology}
\begin{figure}
    \centering
    \includegraphics[width=\linewidth]{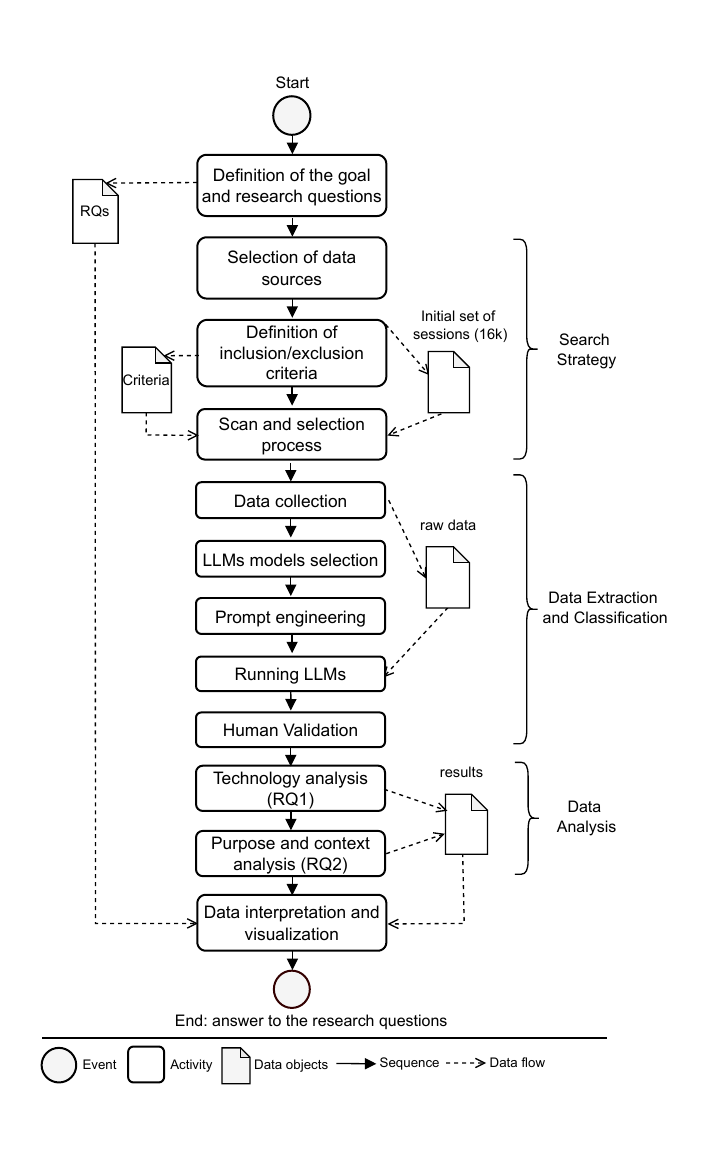}
    \caption{Overview of the Research Process}
    \label{fig:sasprocess}
\end{figure}

In this section, we define the research objective and questions, describe the rationale, introduce the data extraction and processing process, and explain each step we performed (Figure \ref{fig:sasprocess}).

\subsection{Goals and Research Questions}

The \textit{objective} of our study is to analyze the trends in software architecture in the last 5 years \textit{from the perspective of} practitioners. 

Albeit it is widely agreed that academia and industry can benefit from each other, several factors, such as differences in timelines, the perception that research is disconnected from real-world needs, and the demand for academic rigor, have created a noticeable gap between the two worlds~\citep{barroca2018bridging}.

Our goal is to help close this gap by ensuring that our research reflects current industry challenges and priorities. Hence, we define the following two research questions:

\begin{boxC}{\textbf{RQ$_1$}}
\textit{What software architecture trends have emerged from top practitioner conferences over the past five years?}
\begin{itemize}
    \item \textbf{RQ$_{1.1}$}. \textit{Which technologies have been most prominently featured in software architecture field talks at top practitioner conferences over the past five years?}
    \item \textbf{RQ$_{1.2}$}. \textit{What relationships and patterns of co-occurrence exist among the extracted technologies in the software architecture field?}
\end{itemize}
\end{boxC}

Practitioner conferences often present several practical innovations and emerging technologies from industry professionals. \textbf{RQ$_{1.1}$} aims to study the trends of software architecture talks from practitioner conferences to gain insights into the interests and priorities of the practitioner community in the software architecture field.
We attempt to extract these technologies and study their trends. RQ$_{1.1}$ helps ensure that our research integrates academics effectively with the industry, aligning closely with the evolving needs of the industry.

\textbf{RQ$_{1.2}$} focuses on exploring the relationship and co-occurrence between different extracted technologies that they interconnected and often appear together in the title of practitioner conferences. For instance, certain technologies might often co-occur in one practitioner' talk (title) because they represent related components of a larger paradigm, such as Microservices and Containers. \textbf{RQ$_{1.2}$} helps trace the evolution of architectural trends and highlight possible emerging groups of technologies or methodologies that are gaining traction in the industry.

Understanding which technologies are discussed in practitioner' talks offers valuable insight into current trends. However, to fully grasp their role in the software architecture field, it is important to explore why these technologies are adopted and how they are positioned in practice; hence, we ask:

\begin{boxC}{\textbf{RQ$_2$}}
\textit{What are the purposes and context for adopting these extracted technologies in the software architecture field?}
\end{boxC}

Talk titles often convey whether a technology is introduced for onboarding, strategic transformation, quality enhancement, or innovation. We aim to uncover how practitioners frame the role of technologies in real-world architectural settings, offering insights into the driving purpose behind their adoption in the specific context.

\subsection{Search Strategy}
Our search strategy involves the selection of data sources, the definition of inclusion and exclusion criteria, and the selection process (Figure \ref{fig:sasprocess}).

\textbf{Data Sources.} We carefully selected data sources that capture a broad and realistic view of how software architecture is discussed in practice. Therefore, we focused on the most prominent practitioner-oriented software engineering conferences (Table~\ref{tab:IndustryConferences}). 
Such conferences are renowned for their large-scale attendance and prominence among industry professionals. Their wide recognition demonstrates their centrality in disseminating industry practices.

\begin{table}
    \centering
    \scriptsize
    \resizebox{\linewidth}{!}{
    \begin{tabular}{l|l} \hline 
    \textbf{Industrial Conferences}  & \textbf{ Link} \\ \hline 
    Alibaba          & \url{https://www.alibaba.com/}\\
    Amazon REinvent          & \url{https://reinvent.awsevents.com/}\\
    Gartner          & \url{https://www.gartner.com/en/conferences}\\
        Global Azure          &\url{https://globalazure.net/} \\
    Google Cloud Next          & \url{https://cloud.withgoogle.com/next}\\
    InfoQ          & \url{https://www.infoq.com/}\\
    KubeCon         & \url{https://www.cncf.io/kubecon-cloudnativecon-events/}\\
    QCon          &  \url{https://qconferences.com/}\\
     \hline 
    \end{tabular}
}
\caption{Industry conferences included in the study}
\label{tab:IndustryConferences}
\end{table}

Unlike academic conferences, many industry conferences are held at multiple locations within the same year. For example, KubeCon is usually held in Europe, North America, and China every year, and each edition has a different website. For each conference in Table \ref{tab:IndustryConferences}, we searched for their editions from 2020 to 2024 and compiled the corresponding links for each edition. Considering a few conferences where their websites were rewritten and deleted in previous years (e.g., Amazon re: Invent), we found the corresponding events and practitioners' talks on the official YouTube\footnote{\url{https://www.youtube.com/}} channel. In addition, external factors such as COVID-19 have caused some conferences to be canceled or postponed in some years. We recorded these situations to ensure the accuracy of the dataset.

We translated the titles of talks at the few conference editions held in Chinese as the first author is a native Chinese speaker. We provide all raw data in the replication package (Section~\ref{Replicability}).

\textbf{Inclusion and Exclusion Criteria.} We defined the Inclusion (I) and Exclusion (E) criteria (Table~\ref{tab:Criteria}) according to our RQs goals. 

\begin{table}
\centering
\scriptsize
\caption{Inclusion and Exclusion Criteria} 
\label{tab:Criteria}
\begin{tabular}{c|p{7cm}}
\hline 
\textbf{I/E} & \textbf{Criteria} \\ 
\hline 
I$_1$ & The practitioners' talk is related to the software architecture field  \\ \hline
E$_1$ & practitioners' talk is not the presentation  \\
E$_2$ & Out of topic (Practitioner' talk is not related to the software architecture field) \\
E$_3$ & Title of the practitioners' talk do not indicate any related terms in the software architecture field \\
\hline 
\end{tabular}
\end{table}

We included all practitioners' talks that clearly focused on software architecture from the selected practitioner conferences (I$_1$). To keep our dataset relevant, we applied a few exclusion criteria. First, we filtered out non-presentation practitioners' talks such as badge pickups or demo showcases, which often appear in conference schedules but fall outside our scope (E$_1$). Next, since not all conferences listed in Table~\ref{tab:IndustryConferences} focus solely on software architecture, we excluded any practitioner' talks unrelated to the field (E$_2$). Finally, we removed talks whose titles didn’t reference any software architecture-related terms, as they lacked clear relevance (E$_3$).

\textbf{Selection Process.} The search was conducted from October 2024 to December 2024 and included all the practitioners' talks available from 2020 to the end of 2024. We analyzed the \textbf{16778 practitioners' talks retrieved}, applying the above criteria for consistency and adequacy. We first tested the applicability of the inclusion and exclusion criteria to verify their validity~\citep{kitchenham2013systematic}. We randomly selected 100 practitioners' talks from the collected practitioners' talks as a sample and assigned them to two authors for testing. All authors discussed the results of the test. Ultimately, this step did not cause any revisions to the inclusion and exclusion criteria, indicating the applicability of the criteria.

\begin{table*}[t]
\centering
\footnotesize
\caption{Distribution of Included Practitioner' Talks by Inclusion/Exclusion Criteria} 
\label{tab:SelectionResults} 
\begin{tabular}{@{}l|l|r|r|r|r|r|r|r@{}} \hline 
\multirow{2}{*}{\textbf{Conferences}} & \multirow{2}{*}{\textbf{Edition}}  & \multicolumn{2}{c}{\textbf{\# Talks}} & \multicolumn{5}{|c}{\textbf{Year}} \\\cline{3-9}
& & \textbf{Retrieved}	&	\textbf{Included}  &	\textbf{2020}	&	\textbf{2021}	&	\textbf{2022}	&	\textbf{2023}	&	\textbf{2024}	\\ \hline 
\multirow{2}{*}{Alibaba}	&	Alibaba apsara conference	&	1490	&	350	&	10	&	20	&	72	&	117	&	131	\\
	&	Alibaba cloud global summit	&	182	&	41	&	14	&	11	&	9	&	4	&	3	\\ \hline 
Amazon	&	Amazon REinvent	&	5641	&	1667	&	404	&	235	&	311	&	367	&	350	\\ \hline 
Azure	&	Global Azure Con	&	308	&	133	&		&		&		&		&	133	\\ \hline 
Garter	&	Garter Cloud Con	&	341	&	95	&		&		&		&		&	95	\\ \hline 
Google Cloud Next	&	Google cloud next	&	2007	&	419	&	106	&	57	&	54	&	72	&	130	\\ \hline 
\multirow{3}{*}{InfoQ}	&	InfoQ ArchSummit	&	600	&	188	&	1	&	54	&	58	&	56	&	19	\\
	&	InfoQ Dev Summit	&	54	&	8	&		&		&		&		&	8	\\
	&	InfoQ Dive	&	60	&	20	&		&		&	20	&		&		\\ \hline 
\multirow{3}{*}{KubeCon}	&	KubeCon\_China	&	437	&	263	&		&	79	&		&	84	&	100	\\
	&	KubeCon\_Europe	&	1984	&	1108	&	248	&	215	&	198	&	191	&	256	\\
	&	KubeCon\_North America	&	1895	&	1056	&	216	&	196	&	198	&	222	&	224	\\ \hline 
\multirow{6}{*}{QCon}	&	QCon\_Beijing	&	424	&	88	&	12	&	33	&		&	30	&	13	\\
	&	QCon\_Guangzhou	&	60	&	9	&		&		&		&	9	&		\\
	&	QCon\_London	&	479	&	86	&	31	&		&	27	&	13	&	15	\\
	&	QCon\_NewYork	&	79	&	18	&		&		&	18	&		&		\\
	&	QCon\_SanFrancisco	&	344	&	56	&		&		&	19	&	20	&	17	\\
	&	QCon\_Shanghai	&	393	&	72	&	9	&	28	&	11	&	12	&	12	\\ \hline 
	\multicolumn{2}{l}{\textbf{Sum }} 	&	\textbf{16778} &	\textbf{5677}	&	\textbf{1051}	&	\textbf{928}	&	\textbf{995}&	\textbf{1197}	&	\textbf{1506}	\\\hline 
\end{tabular}%
\end{table*}

We applied the final inclusion and exclusion criteria to the remaining practitioners' talks collected. Two authors read the title of each practitioner's talk and extracted the included information. A third author revised the first two authors' decisions to confirm the accuracy of the results.
To validate the selection process, three other authors considered a sample that was 95\% statistically significant, stratified with a 5\% error margin of the 16778 practitioners' talks, which gave us 376 practitioners' talks to be verified.
Based on the inclusion and exclusion criteria application, we finally included \textbf{5677 practitioners' talks}. The distribution of included practitioners' talks by Inclusion/Exclusion criteria is shown in Table~\ref{tab:SelectionResults}.

\begin{boxC}{\textbf{Data Collected}}
We retrieved \textbf{16778 practitioner' talks} from the selected industrial conferences, and after applying the inclusion and exclusion criteria, we included \textbf{5677 practitioner' talks}. 
\end{boxC}

\subsection{Data Extraction and Classification }
\label{sec:DataExtraction}
Once we completed the selection process, we proceeded with the data extraction to collect the technologies, purpose, and context from the practitioners' talk titles. 

Classification and rating tasks are among the most time-consuming and error-prone activities in empirical research, often suffering from inconsistencies, bias, and learning effects in human judgment \citep{esposito_large_2024, esposito_call_2025}. To enhance objectivity, scalability, and reproducibility, we leverage \textbf{LLMs as judges}. The widespread availability and increasing capabilities of LLMs have opened new opportunities for the Software Engineering (SE) community, enabling accurate and efficient support for tasks traditionally performed by humans. Recent studies have shown that LLMs can assist both researchers and practitioners across various SE activities, including classification, summarization, and decision-making, with promising levels of reliability \citep{esposito_generative_2025}.


In our study, we analyzed the titles of practitioner conferences to extract key information about the technologies proposed, the purposes of the practitioner' talks, and the contexts in which they are used. 
The methodology follows five main steps: (1) use a large reasoning model to extract technologies, purposes, and contexts from practitioner talk titles using prompt-based classification; (2) validate the outputs with three LLMs and human reviewers; (3) clean and unify the extracted terms; (4) classify technologies across DevOps phases, deployment types, and cloud providers; and (5) analyze co-occurrence patterns and map relationships using network analysiss.

\subsubsection{LLM Integration, Experimental Design and Validation}






\begin{table*}[hbtp]
  \centering
  \footnotesize
  \caption{Overview of Selected LLMs}
  \begin{tabular}{p{2cm} c p{4cm} p{6cm}}
    \hline
    \textbf{Model} & \textbf{Parameters/Quantization} & \textbf{Details} & \textbf{Highlights} \\
    \hline
    Marco o1\textsuperscript{a} & 7.6B, not quantized & Inspired by OpenAI’s o-1 & Fine-tuned on CoT datasets, uses MCTS + softmax scoring, excels at math, coding, and logic tasks\\
    
    Mistral NeMo Instruct 2407\textsuperscript{b} & 12.2B, not quantized & Fine-tuned version of Mistral-Nemo-Base-2407 & Alignment fine-tuned, supports 128K tokens, outperforms similarly-sized models \\
    
    Qwen2.5 14B\textsuperscript{c} & 14.8B, not quantized & Decoding Transformer-based & Fine-tuned with enhanced instruction-following, excels in math, programming, and dialogue \\
    
    Llama 3.1 8B\textsuperscript{d} & 8B, not quantized & Meta-developed decoding architecture & High performance, strong language understanding and generation, lightweight \\
    \hline

   \multicolumn{4}{l}{\textsuperscript{a} \url{https://huggingface.co/AIDC-AI/Marco-o1}}\\
   \multicolumn{4}{l}{\textsuperscript{b} \url{https://huggingface.co/mistralai/Mistral-Nemo-Instruct-2407}}\\
   \multicolumn{4}{l}{\textsuperscript{c} \url{https://huggingface.co/Qwen/Qwen2.5-14B-Instruct} }\\
   \multicolumn{4}{l}{\textsuperscript{d} \url{https://huggingface.co/meta-llama/Llama-3.1-8B-Instruct}}\\

  \end{tabular}
  \label{tab:model_overview}
\end{table*}


\paragraph{LLM Model Roles}
Table~\ref{tab:model_overview} presents an overview of the selected LLMs. We assigned a specific role to each model based on its responsibility in the data analysis pipeline as follows:

\begin{itemize}
    \item \textbf{Large Reasoning Model (LRM):} Marco-o1 is responsible to extract the required information. For each prompt, it receives input data (e.g., practitioner' talk titles), classifies the data into predefined categories, and provides detailed explanations. Its output formed the basis for further verification and was referred to as LRM responses throughout the study.
    \item \textbf{Validation Models (V1/V2/V3):} Mistral-NeMo, Qwen, and Llama are allocated to the validation task. They receive the same input as the LRM and are additionally asked to evaluate the LRM's reasoning and output. Their role is to independently assess the correctness of the LRM’s decision (agreement or disagreement).
\end{itemize}

All the information regarding the selected models are reported in Table  \ref{tab:model_overview}

\paragraph{Prompting Techniques} 
We employed prompt engineering techniques to guide LLMs in classifying refactoring motivations. According to the state-of-the-art, in-context learning through chat-based prompting provides similar or better results than the more computationally expensive fine-tuning process \citep{esposito_beyond_2024}. During the in-context learning phase, each prompt included two components: a \textit{system message}, which established the assistant’s role and specified the expected output format, and a \textit{user message}, which provided the contextual input. The user messages the current data at hand,e.g., the practitioner' talk title. The expected model output consisted of a structured JSON response containing: (i) classified category, (ii) a concise \texttt{motivation description}, and (iii) the \texttt{reasoning}. We adopted  \textit{Chain of Thought} (CoT) prompting~\citep{wei2022chain}, with few-shot learning.

\paragraph{Running LLM} To efficiently perform this large-scale analysis, we ran our model using vLLM as an sbatch job on the Mahti supercomputer, hosted by CSC, the Finnish IT Center for Science\footnote{\url{http://csc.fi}}. Mahti is a high-performance computing system designed for compute- and data-intensive research, featuring over 180,000 CPU cores and a high-speed interconnect network. Our jobs utilized up to four NVIDIA A100 GPUs, enabling fast and memory-efficient inference for handling a large volume of data. 

\paragraph{Human Validation} Since no previous study reported findings on the accuracy of LLMs for the task at hand, we designed a validation involving three human experts mimicking the LLM validator's roles. The goal was to assess the quality of the model-generated motivations and identify which models consistently produced reasonable outputs.
Our validation followed a three-step protocol. 
\begin{itemize}
    \item One expert independently reviewed the same input provided to the LRM and the three validation models (V1–V3), and manually evaluated the correctness of each model's motivation. For each case, the expert indicated whether they \texttt{agreed} or \texttt{disagreed} with the LRM’s motivation, noted the majority decision among the validation models, and identified the models they considered correct.
    \item  A second expert repeated the same evaluation independently and documented their level of agreement with the first expert’s judgments. 
    \item  In cases of disagreement between the first two reviewers, a third expert was brought in to independently assess the same outputs. Final decisions were made through majority voting among the three validators.   
\end{itemize}

\subsubsection{LLM Terms Classification}
\label{LLMFirst}

To answer \textbf{RQ$_1$} and \textbf{RQ$_2$}, 
We first employed the LLMs described above, followed the above rules and steps, to extract from the practitioner' talk title the technology used, the context in which it was used, and the purpose of the practitioner' talk.
We randomly selected 400 titles among the included 5677 practitioner' talks, which represents a sample 95\% statistically significant, stratified with a 5\% error margin, and split them into 2 groups. Four authors split into two groups to manually analyze and annotate each group's 200 titles. For the analysis and annotation of the title, we followed the aspects below: 
Not all titles contain the complete components, so we asked LLMs to annotate ``N/A'' when we found a lack of certain structures or the meaning was unclear.

 When dealing with LLMs via API, it is possible to customize the model to strengthen its alignment to the task at hand; it is also possible to send a message with two different roles: \texttt{system} and \texttt{user}. The system role allows for specifying the system prompt of the LLM, and it is usually used to define its \texttt{persona}, i.e., how to act in the following conversation and how to respond. We detailed in the system message or the instruction the LRM needs to follow to answer the user message and formalize the output, requesting it to reply in a JSON-structured output to facilitate automated analysis of the model response. More specifically, we asked the model to extract the following information from the titles of the included practitioner' talks:

\begin{itemize}
    \item \textit{Technology.} Any specific technologies, programming languages, platforms, frameworks, or tools used in the title to implement or validate the proposed research. 
    \item \textit{Purpose.} The main problem or phenomenon the practitioner' talk addresses. It highlights what the author aims to explore, explain, investigate, solve, or improve.
    \item \textit{Context.} The application domain or real-world setting in which the research is situated. It answers questions like where the research applies, why it’s relevant, and who benefits from it.
\end{itemize}

We first obtained the results by LRM (Marco); then asked Validation Models to evaluate the LRM’s reasoning and output that dependently assess the correctness of the LRM’s decision (agreement or disagreement); finally, did the human validation to get Marco's overall accuracy. The entire process was strictly followed the LLM experimental setup described above. The detailed prompts, scripts, and results are in the replication package (Section~\ref{Replicability}), and Table~\ref{tab:LLM1} presents the information about the results generated by 4 LLMs. To measure the agreement among the validation models and the LRM, we selected samples with 95\% statistical significance, stratified with a 5\% error margin for human validation. Finally, we calculated the LRM's (Marco) overall accuracy. Table~\ref{tab:LLM1} shows the high Marco overall accuracy (on average $  \ge 90\% $ ) and means we can trust the results generated by Marco.

\begin{table*}[t]
\centering
\footnotesize
\caption{LLM -1 (Extracted Tech, Purpose, Context)} 
\label{tab:LLM1} 
\resizebox{1\linewidth}{!}{
\begin{tabular}{r|rrrrrrrrr}
\hline
\multicolumn{1}{l|}{\multirow{2}{*}{}}                                                                              & \multicolumn{3}{c|}{\textbf{Qwen}}                                                                                       & \multicolumn{3}{c|}{\textbf{Mistral}}                                                                                    & \multicolumn{3}{c}{\textbf{Llama}}                                                                                      \\ \cline{2-10} 
\multicolumn{1}{l|}{}                                                                                               & \multicolumn{1}{l|}{\textbf{Technology}} & \multicolumn{1}{l|}{\textbf{Purpose}} & \multicolumn{1}{l|}{\textbf{Context}} & \multicolumn{1}{l|}{\textbf{Technology}} & \multicolumn{1}{l|}{\textbf{Purpose}} & \multicolumn{1}{l|}{\textbf{Context}} & \multicolumn{1}{l|}{\textbf{Technology}} & \multicolumn{1}{l|}{\textbf{Purpose}} & \multicolumn{1}{l}{\textbf{Context}} \\ \hline
\textbf{Results from Marco (Num)}                                                                                    & \multicolumn{9}{c}{5677}                                                                                                                                                                                                                                                                                                                                                      \\ \hline
\textbf{LLM Agreement}                                                                                               & \multicolumn{1}{r|}{4218}                & \multicolumn{1}{r|}{3100}             & \multicolumn{1}{r|}{1369}             & \multicolumn{1}{r|}{5302}                & \multicolumn{1}{r|}{4342}             & \multicolumn{1}{r|}{3133}             & \multicolumn{1}{r|}{4888}                & \multicolumn{1}{r|}{4907}             & 4621                                  \\ \hline
\textbf{LLM Agreement Percent}                                                                                       & \multicolumn{1}{r|}{74.30\%}             & \multicolumn{1}{r|}{54.61\%}          & \multicolumn{1}{r|}{24.11\%}          & \multicolumn{1}{r|}{93.39\%}             & \multicolumn{1}{r|}{76.48\%}          & \multicolumn{1}{r|}{55.19\%}          & \multicolumn{1}{r|}{86.10\%}             & \multicolumn{1}{r|}{86.44\%}          & 81.40\%                               \\ \hline
\textbf{LLM Disagreemnt}                                                                                             & \multicolumn{1}{r|}{1459}                & \multicolumn{1}{r|}{2577}             & \multicolumn{1}{r|}{4308}             & \multicolumn{1}{r|}{375}                 & \multicolumn{1}{r|}{1335}             & \multicolumn{1}{r|}{2544}             & \multicolumn{1}{r|}{789}                 & \multicolumn{1}{r|}{770}              & 1056                                  \\ \hline
\textbf{\begin{tabular}[c]{@{}r@{}}Sample Size Selection\\ Confidence level 95\%\\ Margin of error 5\%\end{tabular}} & \multicolumn{1}{r|}{305}                 & \multicolumn{1}{r|}{336}              & \multicolumn{1}{r|}{354}              & \multicolumn{1}{r|}{191}                 & \multicolumn{1}{r|}{299}              & \multicolumn{1}{r|}{335}              & \multicolumn{1}{r|}{259}                 & \multicolumn{1}{r|}{257}              & 283                                   \\ \hline
\textbf{\begin{tabular}[c]{@{}r@{}}Disagreemnt for Sample\\ (Still think Marco corrent)\end{tabular}}                & \multicolumn{1}{r|}{87.57\%}             & \multicolumn{1}{r|}{82.49\%}          & \multicolumn{1}{r|}{79.10\%}          & \multicolumn{1}{r|}{80.00\%}             & \multicolumn{1}{r|}{75.22\%}          & \multicolumn{1}{r|}{87.46\%}          & \multicolumn{1}{r|}{90.11\%}             & \multicolumn{1}{r|}{81.63\%}          & 80.92\%                               \\ \hline
\textbf{Marco Overall Accuracy}                                                                                     & \multicolumn{1}{r|}{\textbf{96.81\%}}    & \multicolumn{1}{r|}{\textbf{92.05\%}} & \multicolumn{1}{r|}{\textbf{84.14\%}} & \multicolumn{1}{r|}{\textbf{98.68\%}}    & \multicolumn{1}{r|}{\textbf{94.17\%}} & \multicolumn{1}{r|}{\textbf{94.38\%}} & \multicolumn{1}{r|}{\textbf{98.63\%}}    & \multicolumn{1}{r|}{\textbf{97.51\%}} & \textbf{96.45\%}                      \\ \hline
\end{tabular}
}
\end{table*}

We obtained 1387 technologies and 1556 contexts in total. For these extracted results, many duplicates have the same meaning but are in different forms. Thus, we performed the unified and merged process, and then excluded the technologies and contexts whose frequency was less than 2 times to eliminate the noise. Finally, we obtained 450 technologies and 232 contexts (Frequency $  \ge 2 $ in the last 5 years). Specifically, the results of the purpose are too broad and difficult to count the frequency. However, Marco-o1's results are correct, so we asked Marco-o1 model to categorize these purposes into some general purpose categories and validate by humans. Finally, we obtained 11 purpose categories in total.

\begin{boxC} {\textbf{Terms Classification}}
We obtained \textbf{450 technologies, 232 contexts} (Frequency $  \ge 2 $), and \textbf{11 general purpose categories}. 
\end{boxC}

\subsubsection{LLM Technology Classification}
\label{LLMSecond}
To answer \textbf{RQ$_{1.1}$} specifically, we secondly employed the LLMs descrived as above, followed the above rules and steps, to classify 450 technologies extracted in Section~\ref{LLMFirst}. 

Considering our study aims to aid practitioners in performing educated guess when choosing a specific technology in the domain of software architecture. To such aim, we classified each technology based on three dimensions, in which phases they are used in the DevOps pipeline, the deployment environment, and cloud providers as follows:

\begin{itemize}
    \item \textit{DevOps Phases.} This dimension is further divided into technology based on the eight phases of the DevOps framework: plan, code, build, test, release, deploy, operate, and monitor~\citep{bass2015devops}. These stages represent the whole process of activities in software development and operations, and each phase is defined in Table~\ref{tab:devops}. One technology can be classified in more than one DevOps phase. It is interdependent with software architecture, and each stage is closely related to architectural decisions, and the impact of architecture also runs through all stages of DevOps~\citep{bass2015devops,bolscher2019designing, sen2022using}. This classification reflects the operational realities and evolving requirements of the software architecture field.
    \item \textit{Cloud vs. On-premise vs. Both.} Technologies are then classified depending on whether they are applied in cloud development, specific to on-premise, or used in both situations. This dimension reflects the technical environment of each technology. The motivation is currently that the industry is migrating from traditional on-premise systems to cloud-centric architectures as cloud services become more popular~\citep{jamshidi2013cloud,andrikopoulos2013adapt}. A typical example of this evolution is the migration of enterprises from monolithic to microservices~\citep{taibi2017processes}. This classification provides insight into how practitioners' interests have evolved, especially with the growing popularity of cloud applications, and gives us an in-depth understanding of this evolution in the software architecture field.
    \item \textit{Cloud Providers.} Technologies are further classified based on the cloud providers: Amazon Web Services (AWS), Google Cloud (GCP), Microsoft Azure (Azure), and Other. AWS, GCP, and Azure are three main industry cloud providers, and Other means the technology belongs to other cloud providers (e.g, Alibaba) or this technology does not have a specific cloud provider. This dimension can help us understand the cloud providers' distribution, unique preferences or requirements among different cloud platforms, and technical dynamics in the software architecture community.
\end{itemize}

\begin{table}[h]
\centering
\caption{Definition of Eight DevOps Phases from~\citep{bass2015devops}} 
\label{tab:devops}
\resizebox{\linewidth}{!}{
\begin{tabular}{l|p{8cm}}
\hline 
\textbf{Phase} & \textbf{Definition} \\ 
\hline 
\textit{Plan} & Establish objectives, requirements, user stories, and iteration plans. Activities often include backlog creation, roadmap definition, and initial design considerations to guide development.  \\ \hline
\textit{Code} & Implement features and fixes in the codebase. This phase includes version control usage, code reviews, and adherence to coding standards.  \\ \hline
\textit{Build} & Convert source code into a runnable or deployable artifact. Continuous Integration (CI) is central here: automated builds, linting, and initial tests ensure code correctness and integration stability. \\ \hline
\textit{Test }& Validate code quality through automated testing (unit, integration, acceptance), performance tests, and security scans. The objective is to detect regressions and defects early. \\ \hline
\textit{Release} & Package the validated build and prepare for production. This often involves versioning, release notes, and compliance checks.  \\ \hline
\textit{Deploy} & Move the release artifact into production or staging environments. Emphasis is on automation (e.g., container orchestration, infrastructure as code) to minimize human error.  \\ \hline
\textit{Operate} & Keep the system running smoothly in production, handling configuration, infrastructure management, and reliability. Involves runbooks, on-call rotation, and capacity planning.  \\ \hline
\textit{Monitor} & Collect metrics, logs, and other operational data to inform performance analysis, capacity planning, and issue detection. Feedback from monitoring feeds back into planning.  \\ \hline
\end{tabular}
}
\end{table}

\begin{table*}[]
\centering
\footnotesize
\caption{LLM results (Classification - DevOps Phase, Cloud/On-Premise/Both, Cloud Providers)} 
\label{tab:LLM2} 
\resizebox{1\linewidth}{!}{
\begin{tabular}{r|rrrrrrrrr}
\hline
\multicolumn{1}{l|}{\multirow{2}{*}{}}                                                                              & \multicolumn{3}{c|}{\textbf{Qwen}}                                                                                                       & \multicolumn{3}{c|}{\textbf{Mistral}}                                                                                                    & \multicolumn{3}{c}{\textbf{Llama}}                                                                                                      \\ \cline{2-10} 
\multicolumn{1}{l|}{}                                                                                               & \multicolumn{1}{l|}{\textbf{DevOps Phase}} & \multicolumn{1}{l|}{\textbf{Cloud Category}} & \multicolumn{1}{l|}{\textbf{Cloud Provider}} & \multicolumn{1}{l|}{\textbf{DevOps Phase}} & \multicolumn{1}{l|}{\textbf{Cloud Category}} & \multicolumn{1}{l|}{\textbf{Cloud Provider}} & \multicolumn{1}{l|}{\textbf{DevOps Phase}} & \multicolumn{1}{l|}{\textbf{Cloud Category}} & \multicolumn{1}{l}{\textbf{Cloud Provider}} \\ \hline
\textbf{Results from Marco (Num)}   & \multicolumn{1}{r|}{3600}      & \multicolumn{1}{r|}{450}     & \multicolumn{1}{r|}{450}    & \multicolumn{1}{r|}{3600}      & \multicolumn{1}{r|}{450}     & \multicolumn{1}{r|}{450} & \multicolumn{1}{r|}{3600}      & \multicolumn{1}{r|}{450}     & \multicolumn{1}{r}{450}                                                                                                                                                                                                                                                                                                                                                                                                   \\ \hline
\textbf{LLM Agreement}                                                                                               & \multicolumn{1}{r|}{2981}                  & \multicolumn{1}{r|}{391}                     & \multicolumn{1}{r|}{428}                     & \multicolumn{1}{r|}{3092}                  & \multicolumn{1}{r|}{310}                     & \multicolumn{1}{r|}{282}                     & \multicolumn{1}{r|}{3457}                  & \multicolumn{1}{r|}{429}                     & 343                                          \\ \hline
\textbf{LLM Agreement Percent}                                                                                       & \multicolumn{1}{r|}{82.81\%}               & \multicolumn{1}{r|}{86.89\%}                 & \multicolumn{1}{r|}{95.11\%}                 & \multicolumn{1}{r|}{85.89\%}               & \multicolumn{1}{r|}{68.89\%}                 & \multicolumn{1}{r|}{62.67\%}                 & \multicolumn{1}{r|}{96.03\%}               & \multicolumn{1}{r|}{95.33\%}                 & 76.22\%                                      \\ \hline
\textbf{LLM Disagreemnt}                                                                                             & \multicolumn{1}{r|}{619}                   & \multicolumn{1}{r|}{59}                      & \multicolumn{1}{r|}{22}                      & \multicolumn{1}{r|}{508}                   & \multicolumn{1}{r|}{140}                     & \multicolumn{1}{r|}{168}                     & \multicolumn{1}{r|}{143}                   & \multicolumn{1}{r|}{21}                      & 107                                          \\ \hline
\textbf{\begin{tabular}[c]{@{}r@{}}Sample Size Selection\\ Confidence level 95\%\\ Margin of error 5\%\end{tabular}} & \multicolumn{1}{r|}{238}                   & \multicolumn{1}{r|}{52}                      & \multicolumn{1}{r|}{21}                      & \multicolumn{1}{r|}{220}                   & \multicolumn{1}{r|}{103}                     & \multicolumn{1}{r|}{118}                     & \multicolumn{1}{r|}{105}                   & \multicolumn{1}{r|}{20}                      & 84                                           \\ \hline
\textbf{\begin{tabular}[c]{@{}r@{}}Disagreemnt for Sample\\ (Still think Marco corrent)\end{tabular}}                & \multicolumn{1}{r|}{75.52\%}               & \multicolumn{1}{r|}{76.92\%}                 & \multicolumn{1}{r|}{23.81\%}                 & \multicolumn{1}{r|}{35.45\%}               & \multicolumn{1}{r|}{74.76\%}                 & \multicolumn{1}{r|}{64.41\%}                 & \multicolumn{1}{r|}{80\%}                  & \multicolumn{1}{r|}{52.38\%}                 & 90.48\%                                      \\ \hline
\textbf{Marco Overall}                                                                                               & \multicolumn{1}{r|}{\textbf{95.79\%}}      & \multicolumn{1}{r|}{\textbf{96.97\%}}        & \multicolumn{1}{r|}{\textbf{96.28\%}}        & \multicolumn{1}{r|}{\textbf{90.89\%}}      & \multicolumn{1}{r|}{\textbf{92.15\%}}        & \multicolumn{1}{r|}{\textbf{86.71\%}}        & \multicolumn{1}{r|}{\textbf{99.21\%}}      & \multicolumn{1}{r|}{\textbf{97.78\%}}        & \textbf{97.74\%}                             \\ \hline
\end{tabular}
}
\end{table*}

We performed the same approach to obtain the classification results from LRM, evaluation correctness results from validation LLMs, and human validation to get Marco’s overall accuracy. The detailed prompts, scripts, and results are also in the replication package (Section~\ref{Replicability}), and Table~\ref{tab:LLM2} presents the information about the classification results generated by 4 LLMs. To measure the agreement among the validation models and the LRM, we selected samples with 95\% statistical significance, stratified with a 5\% error margin for human validation. Finally, we calculated the LRM's (Marco) overall accuracy. Table~\ref{tab:LLM2} shows the high Marco overall accuracy (on average $  \ge 90\% $ ) and means we can trust the classification results generated by Marco.

\begin{boxC} {\textbf{Technology Classification}}
Through this part, we obtained DevOps Phases, Cloud/On-premise/Both, and Cloud Providers Distributions among 450 Technologies to answer RQ$_{1.1}$. 
\end{boxC}

\subsection{Data Analysis}
\label{sec:Dataanalysis}
In this Section, we report how we analyzed the data to answer our RQs. 

\textbf{Software architecture trends in the past five years (RQ$_1$)}. To answer \textbf{RQ$_{1.1}$}, we considered 450 technologies whose frequency appeared at least twice in the past five years. First, we want to understand these technologies' popularity and importance in the software architecture field over the past five years. We split the technologies into quartiles based on the distribution and focused on the fourth quartile, which includes the most frequent technologies whose frequency is within the 75\%-100\% interval. We can identify the most frequently mentioned technologies over the past five years and the most frequently mentioned technologies by year, so that we can clearly understand the trends and evolutions regarding the popularity and importance of technologies.

Moreover, to better understand the specific roles and ecosystem of these technologies in the software architecture field, as reported in Section~\ref{sec:DataExtraction}, we explored each technology's classification: DevOps Phases, Cloud/On-premise/Both, and Cloud Providers. Through this classification, we can clearly understand these technologies' DevOps applicability phases, deployment environments, and cloud providers. This classification distribution can also help us know the application distribution situation of applicable technologies in the software architecture field, enabling practitioners to make informed decisions when selecting specific technologies in the software architecture field.

Practitioner' talk titles may contain multiple technologies, and the relationship between these technologies of a given practitioner' talk may be either parallel or interconnected. Thus, to answer \textbf{RQ$_{1.2}$}, we explored the connections and co-occurrences between these technologies adopting three methods as follows (more details are presented in Appendix~\ref{sec:Background}):

\begin{itemize}
    \item \textit{Gephi network and visualization analysis} to calculate and visualize the results of technology co-occurrence relationships ~\citep{bastian2009gephi};
    \item \textit{Centrality metrics}  to analyze the technology's centrality in the network created by Gephi (Section~\ref{centralitymeasure});
    \item \textit{Louvain Method for Community Detection}  to detect communities in large networks to obtain the technology' communities~\citep{blondel2008fast} (Section~\ref{louvain}).
\end{itemize}

We used \textit{Gephi}~\citep{bastian2009gephi} to calculate and visualize the results of technology co-occurrence relationships. Gephi generates the co-occurrence network diagram to see the co-occurrence relationship of technology directly; showing the degree of technology based on the font size of technology and thickness of lines to present the connection breadth of technology in the diagram, calculating metrics such as degree, weighted degree, closeness centrality, and betweenness centrality; and performing the community detection to find possible subtopics or subfields in the software architecture field. Considering the technology co-occurrence relationship has no directionality, we planned to generate the undirected diagram. Therefore, a co-occurrence network is a graph where:

    \begin{itemize}
        \item \textbf{Nodes} represent terms (e.g., technologies).
        \item \textbf{Edges} represent how often those terms appear together in the same context (e.g., practioner's talks).
    \end{itemize}

To analyze the technology's centrality, we adopted the three classic \textit{centrality metrics} described in Section~\ref{centralitymeasure} as suggested by \citep{brandes2016maintaining}. 
The weighted degree measures the strength of the vertex's connection. It shows how strongly a technology is connected to other technologies in the network. The technology with a high weighted degree tends to represent that this technology is closely related and frequently co-occurs with multiple technologies, and maybe a core technology with high frequency or relevance. 
Closeness centrality measures the distance of a node in the network. The technology with a high closeness centrality can connect to other technologies faster and often refers to the core concept of the network. 
Betweenness centrality measures the effect of a technology as a ``bridge'' in the shortest path. The technology with high-betweenness centrality indicates that this technology is usually located between different technology groups, which may be the link between different topics. 
Thus, three metrics can show technology's relative importance and centrality from three dimensions: connection strength, proximity, and bridge influence. Combined with the connection breadth (degree) shown in the diagram, we can comprehensively analyze which technologies are more critical and closely related to other technologies in the technology of co-occurrence relationship, and are considered core technologies.

We leveraged the \textit{Louvain Method for Community Detection} (Section~\ref{louvain}), an algorithm to detect communities in large networks, to obtain the technology' communities~\citep{blondel2008fast, DeMeo2011, Sattar2022Scalable}. Louvain's method is based on modularity optimization and can automatically divide technology into different sub-communities. The technology in each sub-community is closely connected. In this study, we considered each \textit{sub-community as a genre}, and each genre has different possible research fields or subtopics in the software architecture field. For instance, if some technologies always appear together, they may belong to or be applied in the same research subfield.

We reviewed the included practitioner' talks and labeled titles that contained multiple extracted technologies. We took the 450 extracted technology sets as the screen criteria to screen the labeled practitioner' talk titles. To better understand the co-occurrence relationship in the software architecture field to answer RQ$_{1.2}$, we finally excluded general purpose fot the technologies like ``AI'' or ``Cloud''. Finally, we found 112 technologies that exist in the co-occurrence relationship.

\textbf{Purposes and context for adopting the extracted technology (RQ$_2$)}.
To address \textbf{RQ$_2$}, we investigated the possible purpose and context of adopting extracted technology in the practitioner' talk titles labeled. As reported in Section~\ref{sec:DataExtraction}, we obtained 11 general purposes and 232 contexts (Frequency $  \ge 2 $). First, we analyzed the purpose and context separately from a single perspective. For the purpose, we analyzed 11 purpose categories definitions and their frequency distribution, and investigated the percentage of each purpose category by year. This allows us to gain an initial understanding of the distribution and proportions of different purpose categories in the included practitioner' talks. For the context, we want to understand the popularity and importance of these contexts in the software architecture field over the past five years. We split the contexts into quartiles based on the distribution and focused on the fourth quartile, which includes the most frequent contexts whose frequency is within the 75\%-100\% interval. We can identify the most frequently mentioned contexts over the past five years and the most frequently mentioned contexts by year, so that we can clearly understand the trends and evolutions regarding the popularity and importance of contexts.

Moreover, we want to investigate the purpose and context of the top technologies in the practitioner' talks. Thus, we planned to use the Sankey diagram to visualize the interconnections and flow between technology, purpose, and context. Specifically, we focused on the top 10 technologies in the fourth quartile (top 25\%) of technology frequency with their purposes and contexts. Considering these technologies are related to many context scenarios, we only focused on the contexts most closely related to the technologies.

\section{Results}
\label{sec:Results}
In this Section, we report the results of the empirical study, discussing them by research question (RQ).
\subsection{Technology Trends (RQ$_{1.1}$)}

For the 450 extracted technologies, we have split the technologies into quartiles based on the frequency distribution and only presented the fourth quartile (the most frequent technologies whose frequency is within the 75\%–100\% interval). We got 120 technologies in the fourth quartile, and the complete list is shown in Appendix A. Table~\ref{tab:top10_quartile_tech} lists the top 10 technologies in the fourth quartile (top 25\%) of the frequency distribution of technologies by year.

\begin{table*}[htb]
  \centering
  \scriptsize 
  \caption{Top 10 technologies in the fourth quartile (top 25\%) of Technologies Frequency Distribution by Year (RQ$_{1.1}$)}
\resizebox{0.95\linewidth}{!}{%
  \begin{tabularx}{\textwidth}{c|X|c|X|c|X|c|X|c|X|c|X|c}
    \hline
    \textbf{Ranking} & \textbf{Total} & \textbf{\#} & \textbf{2020} & \textbf{\#} & \textbf{2021} & \textbf{\#} & \textbf{2022} & \textbf{\#} & \textbf{2023} & \textbf{\#} & \textbf{2024} & \textbf{\#} \\
    \hline
    1 & Kubernetes    & 1054 & Kubernetes    & 202 & Kubernetes    & 200 & Kubernetes    & 161 & Kubernetes    & 230 & Kubernetes    & 261 \\
    2 & Cloud Native  & 584  & Cloud Native  & 71  & Cloud Native  & 124 & Cloud Native  & 129 & Cloud Native  & 146 & Cloud Native  & 114 \\
    3 & Serverless    & 325  & Serverless    & 63  & Serverless    & 47  & Serverless    & 59  & Serverless    & 61  & Cloud         & 102 \\
    4 & Cloud         & 297  & Cloud         & 55  & Container     & 38  & Cloud         & 58  & Cloud         & 58  & Serverless    & 95  \\
    5 & Container     & 203  & Container     & 36  & Service Mesh  & 25  & Container     & 31  & Container     & 37  & Container     & 61  \\
    6 & AI             & 97   & AWS SageMaker & 22  & Cloud         & 24  & Microservices & 23  & AI            & 34  & AI             & 54  \\
    7 & Google Cloud  & 95   & AWS S3        & 19  & Google Cloud  & 14  & Google Cloud  & 16  & Generative AI  & 28  & Google Cloud  & 34  \\
    8 & Opentelemetry & 82   & Anthos        & 18  & Microservices & 13  & WebAssembly   & 15  & Opentelemetry & 21  & WebAssembly   & 33  \\
    9 & Service Mesh  & 81   & Service Mesh  & 17  & AWS SageMaker & 12  & AWS S3        & 14  & WebAssembly   & 18  & Opentelemetry & 32  \\
    10 & Microservices & 80  & Microservices & 15  & AWS S3        & 12  & Prometheus    & 14  & Google Cloud  & 17  & API           & 28  \\
    \hline
  \end{tabularx}
  }
  \label{tab:top10_quartile_tech}
\end{table*}

Kubernetes leads with 1,054 mentions, showing consistent usage across all years and a notable peak in 2024. Following it are Cloud Native, Serverless, Cloud, and Container, consistently ranking among the top five nearly every year. Interestingly, technologies like AWS SageMaker and AWS S3 received many mentions in the early years (2020, 2021), but their popularity began to drop in 2022. They have been overshadowed by emerging technologies such as Generative AI and WebAssembly. AI technology has also seen frequent mentions since 2023, indicating that the industry is placing more focus on the combined use of AI and software architecture. Additionally, other technologies like Microservices, and Service Mesh maintain steady interest without significant fluctuations.

Moreover, we provide a complementary perspective, by classifying 450 technologies according to their roles in the DevOps Phases and their deployment context (Cloud, On-Premise, or Both). The classification distribution is reported in Table~\ref{tab:450Tech}. Interestingly, a large number of technologies are versatile, i.e., 364 of them support at least four, and up to eight, out of eight DevOps phases.

\begin{table}
    \centering
    \scriptsize
    \caption{DevOps Phases, Cloud/On-premise/Both, and Cloud Providers Distributions - 450 Technologies (RQ$_{1.1}$)}
    \label{tab:450Tech}
    \begin{tabular}{l|r r }\hline 
\textbf{DevOps Phases*} & \textbf{\#} & \textbf{\%} \\ \hline 
Plan &13 & 2.89 \\
Code &94 & 3.11\\
Build &365 & 81.11\\
Test &69& 15.33\\
Release &23& 5.11\\
Deploy &424& 94.22\\
Operate &432& 96\\
Monitor &446& 99.11\\\hline 
\textbf{Cloud/On-premise/Both} & \textbf{\#} & \textbf{\%}\\ \hline 
Cloud Specific (C) & 66 & 14.67\\
On-Premise (O) & 12 & 2.67\\
Both (B) & 372 & 82.67\\ \hline
\textbf{Cloud Providers} & \textbf{\#} & \textbf{\%}\\ \hline 
AWS & 90 & 20.00\%\\
GCP & 16 & 3.56\%\\
Azure & 13 & 2.89\%\\
Other & 331 & 73.56\%\\ \hline
\multicolumn{3}{l}{\scriptsize{*One technology can be used in more than one DevOps phase}}
\end{tabular}
\end{table}

More specifically,  most of the technologies relate to the Build (81.11\%), Deploy (94.22\%), Operate (96\%), and Monitor (99.11\%) phases, which are DevOps automation hotspots (Table~\ref{tab:450Tech}). For instance, Kubernetes, the most mentioned tool in general (1054), comes up frequently during the Build, Deploy, Operate, and Monitor stages, but never for the Code and Test stages. Similarly, Docker-centric tools like Containerd, Harbor, or Helm are strongly linked with deployment pipelines. while Monitor features Prometheus, OpenTelemetry, and Jaeger for monitoring and telemetry.

The test phase gathers around 15\% of the technologies. For testing, with Selenium equivalents like Falco, Fluent Bit, and even the newer ones like Keptn being included. However, Plan, Code, and Release is the clear laggard (each less than 10\%). Few technologies are explicitly linked to these three phases, indicating a tooling shortfall or a lessened focus in recent years.

Regarding the environment in which the technologies belong, most of them (82.67\%) refer to both cloud-specific and on-premise situations (Table~\ref{tab:450Tech}). Even some established cloud-native products like AWS S3, AWS EMR, or Anthos are deemed appropriate in hybrid environments. Notably, cloud-specific tools are minimal (14.67\%), i.e., like AWS Lambda, AWS ECS, Azure OpenAI, AWS Redshift, or AWS Graviton, and exclusive on-premise references are virtually non-existent ($<$ 2.67\%), highlighting how the DevOps community immensely favors flexibility and cross-deployment compatibility.

For readability's sake, we do not display tables or figures for all 450 technologies. Instead, we grouped the words into quartiles based on their frequency and present only the fourth quartile, i.e., the top 25 most frequently occurring words (within the 75\%–100\% range). The complete list of these technologies, along with their classification, is available in Appendix A.

\begin{figure}[htb]
    \centering
    \includegraphics[width=\linewidth]{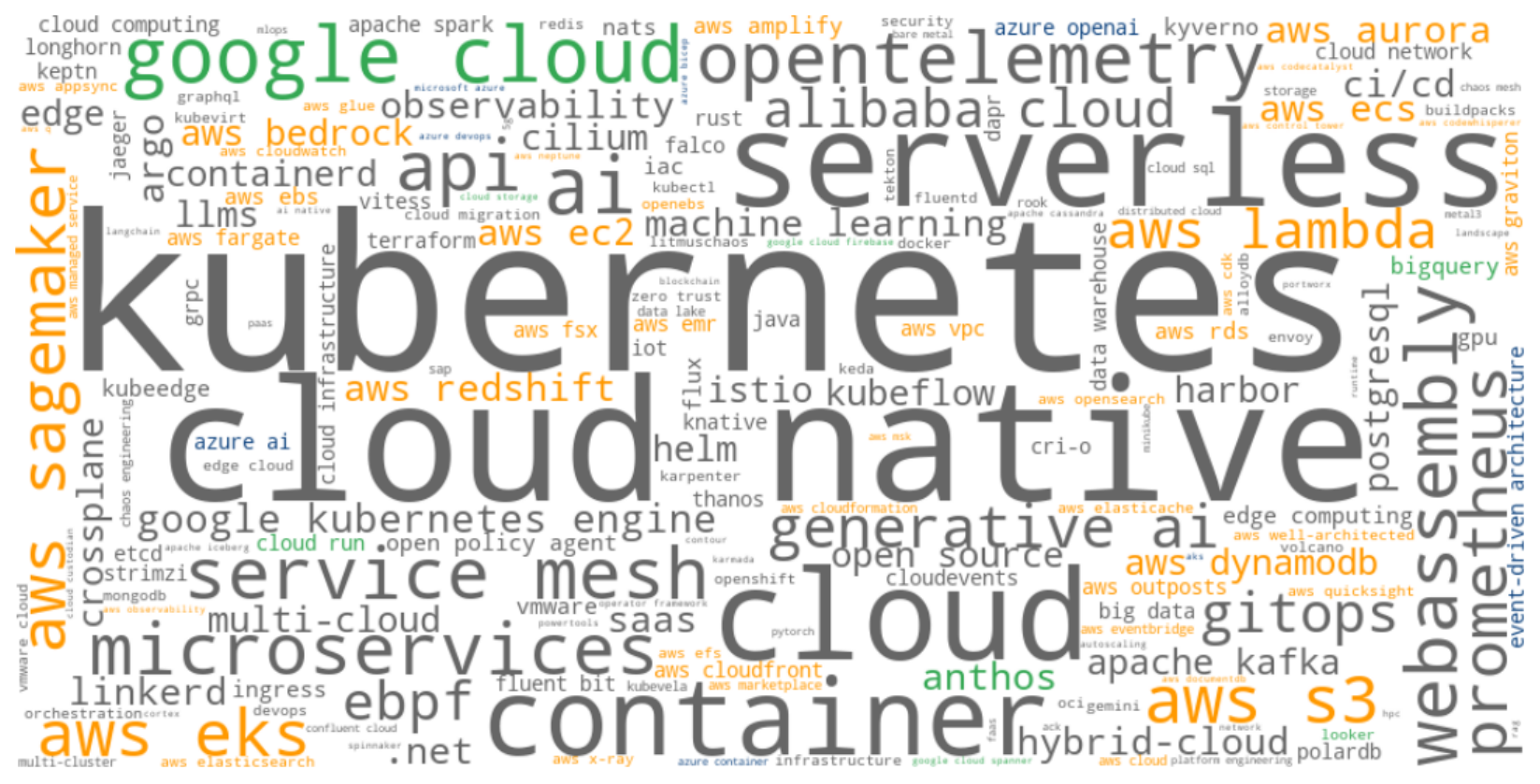}
    \caption{WordCloud for the Technologies (Frequency $  \ge 2 $ -  Orange: AWS; Green: GCP; Blue: Azure, Gray: Other)}
    \label{fig:techfreRQ1}
\end{figure}

We also classified each technology based on the cloud providers: the three largest cloud providers, AWS, GCP, and Azure, and others. Figure~\ref{fig:techfreRQ1} shows a WordCloud that includes the 450 technologies we extracted. The colors in the WordCloud represent their cloud providers: AWS is orange, GCP is green, Azure is blue, and others are gray. The font size of each technology corresponds to its frequency. Kubernetes, Cloud Native, Serverless, Cloud, and Container stand out in the visualization and highlight their significant role and wide use. According to Table~\ref{tab:450Tech}, AWS-specific technologies make up 20\%. On the one hand, technologies such as AWS SageMaker, AWS EKS, AWS S3, and AWS Lambda, show AWS’s influence and overall adotpion. On the other hand, Google Cloud, and Azure, appear less often. Most technologies fall under the "Other" category, meaning they do not belong to the three main cloud service providers thus being cloud agnostic, open source, architectural, or basic technical components.

\begin{figure*}[htb]
    \centering
    \includegraphics[width=\linewidth]{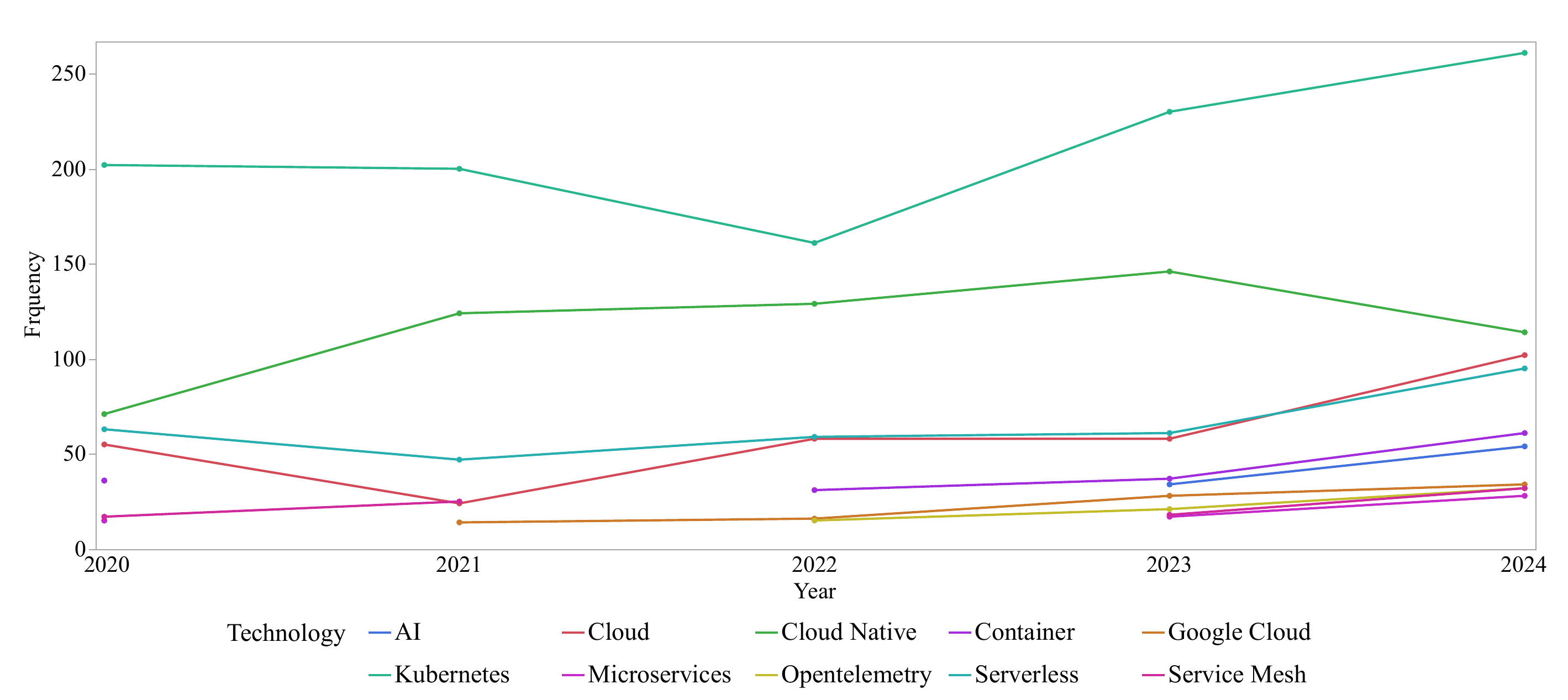}
    \caption{Top 10 technologies in the fourth quartile (top 25\%) by Year (RQ$_{1.1}$)}
    \label{fig:missingTredns}
\end{figure*}

Finally, we analyzed the trends of the top 10 technologies (Figure~\ref{fig:missingTredns}). For each technology, if it was not mentioned in a given year, we used a dotted line to visually highlight the gap in mentions connecting the last year it appeared to the next year in which it was again mentioned.

 Therefore, we observe that \textbf{Kubernetes} dominates across all years, maintaining high and steadily increasing frequency, from 202 in 2020 to 261 in 2024. This suggests it remains a foundational technology in modern software infrastructure while \textbf{Cloud Native} also shows strong and consistent growth, peaking in 2023 with 146 mentions before a slight dip in 2024.
Moreover, \textbf{Serverless} and \textbf{Cloud} technologies follow similar trends: consistently present, but with more modest year-to-year variation. Notably, \textit{Serverless} peaked again in 2024 after a small mid-period decline.

Furthermore, \textbf{AI} emerges only from 2023 onward, with growing frequency (34 in 2023, 54 in 2024), indicating a late but accelerating adoption trend in this technological context. Conversely, \textbf{Google Cloud}, \textbf{Opentelemetry}, and \textbf{Microservices} show \textbf{intermittent patterns}, with zero mentions in some early years and moderate growth later, suggesting either delayed traction or shifting usage contexts. Finally, \textbf{Service Mesh} and \textbf{Container} technologies show earlier relevance (from 2020 onward), but with modest growth, potentially pointing to maturation or niche adoption.

\begin{keyTakeAways}[\textbf{Most Prominent Technologies}]
The most prominent technologies are Kubernetes, Cloud Native, Serverless, and Container, consistently ranking highest in frequency. Recent years show a clear rise in Generative AI and WebAssembly, indicating a growing interest in intelligent and scalable architectures. Most technologies align with the Build, Deploy, Operate, and Monitor DevOps phases. A vast majority (82.7\%) are hybrid-compatible, showing the field’s emphasis on flexibility, with AWS-specific tools leading among cloud providers.
\end{keyTakeAways}

\subsection{Software Technology co-occurrences (RQ$_{1.2}$)}

In this section, to understand the relationships and co-occurrences between the extracted technologies to answer RQ$_{1.2}$, we analyzed the technology's co-occurrence network, the technology's centrality, and the potential subfield or subtopic (genre) in the software architecture field.

\subsubsection{Software Technology co-occurrence network}

\begin{figure*}
    \centering
    \includegraphics[width=\linewidth]{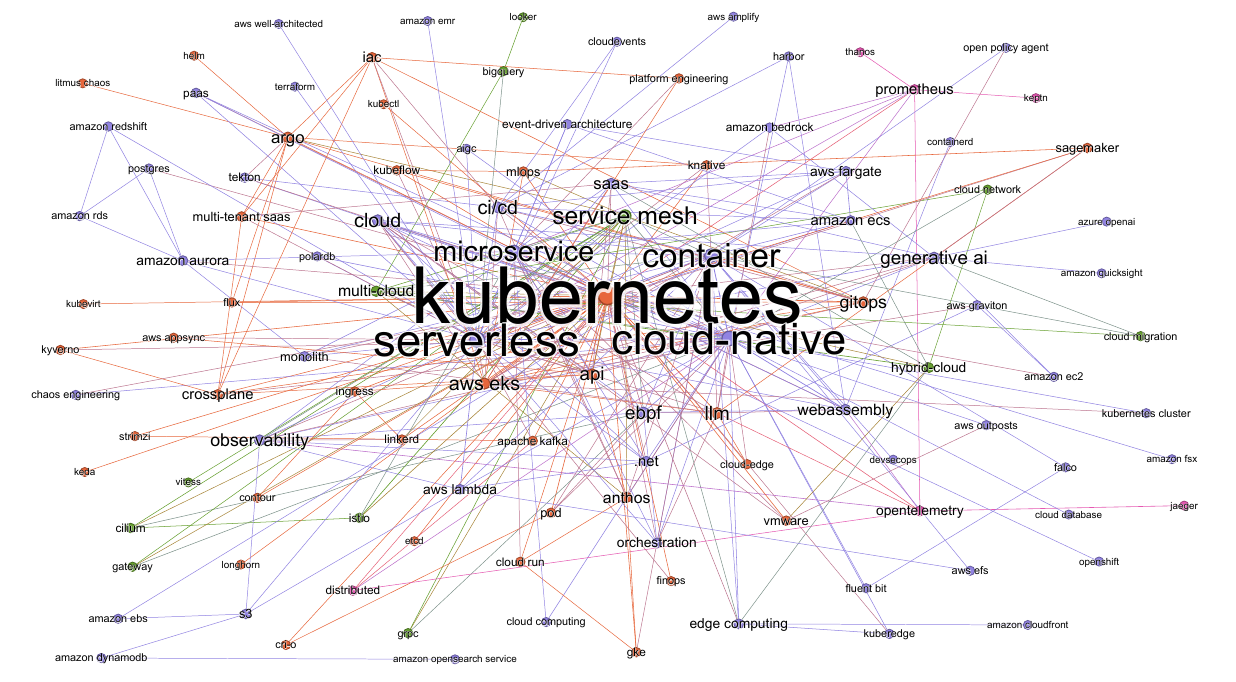}
    \caption{Technology co-occurrence network (RQ$_{1.2}$)}
    \label{fig:technologiesnetwork}
\end{figure*}

We utilized Gephi to visualize the technology co-occurrence network (Figure~\ref{fig:technologiesnetwork}) and to analyze the degree centrality and connectivity patterns among technologies. The network exhibits typical characteristics for its size, with a diameter of 6, an average path length of 2.504, and a graph density of 0.055. The average degree across nodes is 6.054, while the average weighted degree reaches 13.071. Notably, Kubernetes emerges as the most connected technology, with the highest degree (64) and highest weighted degree (217). 



The \textbf{font size of each technology label} in Figure~\ref{fig:technologiesnetwork} reflects its \textbf{degree}, which represents the number of direct connections a technology has with others, indicating its \textbf{connection breadth} within the network. In contrast, the \textbf{weighted degree} accounts for the \textbf{frequency} of co-occurrence (edge weights), thereby capturing the \textbf{strength} of associations. Additionally, the \textbf{thickness of the connecting lines} between nodes visualizes co-occurrence frequency, while the \textbf{coloring is purely aesthetic} and carries no analytical meaning. The layout follows a force-directed arrangement, placing \textbf{high-degree technologies toward the center} and lower-degree ones toward the periphery.

From this network, several technologies stand out as \textbf{central hubs}, including \textit{Kubernetes}, \textit{Serverless}, \textit{Cloud Native}, and \textit{Container}. Their prominent font size and extensive connections suggest they serve as \textbf{core topics} in practitioner discussions, especially in terms of connectivity across the technology landscape. Surrounding these are \textbf{moderately prominent technologies} such as \textit{Microservice}, \textit{Service Mesh}, and \textit{Cloud}, which exhibit strong ties to the core set. Their relatively high degrees suggest they frequently co-occur in contexts involving \textbf{cloud infrastructure} and \textbf{software architecture}.

Interestingly, certain technologies, such as \textit{Prometheus} and \textit{AWS S3}, while not exhibiting high co-occurrence frequencies overall, are \textbf{directly linked to core technologies}. This pattern implies a specialized relevance; for example, \textit{Prometheus} is a widely adopted monitoring tool tightly integrated with \textit{Kubernetes}.

Moreover, the network reveals a cluster of technologies prefixed with ``\textbf{Amazon}'' or ``\textbf{AWS},'' such as \textit{Amazon EBS} and \textit{AWS EKS}. These are offerings from \textbf{Amazon Web Services (AWS)} that support specific subdomains of the cloud ecosystem, highlighting AWS’s extensive role in shaping cloud-native architectures.

Finally, the graph includes a subset of \textbf{general-purpose technologies} that, while not central in terms of frequency, play important roles across domains. For instance, \textit{eBPF} is a versatile technology used in \textbf{cloud-native systems} to optimize networking and enhance security strategies, but its applicability also extends to other fields beyond software architecture.

\begin{keyTakeAways}[\textbf{Technology Co-Occurrence Network}]
\textit{Kubernetes, Cloud Native, Serverless, and Container are the core technologies in the cloud network.}
\end{keyTakeAways}

\subsubsection{Software Technology Centrality Analysis}
\label{rq2}

\begin{figure*}[!ht]
    \centering
    \resizebox{\linewidth}{!}{%
    \begin{subfigure}[t]{0.45\linewidth}
    \centering
    \begin{tikzpicture}
    \begin{axis}[
        width=\linewidth,
        height=6cm,
        ybar,
        bar width=7pt,
        enlarge x limits=0.03,
        symbolic x coords={kubernetes,cloud-native,serverless,container,microservice,service mesh,llm,api,webassembly,ci/cd,cloud,ebpf,gitops,aws eks,generative ai},
        xtick=data,
        xticklabel style={rotate=45, anchor=east,},
        ymin=0,
        ymax=250,]
    
    \addplot+[style={googleblue, fill=googleblue}] 
    coordinates {
        (kubernetes,217)
        (cloud-native,155)
        (serverless,107)
        (container,86)
        (microservice,57)
        (service mesh,48)
        (llm,29)
        (api,29)
        (webassembly,29)
        (ci/cd,24)
    };
    \end{axis}
    \end{tikzpicture}
    \caption{Ranking of Technology’s Weighted Degree ($D_W(v)$)}
    \label{fig:weighteddegree}
    \end{subfigure}
    \hfill
    \begin{subfigure}[t]{0.45\linewidth}
    \centering
    \begin{tikzpicture}
    \begin{axis}[
        width=\linewidth,
        height=6cm,
        ybar,
        bar width=7pt,
        enlarge x limits=0.03,
        symbolic x coords={kubernetes,serverless,cloud-native,container,microservice,service mesh,llm,aws eks,ci/cd,cloud,api,generative ai,observability,webassembly,ebpf,saas},
        xtick=data,
        xticklabel style={rotate=45, anchor=east},
        ymin=0,
        ymax=0.8,]
    
    \addplot+[style={googleblue, fill=googleblue}] 
    coordinates {
        (kubernetes,0.68)
        (serverless,0.6)
        (cloud-native,0.57)
        (container,0.54)
        (microservice,0.53)
        (service mesh,0.51)
        (llm,0.5)
        (aws eks,0.5)
        (ci/cd,0.5)
        (cloud,0.5)
    };
    \end{axis}
    \end{tikzpicture}
    \caption{Ranking of Technology’s Closeness Centrality ($C_C(v)$)}
    \label{fig:closeness}
    \end{subfigure}
    \hfill
    \begin{subfigure}[t]{0.45\linewidth}
    \centering
    \begin{tikzpicture}
    \begin{axis}[
        width=\linewidth,
        height=5.7cm,
        ybar,
        bar width=7pt,
        enlarge x limits=0.03,
        symbolic x coords={kubernetes,serverless,cloud-native,container,cloud,generative ai,microservice,service mesh,prometheus,argo,s3,ci/cd,aws eks,webassembly,edge computing},
        xtick=data,
        xticklabel style={rotate=45, anchor=east},
        ymin=0,
        ymax=3000,]
    
    \addplot+[style={googleblue, fill=googleblue}] 
    coordinates {
        (kubernetes,2557)
        (serverless,1381)
        (cloud-native,891)
        (container,686)
        (cloud,337)
        (generative ai,283)
        (microservice,270)
        (service mesh,269)
        (prometheus,238)
        (argo,234)
    };
    \end{axis}
    \end{tikzpicture}
    \caption{Ranking of Technology’s Betweenness Centrality ($C_B(v)$)}
    \label{fig:betweenness}
    \end{subfigure}%
    }

    \caption{Comparison of Centrality Measurement Rankings (RQ\textsubscript{1.2}).}
    \label{fig:centrality_comparison}
\end{figure*}
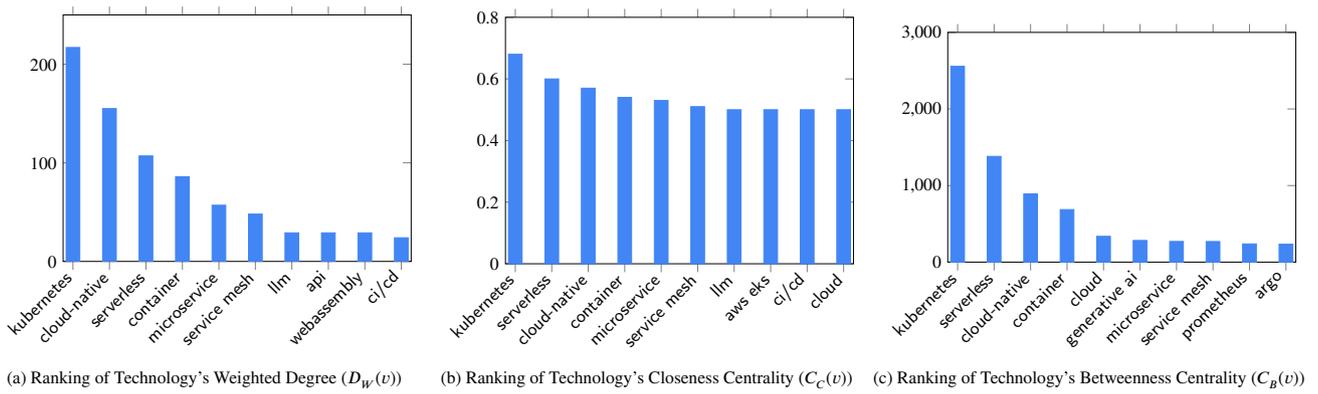

Following the construction of the technology co-occurrence network, we computed three centrality metrics, Weighted Degree ($D_W(v)$), Closeness Centrality ($C_C(v)$), and Betweenness Centrality ($C_B(v)$), to assess the relative importance of technologies from multiple perspectives (Figure~\ref{fig:centrality_comparison}). 

The \textbf{Weighted Degree} ($D_W(v)$) ranking (Figure~\ref{fig:weighteddegree}) captures the \textbf{strength of each technology’s co-occurrence with others} in the network. Kubernetes emerges as the most frequently co-occurring technology, with the highest weighted degree (217.0), followed by Cloud Native (155.0), Serverless (107.0), and Container (86.0). These values highlight strong correlations among these technologies, underscoring their prominence as central topics in practitioner discussions at industry conferences. Other highly ranked technologies fall into two broad categories: foundational architectural concepts such as Microservice and Service Mesh, and enabling technologies like LLM, which support software architecture tasks such as natural language interaction and code generation.

The \textbf{Closeness Centrality} ($C_C(v)$) ranking (Figure~\ref{fig:closeness}) measures each technology’s \textbf{proximity to all others in the network}, indicating how efficiently it can reach the rest of the graph. Once again, Kubernetes leads with a closeness score of 0.677, followed by Serverless (0.594), ``\textbf{Cloud Native}'' (0.569), and ``\textbf{Container}'' (0.541). These high values suggest that these technologies are centrally positioned and strongly interrelated with others. Notably, AWS EKS and Cloud show a marked increase in this ranking despite not appearing in the weighted degree ranking, indicating that while they may not be frequently mentioned, they maintain high proximity to a broad range of technologies, suggesting structural importance. Conversely, API and WebAssembly appear in the weighted degree ranking but not in the closeness centrality list, implying that although they co-occur frequently, they are less central in terms of network reachability.

The \textbf{Betweenness Centrality} ($C_B(v)$) ranking (Figure~\ref{fig:betweenness}) evaluates each technology’s role as a \textbf{bridge or intermediary between different parts of the network},i.e. different technology deployments and purpose. As expected, Kubernetes dominates, with a betweenness score of 2557.42, substantially higher than any other technology. ``Serverless'' (1381.23), ``Cloud Native'' (891.49), and ``Container'' (686.17) also demonstrate substantial bridging influence, linking diverse technology clusters. Interestingly, Generative AI, Prometheus, and Argo appear uniquely in this ranking, despite not being featured in the weighted degree or closeness rankings. This suggests their primary role lies in connecting otherwise disparate technologies, rather than frequent or proximal interactions.

Taken together, these three centrality measures consistently identify Kubernetes as the most central technology in the network. Cloud Native, Serverless, and Container also demonstrate strong influence across all dimensions, connection strength, proximity, and bridging capability. More specifically, such four \textbf{core technologies} are connected to 89 of the other 108 technologies. Therefore according to \cite{wen2011improving}, the community analysis would focus on all the other technology excluding the four core ones to avoid skewing the modularity of the graph.  Along with the four core technologies we also removed 13 additional technologies that were exclusively linked to these core nodes, their removal yielded a refined graph comprising 95 technologies and 175 co-occurrence edges.

\begin{figure}
    \centering
    \includegraphics[width=\linewidth]{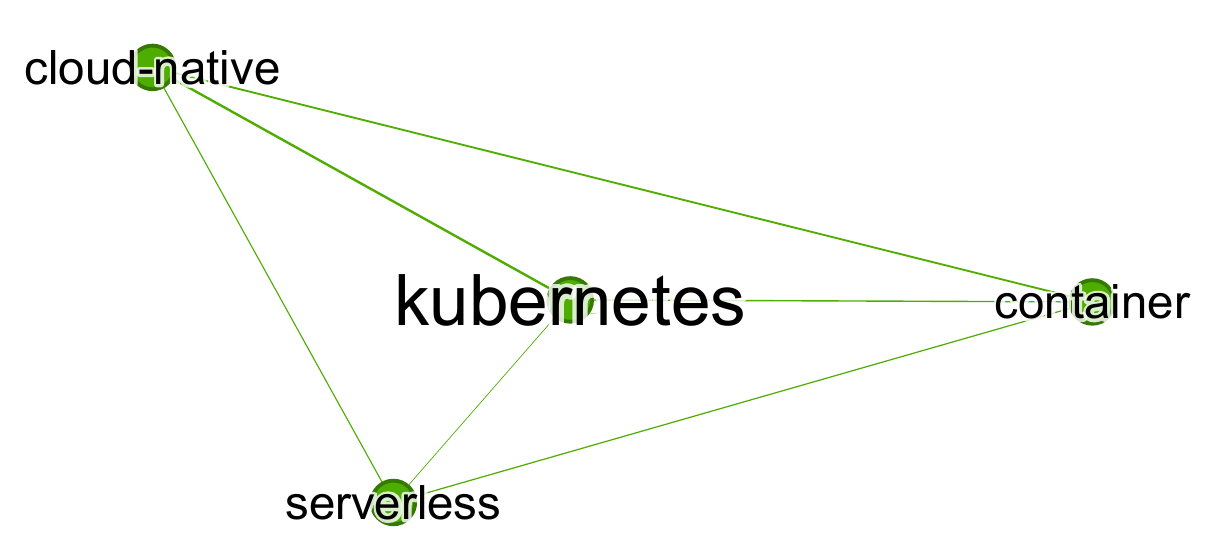}
    \caption{The Connections Between Four Core technologies in the Network (RQ$_{1.2}$)}
    \label{fig:fourcoretags}
\end{figure}technologies and provide data support for the community detection process below.

\begin{keyTakeAways}[\textbf{Centrality Analysis}]
\textit{The core technologies (Kubernetes, Cloud Native, Serverless, and Container) are connected to over 80\% of technologies in the network.}
\end{keyTakeAways}

\subsubsection{Community Detection}
To identify latent technology cliques, i.e., cohesive communities, within practitioner' talks on SA, we employed the Louvain method for community detection. We base our analysis on the previously refined graph that achieved a modularity score of $Q = 0.551$ across 100 iterations, significantly higher than the original score of $Q = 0.262$, indicating the presence of well-defined and robust communities~\citep{fortunato2007resolution,newman2006modularity}. Although the resulting communities are not entirely disconnected, their internal link density exceeds that of inter-community connections, reflecting the formation of coherent clusters.

To characterize each of the five communities, we analyzed their most central technologies using three metrics: weighted degree ($D_W$), closeness centrality ($C_C$), and betweenness centrality ($C_B$). More specifically, $D_W$ captures how strongly a technology is connected within its community, $C_C$ reflects how quickly a technology can reach others across the network, and $C_B$ identifies technologies that act as bridges between otherwise separate clusters.

\textbf{Community 1: Architecture Deployment and Infrastructure Automation}

\begin{figure}
    \centering
    \includegraphics[width=\linewidth]{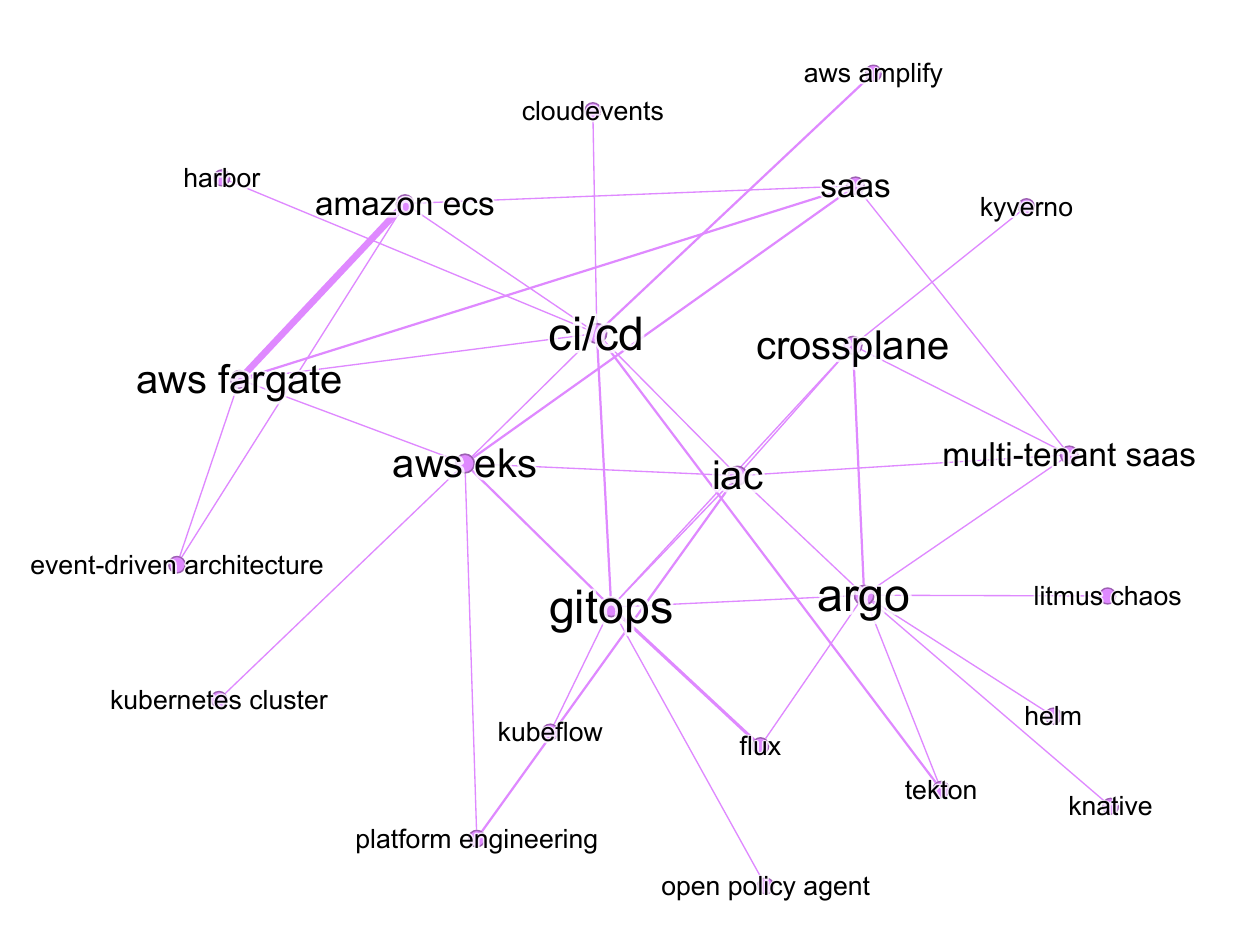}
    \caption{C1 - Architecture Deployment and Infrastructure Automation Community (RQ$_{1.2}$)}
    \label{fig:genre1}
\end{figure}

The first community (\textbf{C1}) includes 24 technologies, making up 25.3\% (Figure~\ref{fig:genre1}) of the total, along with 40 co-occurrence links. We present the ranks of the top ten technologies based on the three centrality metrics in Table~\ref{tab:genre1}.

C1 focuses on providing infrastructure, orchestrating resources, and automating deployments. Technologies like AWS EKS, GitOps, and CI/CD rank highly in all three centrality metrics, showing their key role in practitioner' talks. This importance goes beyond mere frequency; it highlights their structural significance. AWS EKS and GitOps demonstrate both high weighted centrality and betweenness centrality, hinting at their roles as main ideas and connections in automated deployment discussions. CI/CD, Argo, and SaaS also serve as important links in various conversations, showing their contribution to integrating and amplifying deployment practices.

Interestingly, while they appear less frequently, Crossplane, Harbor, and Event-Driven Architecture display a notably high betweenness. This suggests they connect, i.e., acting as bridges, otherwise separate subtopics. These are not leading trends but facilitate shifts between themes such as cloud-native infrastructure and security.

All in all, C1 highlights a genre that stresses automation at different levels: infrastructure-as-code, container orchestration, deployment pipelines, and multi-tenant scaling.

\begin{table}
    \centering
    \caption{Architecture Deployment and Infrastructure Automation: Rank of Technology's Centrality Measurement (RQ$_{1.2}$)}
    \label{tab:genre1}
    \resizebox{\linewidth}{!}{
    \begin{tabular}{l r|l r|l r}
    \hline
       \textbf{Technology} & \textbf{$D_W(v)$} & \textbf{Technology} & \textbf{$C_C(v)$} & \textbf{Technology} & \textbf{$C_B(v)$} \\
        \hline
        AWS EKS & 14 & GitOps & 0.39 & AWS EKS & 419.62 \\
        GitOps & 14 & AWS EKS & 0.39 & GitOps & 412.59 \\
        CI/CD & 13 & SaaS & 0.39 & Argo & 355.49 \\
        Amazon ECS & 12 & CI/CD & 0.35 & CI/CD & 312.36 \\
        AWS Fargate & 12 & Amazon ECS & 0.35 & SaaS & 164.56 \\
        SaaS & 11 & Platform Engineering & 0.35 & Crossplane & 123.53 \\
        Argo & 11 & Argo & 0.34 & Amazon ECS & 70.48 \\
        IaC & 8 & IaC & 0.34 & IaC & 57.25 \\
        Crossplane & 7 & Crossplane & 0.34 & Platform Engineering & 21.87 \\
        Platform Engineering & 4 & Event-driven & 0.32 & Harbor & 17.17 \\
        \hline
    \end{tabular}
}
\end{table}

\textbf{Community 2: Service Communication and Distributed Systems.}

\begin{figure}
    \centering
    \includegraphics[width=\linewidth]{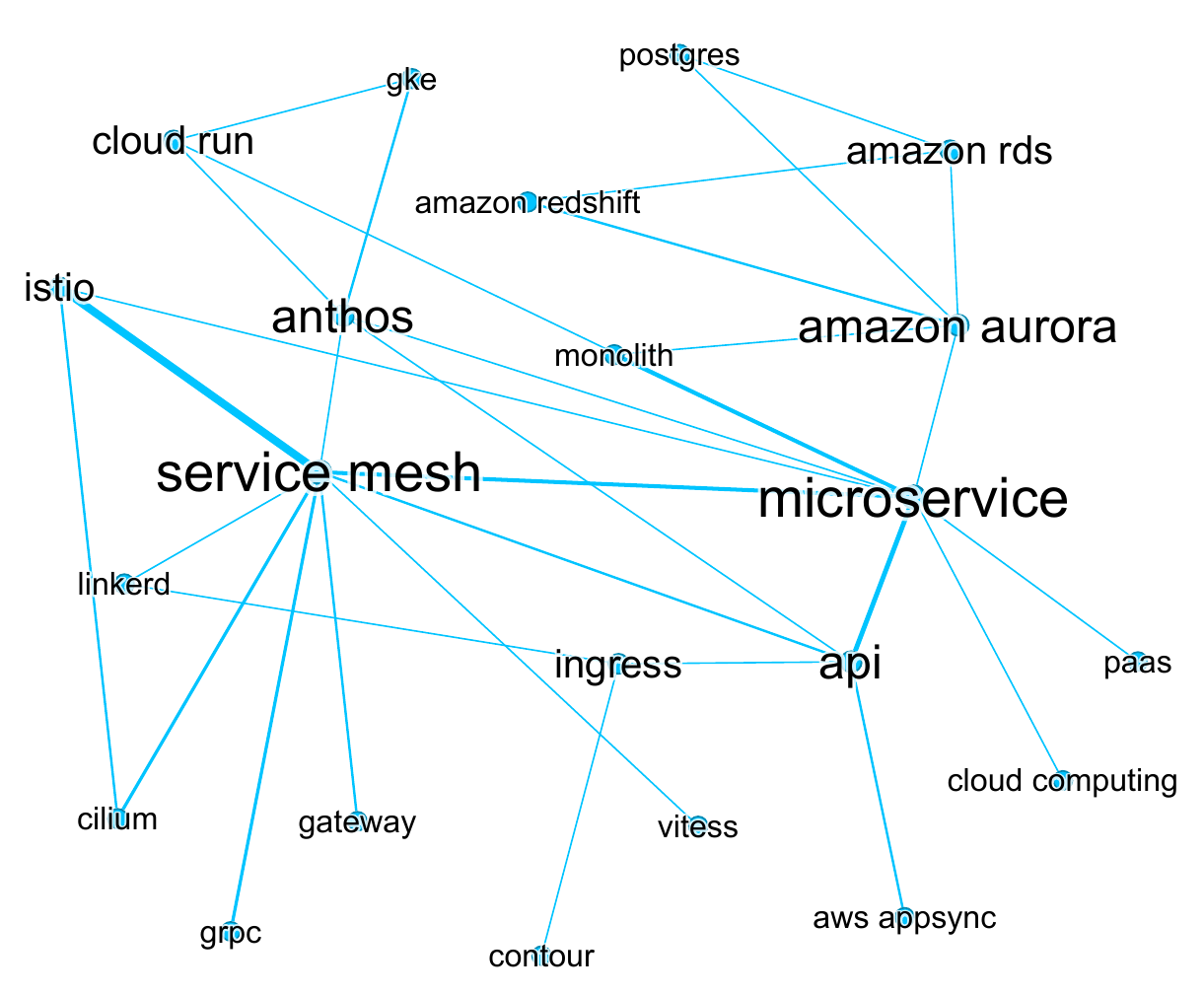}
    \caption{C2 - Service Communication and Distributed Systems Community (RQ$_{1.2}$)}
    \label{fig:genre2}
\end{figure}

The second community, (\textbf{C2}) includes 22 technologies (23.2\%) and 32 co-occurrence links (Figure~\ref{fig:genre2}). We present the ranks of the top ten technologies based on the three centrality metrics in Table~\ref{tab:genre2}. We note that the strongest co-occurrence is between Istio and Service Mesh, with 9 connections.

C2 centers on technologies enabling communication across distributed systems, with Microservice and Service Mesh emerging as structurally dominant. These technologies lead in all three centrality metrics, hinting at their central role in practitioner' talks on modern architectural decomposition and operational coordination. Moreover, API and gRPC also score high across all measures, suggesting a continuous focus on communication layers and interoperability standards.

On the one hand, while Microservice often appears as a standalone architectural concept, its frequent ties to Monolith (moderate $C_B$) suggest continued interest in migration strategies, a theme echoed in several tools like Amazon Aurora, which bridge monolithic storage with distributed scalability. Aurora, with its strong betweenness, reflects its utility as a transitional backend within microservices adoption practitioner' talks.

On the other hand, Istio, Linkerd, and Cilium, although all key service mesh technologies, occupy slightly different roles in the network. Istio stands out for both $D_W$ and $C_B$, marking it as a well-connected and cross-topic technology. Linkerd and Cilium, with lower overall connectivity but non-negligible $C_B$, seem to serve more specific use cases, acting as alternatives or complements within certain subsets of practitioner' talks.

Technologies like Anthos and Cloud Run show up as bridges between communication tooling and cloud-native execution environments. Though not among the highest in $D_W$, their elevated $C_B$ suggests they act as a bridge between infrastructure provisioning and service communication.

Overall, this community reflects a tight conceptual cluster around the mechanics of inter-service communication and orchestration in distributed systems. Our findings highlight the key role of service meshes, APIs, and microservices in shaping practitioner' talks on scalable and maintainable architectures.

\begin{table}[]
    \centering
    \caption{Service Communication and Distributed Systems: Rank of Technology's Centrality Measurement (RQ$_{1.2}$)}
    \label{tab:genre2}
    \resizebox{\linewidth}{!}{
    \begin{tabular}{l r|l r|l r}
    \hline
        \textbf{Technology} & \textbf{$D_W(v)$} & \textbf{Technology} & \textbf{$C_C(v)$} & \textbf{Technology} & \textbf{$C_B(v)$} \\
        \hline
Service Mesh	&	35	&	Microservice	&	0.45	&	Microservice	&	1203.59	\\
Microservice	&	33	&	Service Mesh	&	0.44	&	Service Mesh	&	1008.36	\\
API	&	16	&	API	&	0.41	&	API	&	488.25	\\
Istio	&	12	&	Anthos	&	0.39	&	Amazon Aurora	&	267.5	\\
Anthos	&	8	&	Istio	&	0.34	&	Anthos	&	192.57	\\
Cilium	&	7	&	Monolith	&	0.33	&	Ingress	&	95.62	\\
Monolith	&	6	&	Linkerd	&	0.33	&	Linkerd	&	44.04	\\
Amazon Aurora	&	6	&	Cilium	&	0.32	&	Monolith	&	28.62	\\
gRPC	&	4	&	gRPC	&	0.32	&	Cloud Run	&	17.5	\\
Linkerd	&	3	&	Amazon Aurora	&	0.31	&	Istio	&	5.79	\\
        \hline
    \end{tabular}
    }
\end{table}

\textbf{Community 3: Cloud AI and Serverless Computing.} 

\begin{figure}
    \centering
    \includegraphics[width=\linewidth]{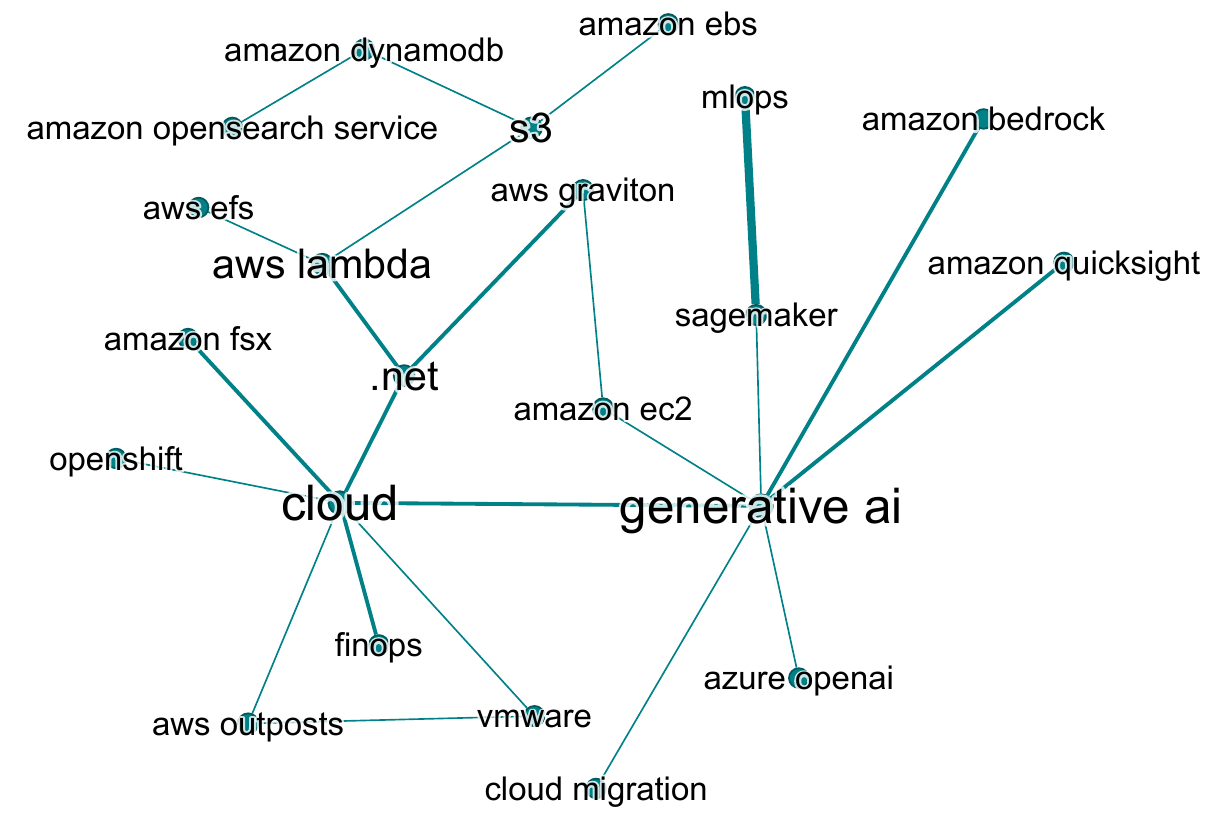}
    \caption{C3 - Cloud AI and Serverless Computing Community (RQ$_{1.2}$)}
    \label{fig:genre3}
\end{figure}

The third community, (\textbf{C3}) includes 22 technologies (23.2\%) and 23 co-occurrence links (Figure~\ref{fig:genre3}). According to Table~\ref{tab:genre3}, C3 is led by Cloud and Generative AI, with a strong link between MLOps and AWS SageMaker.

C3 shows a mix of cloud-native services, AI platforms, and serverless technologies. As expected, ``Cloud'' leads in all three centrality metrics, suggesting its essential role in practitioner' talks on infrastructure, storage, and computation. Surprisingly, in less than five years, ``Generative AI'' ranks second in both $D_W$ and $C_C$. It stands out as a key theme, often mentioned alongside tools that help deploy and scale large models.

Several technologies in this community enable AI workflows in cloud settings. MLOps, AWS SageMaker, and Amazon Bedrock hold central positions not because they are generally popular, but because they often come up in practitioner' talks about automating, training, and integrating AI pipelines. Their prominence reflects the increased effort to make AI usable through scalable, cloud-based solutions.

On the serverless side, AWS Lambda and Amazon DynamoDB show strong betweenness centrality. This reveals their role as flexible connectors between event-driven logic and persistent data management. Technologies like .Net and VMware, while not directly linked to AI, appear in key bridge positions ($C_B$ and $C_C$). This likely happens because they fit into hybrid or enterprise setups where cloud AI workloads get added to existing systems.

Other important technologies acting as bridges are AWS Graviton and AWS S3, which support efficient computing and storage in both AI and serverless areas. Although they are not the most frequent, their position highlights their usefulness in creating multi-service AI architectures.

Overall, the centrality metrics indicate that this community reflects active and ongoing discussions about implementing AI in the cloud, especially through low-infrastructure and event-driven models. 

\begin{table}
    \centering
    \caption{Cloud AI and Serverless Computing: Rank of Technology's Centrality Measurement (RQ$_{1.2}$)}
    \label{tab:genre3}
    \resizebox{\linewidth}{!}{
    \begin{tabular}{l r|l r|l r}
    \hline
        \textbf{Technology} & \textbf{$D_W(v)$} & \textbf{Technology} & \textbf{$C_C(v)$} & \textbf{Technology} & \textbf{$C_B(v)$} \\
        \hline
Cloud	&	16	&	Cloud	&	0.38	&	Cloud	&	461.67	\\
Generative AI	&	13	&	.Net	&	0.36	&	Genrative AI	&	356.79	\\
.Net	&	9	&	Generative AI	&	0.36	&	AWS S3	&	269	\\
MLOps	&	7	&	MLOps	&	0.34	&	.Net	&	125.51	\\
AWS SageMaker	&	6	&	Vmware	&	0.34	&	AWS Lambda	&	105.05	\\
AWS Lambda	&	5	&	AWS SageMaker	&	0.31	&	Amazon Dynamodb	&	91	\\
Vmware	&	5	&	AWS Lambda	&	0.31	&	Vmware	&	61.13	\\
Amazon Bedrock	&	5	&	Amazon Bedrock	&	0.3	&	MLOps	&	61.01	\\
AWS S3	&	4	&	AWS Graviton	&	0.3	&	Amazon Bedrock	&	38.89	\\
AWS Graviton	&	4	&	AWS S3	&	0.3	&	AWS Graviton	&	26.14	\\
        \hline
    \end{tabular}
    }
\end{table}

\textbf{Community 4: Cloud Observability, Security, and Performance Optimization.} 

\begin{figure}
    \centering
    \includegraphics[width=\linewidth]{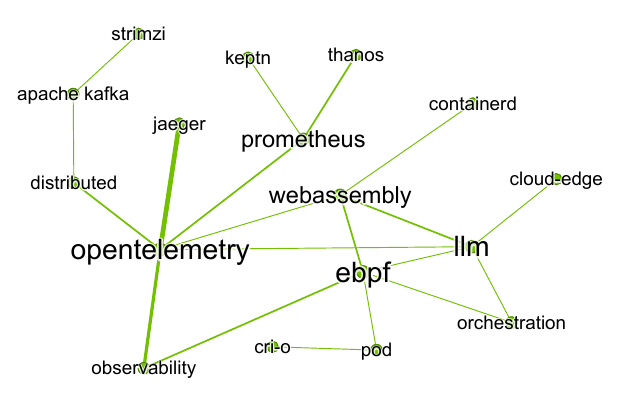}
    \caption{C4 - Cloud Observability, Security, and Performance Optimization Community (RQ$_{1.2}$)}
    \label{fig:genre4}
\end{figure}

The fourth community, (\textbf{C4}) highlighted in green (Figure~\ref{fig:genre4}), includes 17 technologies (17.9\%) and 20 co-occurrence connections.  The pair with the strongest co-occurrence is Jaeger and OpenTelemetry, with 6 links (Table~\ref{tab:genre4}).

C4 focuses on monitoring, securing, and improving performance in cloud-native environments. OpenTelemetry and ``Observability'' lead all centrality measures. This highlights their role as unifying elements in metrics, logs, and distributed tracing conversations. Their strong $C_B$ and $C_C$ suggest that these technologies often appear across different toolchains, linking various topics like performance monitoring, root-cause detection, and system introspection.

On the one hand, eBPF has high values across all centralities, showing its broad application in low-level monitoring, security, and networking. Its presence in discussions about system performance and container orchestration (e.g., Pods, Orchestration) reflects its versatility and importance for modern optimization strategies. LLM, while less central in earlier communities, emerges here as a bridge ($C_B$ = 321.5) that links topics like AI workloads to infrastructure-level issues like container behavior and cloud-edge deployment.

On the other hand, Prometheus and Jaeger also have strong $C_B$ scores, acting as key components in tracing and alerting pipelines. Meanwhile, WebAssembly, Pod, and Apache Kafka are more specialized technologies, each connecting observability tools and platform-specific performance management.

Finally, the operational cornerstone of C4 is Containers, Cloud-Edge, and Orchestration technologies since their moderate centralities suggest that they support more prominent players like OpenTelemetry and eBPF in their roles within distributed systems.

All in all, this community demonstrates a well-organized set of technologies that appear in practitioner' talks aimed at making cloud systems observable, secure, and efficient, particularly in environments that include containers, edge nodes, and AI-driven workloads.
\begin{table}
    \centering
    \caption{Cloud Observability, Security, and Performance Optimization: Rank of Technology's Centrality Measurement (RQ$_{1.2}$)}
    \label{tab:genre4}
    \resizebox{\linewidth}{!}{
    \begin{tabular}{l r|l r|l r}
    \hline
        \textbf{Technology} & \textbf{$D_W(v)$} & \textbf{Technology} & \textbf{$C_C(v)$} & \textbf{Technology} & \textbf{$C_B(v)$} \\
        \hline
OpenTelemetry	&	16	&	Observability	&	0.41	&	Observability	&	706.57	\\
Observability	&	14	&	eBPF	&	0.4	&	OpenTelemetry	&	400.07	\\
eBPF	&	13	&	LLM	&	0.37	&	eBPF	&	360.50	\\
LLM	&	12	&	WebAssembly	&	0.35	&	LLM	&	321.51	\\
WebAssembly	&	9	&	Orchestration	&	0.35	&	Prometheus	&	231.03	\\
Prometheus	&	7	&	OpenTelemetry	&	0.33	&	WebAssembly	&	180.81	\\
Jaeger	&	6	&	Prometheus	&	0.33	&	Distributed	&	180	\\
Orchestration	&	4	&	Pod	&	0.29	&	Pod	&	91	\\
Distributed	&	3	&	Cloud-Edge	&	0.27	&	Apache Kafka	&	91	\\
Pod	&	2	&	Containerd	&	0.26	&	Orchestration	&	13.06	\\
        \hline
    \end{tabular}
    }
\end{table}

\textbf{Community 5: Cross-Cloud Architecture and Cloud-Edge Collaboration.} 

\begin{figure}
    \centering
    \includegraphics[width=\linewidth]{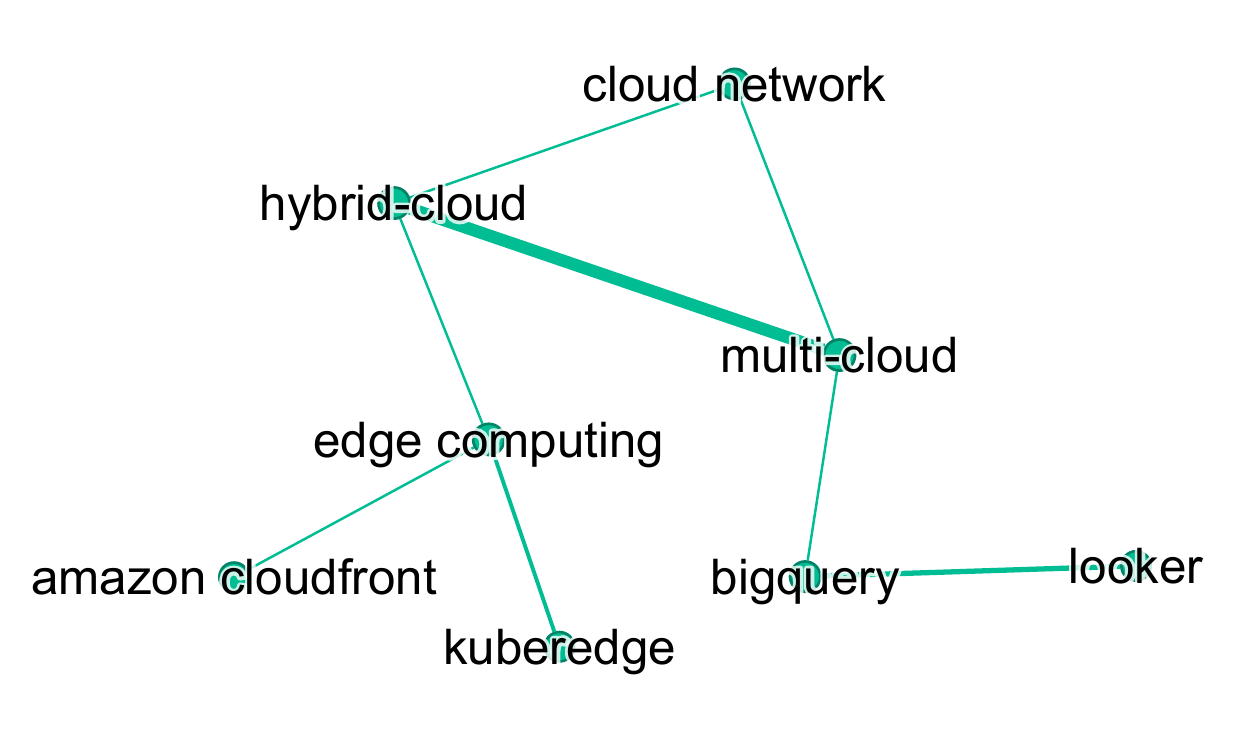}
    \caption{C5 - Cross-Cloud Architecture and Cloud-Edge Collaboration Community (RQ$_{1.2}$)}
    \label{fig:genre5}
\end{figure}

The smallest community, (\textbf{C5}), consists of 8 technologies (8.4\%) and 8 co-occurrence links (Figure~\ref{fig:genre5}).
Multi-Cloud and Hybrid-Cloud dominate across all centrality metrics. The strongest co-occurrence also lies between these two, with 8 links (Table~\ref{tab:genre5}).

Despite its size, this cluster reflects a focused interest in recent discussions about architectural patterns that span multiple cloud environments and integrate with edge nodes. Multi-Cloud and Hybrid-Cloud hold the highest $D_W$, $C_C$, and $C_B$ scores, hinting at their central topics and ability to connect technologies in different scenarios. More specifically, they serve as stable reference points in deployment flexibility, vendor independence, and federated data governance.

Edge Computing, while having smaller total connections, shows relatively high $C_B$, thus bridging cross-cloud strategies with distributed compute scenarios. BigQuery and Looker, which have moderate $D_W$ and $C_C$, act as supporting technologies. They are often discussed in relation to data analytics and visualization in cross-cloud pipelines. Their presence confirms that conversations about architecture focus not only on compute and networking but also on insights and reporting across distributed systems.

Technologies like KubeEdge and Amazon CloudFront have lower overall centrality but are important for discussions about moving from centralized cloud services to edge-capable deployments. Their roles tend to be specific to implementation but point to a growing interest in edge-cloud coordination.

All in all, C5 focused on how modern organizations try to unify cloud resources, analytics platforms, and edge computing through architectural practices that connect providers and topologies. The technologies here form the links between global distribution and local responsiveness.

\begin{table}
    \centering
    \caption{Cross-Cloud Architecture and Cloud-Edge Collaboration: Rank of Technology's Centrality Measurement (RQ$_{1.2}$)}
    \label{tab:genre5}
    \resizebox{\linewidth}{!}{
    \begin{tabular}{l r|l r|l r}
    \hline
        \textbf{Technology} & \textbf{$D_W(v)$} & \textbf{Technology} & \textbf{$C_C(v)$} & \textbf{Technology} & \textbf{$C_B(v)$} \\
        \hline
Multi-Cloud	&	15	&	Multi-Cloud	&	0.35	&	Multi-Cloud	&	293.57	\\
Hybrid-Cloud	&	14	&	Edge Computing	&	0.32	&	Edge Computing	&	189.04	\\
Edge Computing	&	5	&	Hybrid-Cloud	&	0.31	&	Bigquery	&	91	\\
Bigquery	&	4	&	Cloud Network	&	0.27	&	Hybrid-Cloud	&	87.5	\\
Looker	&	3	&	Bigquery	&	0.26	&	Cloud Network	&	0	\\
Cloud Network	&	2	&	KubeEdge	&	0.24	&	KubeEdge	&	0	\\
KubeEdge	&	2	&	Amazon CloudFront	&	0.24	&	Amazon CloudFront	&	0	\\
Amazon CloudFront	&	1	&	Looker	&	0.21	&	Looker	&	0	\\
        \hline
    \end{tabular}
    }
\end{table}

\subsubsection{Emerging Patterns Across Communities (RQ$_1$)}

Across the five identified technology communities, distinct yet interconnected themes emerge that reflect how practitioners have talked about software architecture in recent years. C1 revolves around deployment and automation, emphasizing the role of tools like GitOps and CI/CD in shaping infrastructure practices. C2 on microservices and service communication, with service mesh technologies linking implementation and operation. C3 brings together cloud-native AI tooling and serverless platforms, illustrating how scalable intelligence is becoming deeply embedded in architectural decisions. Similarly, C4 emphasizes observability, performance tuning, and security, driven by technologies like OpenTelemetry and eBPF. Finally, C5 focuses on bridging multiple clouds and edge environments, pointing toward a future of more distributed, cross-boundary architectures. These communities, revealed through centrality and co-occurrence, offer a snapshot of the evolving concerns and strategies in software architecture practice.

From the original selection \textbf{two technologies}, Falco and Fluent Bit, were not assigned to any community by the Louvain method. Both have only one co-occurrence edge, which limits their contribution to internal modularity optimization. Falco, a runtime security tool, and Fluent Bit, a telemetry data collector, likely function as general-purpose or cross-community enablers. Their exclusion suggests they are mentioned across multiple thematic areas but do not have strong enough connections to fit into any specific cluster.

\begin{keyRQAnswer}[Emerging Relationship and Patterns (RQ$_1$)]
The technologies Kubernetes, Cloud Native, Serverless, and Container are the structural core of the co-occurrence network, showing the highest centrality across all metrics. They anchor most connections in practitioner' talks. Surrounding them, technologies like Microservice, Service Mesh, eBPF, and Generative AI act as bridges between subtopics. Five coherent communities emerged, reflecting key concerns in architecture: deployment automation, service communication, cloud AI, observability, and cross-cloud-edge collaboration.
\end{keyRQAnswer}

\subsection{Purposes and context for adopting the extracted technology (RQ$_2$)}
To get a clearer picture of why certain technologies are used in software architecture and how practitioner' talk about them, we looked at both their intended purposes and the contexts in which they appear. In this section, we first break down the main reasons these technologies are adopted, then explore the recurring themes in practitioner' talk titles, and finally show how these elements connect back to the technologies themselves.
\subsubsection{Purpose for Adopting a Technology} 
\begin{figure*}[htb]
    \centering
    \includegraphics[width=\linewidth]{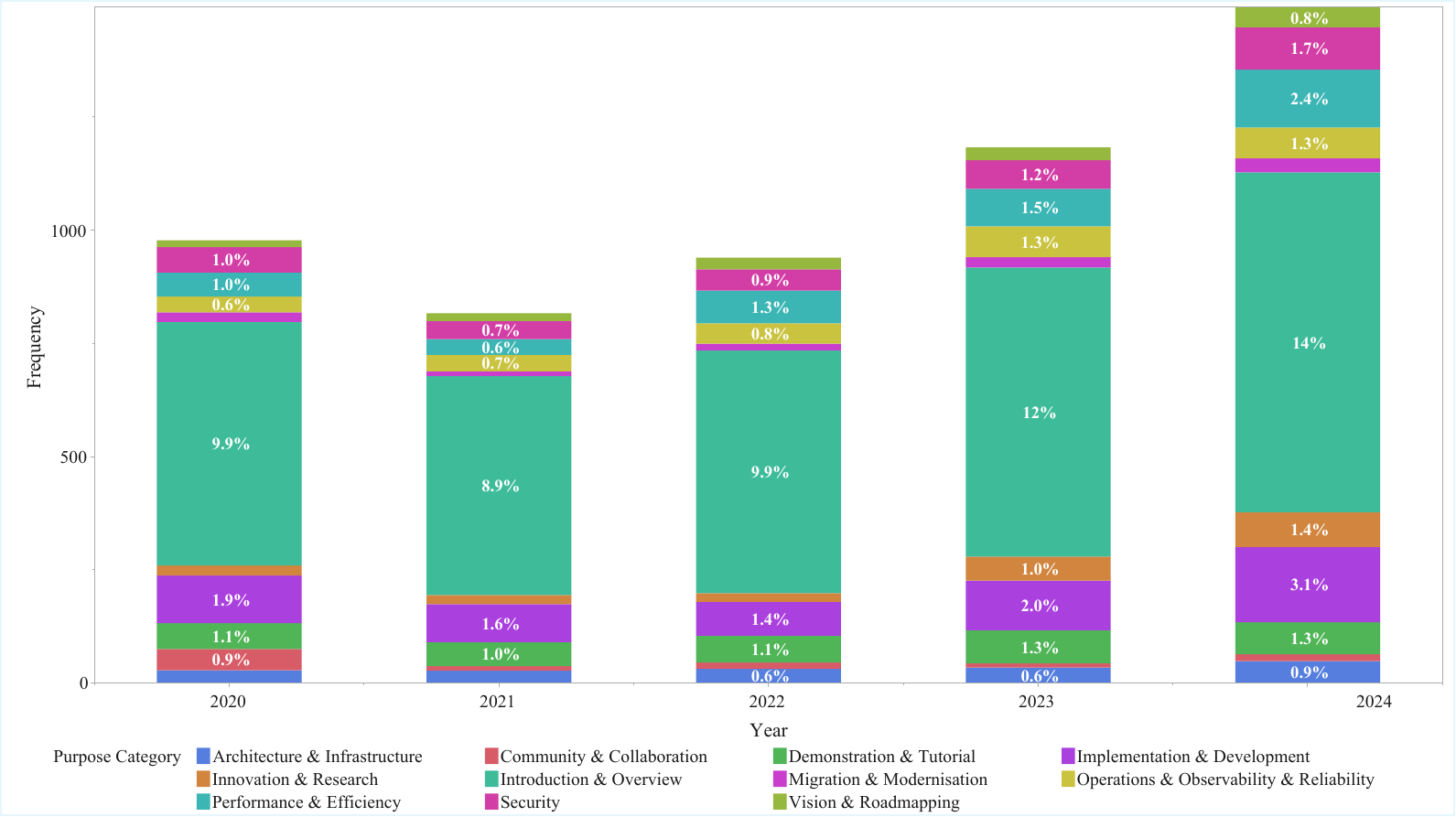}
    \caption{Purpose category distribution by year (RQ$_2$)}
    \label{fig:RQ2purpose}
\end{figure*}

To investigate the practitioners' purpose for adopting a specific technology, we classified the extracted purposes (Section~\ref{sec:DataExtraction}) in Table~\ref{tab:purpose_category_frequency} and Figure~\ref{fig:RQ2purpose} visualizes with definition, frequency, and evolution over the last 5 years.

\begin{table*}[htb]
  \centering
  \caption{Purpose Categories with Frequency by Year (RQ$_2$)}
      \resizebox{\linewidth}{!}{
  \begin{tabular}{l|l|r|r|r|r|r|r}
    \hline
    \textbf{Purpose Category} & \textbf{Definition} & \textbf{Total} & \textbf{2020} & \textbf{2021} & \textbf{2022} & \textbf{2023} & \textbf{2024} \\
    \hline
    Introduction \& Overview & Orient, introduce, summarise a technology or project & 2945 & 538 & 483 & 536 & 638 & 750 \\
    Vision \& Roadmapping & Envision, plan, strategise future directions & 130 & 15 & 17 & 26 & 28 & 44 \\
    Architecture \& Infrastructure & Design, structure, modernise foundational systems & 173 & 29 & 28 & 32 & 35 & 49 \\
    Implementation \& Development & Build, code, integrate, ship software artefacts & 540 & 105 & 84 & 75 & 110 & 166 \\
    Operations \& Observability \& Reliability & Run, monitor, troubleshoot, harden availability & 252 & 35 & 36 & 45 & 68 & 68 \\
    Performance \& Efficiency & Optimise, scale, streamline speed or cost & 371 & 53 & 35 & 72 & 83 & 128 \\
    Security & Secure, safeguard, comply with policies or threats & 299 & 56 & 40 & 47 & 63 & 93 \\
    Innovation \& Research & Experiment, pioneer, validate novel ideas or tech & 191 & 22 & 20 & 39 & 53 & 77 \\
    Community \& Collaboration & Coordinate, share, govern people and processes & 95 & 47 & 10 & 14 & 9 & 15 \\
    Migration \& Modernisation & Move, refactor, upgrade from old environments & 101 & 21 & 11 & 15 & 23 & 31 \\
    Demonstration \& Tutorial & Show, teach, walk through practical usage & 313 & 57 & 53 & 59 & 73 & 71 \\
    \hline
  \end{tabular}
  }
  \label{tab:purpose_category_frequency}
\end{table*}

Among the eleven identified purpose categories, \textbf{Introduction \& Overview} clearly dominates, with a total frequency of 2,945. Its mentions have increased steadily from 538 in 2020 to 750 in 2024, indicating a consistent emphasis on high-level orientation and summary in software architecture practitioner' talks (see Table~\ref{tab:purpose_category_frequency}).

While categories such as \textbf{Vision \& Roadmapping} (130), \textbf{Architecture \& Infrastructure} (173), \textbf{Operations, Observability \& Reliability} (252), \textbf{Innovation \& Research} (191), and \textbf{Demonstration \& Tutorial} (313) appear less frequently overall, they have shown a clear upward trend across the five-year period (Figure~\ref{fig:RQ2purpose}.

By contrast, \textbf{Implementation \& Development} (540) ranks second in total mentions but displays a fluctuating pattern, declining from 2020 to 2022, then rebounding in 2023 and peaking in 2024 (166). A similar trend is seen in \textbf{Performance \& Efficiency} (371), \textbf{Security} (299), \textbf{Community \& Collaboration} (95), and \textbf{Migration \& Modernization} (101), which dipped in the middle years but regained attention recently.

\subsubsection{Context of the Practitioner' Talk} As reported in Section~\ref{sec:DataExtraction}, we extracted the context from the included practitioner' talk titles. We considered 232 contexts that occur more than 2 times. For these 232 extracted contexts, we split the contexts into quartiles based on the frequency distribution and only presented the fourth quartile (the first 25 most frequent technologies whose frequency is within the 75\%–100\% interval). We got 60 contexts in the fourth quartile, and Table~\ref{tab:top10_quartile_context} lists the top 10 contexts in the fourth quartile (top 25\%) of the frequency distribution of contexts by year.

\begin{table*}
  \centering
  \caption{Top 10 Contexts in the fourth quartile (top 25\%) of Contexts Frequency Distribution by Year (RQ$_2$)}
    \resizebox{\linewidth}{!}{
  \begin{tabular}{c|l|r|l|r|l|r|l|r|l|r|l|r}
    \hline
    \textbf{Ranking} & \textbf{Total} & \textbf{\#} & \textbf{2020} & \textbf{\#} & \textbf{2021} & \textbf{\#} & \textbf{2022} & \textbf{\#} & \textbf{2023} & \textbf{\#} & \textbf{2024} & \textbf{\#} \\
    \hline
    1 & Cloud & 320 & Cloud Native & 46 & Cloud Native & 41 & Cloud & 67 & Cloud & 66 & Cloud & 109 \\
    2 & Cloud Native & 269 & Cloud & 45 & Kubernetes & 38 & Software Development & 46 & Cloud Native & 53 & Cloud Native & 86 \\
    3 & Kubernetes & 235 & Kubernetes & 39 & Software Development & 35 & Cloud Native & 43 & Kubernetes & 50 & Kubernetes & 78 \\
    4 & Software Development & 231 & Software Development & 39 & Cloud & 33 & Kubernetes & 30 & Software Development & 42 & Software Development & 69 \\
    5 & AWS & 112 & AWS & 36 & AWS & 30 & Security & 23 & Security & 25 & AI & 57 \\
    6 & Enterprise & 104 & Applications & 29 & Enterprise & 20 & Infrastructure & 21 & Applications & 24 & Enterprise & 30 \\
    7 & Applications & 99 & Machine Learning & 24 & Cloud Computing & 16 & Enterprise & 18 & AI & 24 & Container & 27 \\
    8 & Security & 98 & Cloud Computing & 22 & Database & 14 & Cloud Computing & 17 & AWS & 23 & SaaS & 26 \\
    9 & Cloud Computing & 89 & Enterprise & 17 & Data Storage & 14 & AWS & 14 & Enterprise & 19 & Security & 24 \\
    10 & AI & 87 & Container & 17 & Security & 12 & Applications & 13 & Cloud Computing & 18 & LLMs & 23 \\
    \hline
  \end{tabular}
  }
  \label{tab:top10_quartile_context}
\end{table*}

Terms such as \textbf{Cloud}, \textbf{Cloud Native}, \textbf{Kubernetes}, and \textbf{Cloud Computing} appear consistently, reflecting the ongoing influence of cloud ecosystems on current software architecture research and practice. The frequent mention of \textbf{Software Development} underscores the strong connection between architecture and engineering practices, highlighting the integration of architectural thinking into development workflows and methods. Additionally, contexts such as \textbf{AWS}, \textbf{Applications}, and \textbf{Enterprise} suggest that architectural discussions are often shaped by platform-specific or organizational perspectives, likely influenced by our data source, which consists of industry conferences.

\begin{figure}
    \centering
    \includegraphics[width=\linewidth]{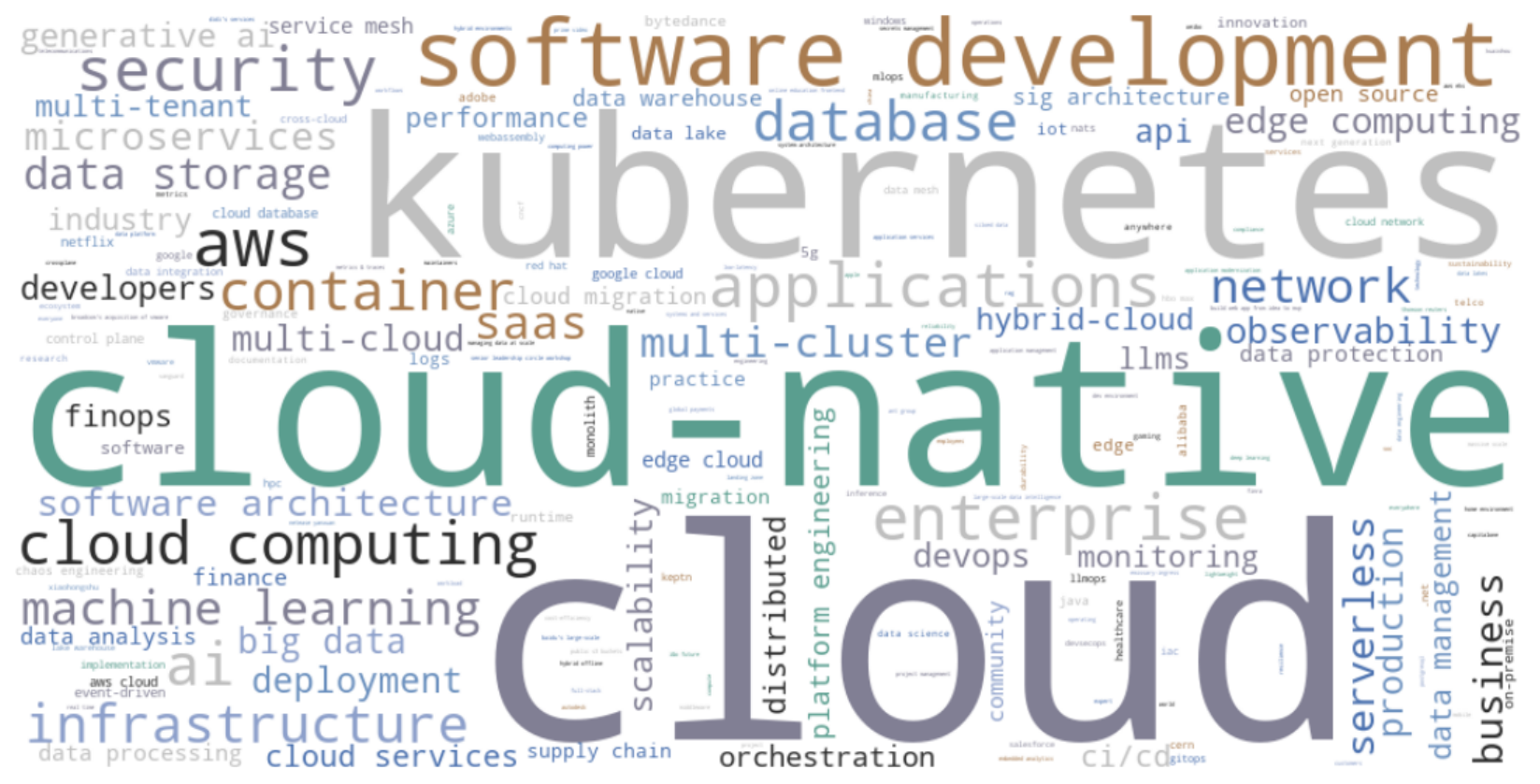}
    \caption{WordCloud for the Context (Frequency $  \ge 2 $) (RQ$_2$)}
    \label{fig:techfreRQ2}
\end{figure}

It is also interesting that an emerging term such as  \textbf{LLMs}, appears in Table~\ref{tab:top10_quartile_context} alongside cornerstone of the current modern architecture such as \textbf{AI}, \textbf{ML}, and \textbf{SaaS}, thus hinting to a shift toward more intelligent and adaptive architectural paradigms driven by artificial intelligence. Finally, we visualize the frequencies of such terms in a word cloud in Figure~\ref{fig:techfreRQ2}. 
All in all, these results illustrate that software architecture practitioner' talks span a range of lenses, from core technologies to enterprise strategy, business needs, and emerging innovations. Given the overlapping nature of these contexts and the diversity of perspectives presented in practitioner' talks, we chose not to merge them into broader categories.

\subsubsection{Connecting Technologies to Purpose and Context} 
\begin{figure*}
    \centering
    \includegraphics[width=\linewidth]{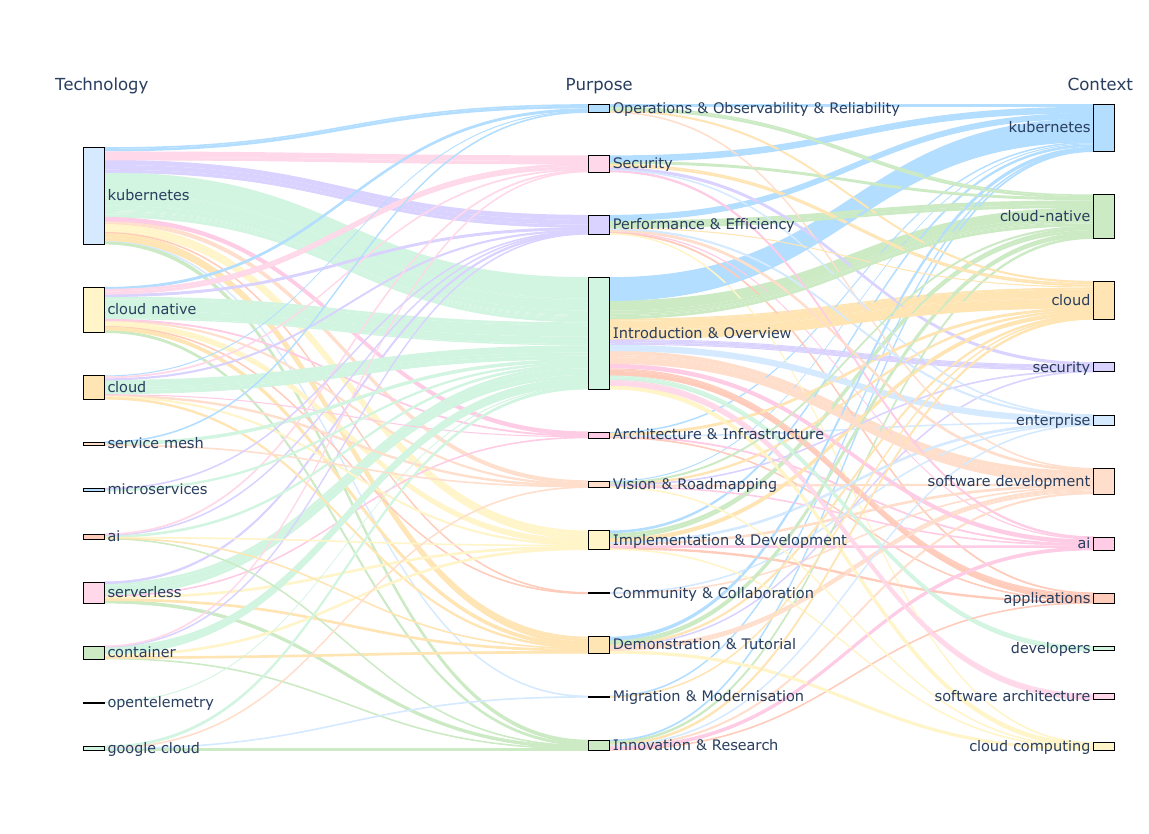}
    \caption{Sankey for the top-10 ``Technologies - Purpose - Context'' (RQ$_2$)}
    \label{fig:RQ2Sankey}
\end{figure*}

Finally, to better understand how technologies are used in practice, we mapped each one to its intended purpose and the context in which it appears. Figure~\ref{fig:RQ2Sankey} presents a Sankey diagram that visualizes the connections between \textbf{Technologies}, \textbf{Purpose Categories}, and \textbf{Contexts}. To keep the visualization focused and readable, we included the top 10 technologies from the fourth quartile (top 25\% by frequency), all eleven purpose categories, and only those contexts with more than five links to these technologies. This selection highlights the most meaningful patterns without overwhelming detail.

On the left side of the diagram, we see the technologies themselves. \textbf{Kubernetes} stands out with the widest and most varied set of connections, often linked to \textit{Introduction \& Overview}, \textit{Performance \& Efficiency}, and \textit{Security}. These, in turn, tie into contexts like \textit{Kubernetes}, \textit{Cloud Native}, and \textit{Cloud}, showing how central Kubernetes is to both the content and framing of architectural discussions. Similarly, \textbf{Cloud Native} and \textbf{Cloud} connect broadly across purposes like \textit{Implementation \& Development} and \textit{Architecture \& Infrastructure}, highlighting their wide relevance in architecture practitioner' talks.

Other technologies, such as \textbf{Serverless} and \textbf{Container}, show more focused flows, mainly toward \textit{Introduction \& Overview} and \textit{Performance \& Efficiency}. This suggests they are often discussed in specific, technical use cases rather than as overarching architectural themes.

At the center of the diagram (Figure~\ref{fig:RQ2Sankey}), \textbf{Purpose} (of the practitioner' talk) categories reveal how technologies are framed in practitioner' talks. \textit{Introduction \& Overview} appears as the most common category, serving as a hub that links nearly every technology to a range of contexts. This indicates a strong emphasis on introducing and explaining tools. In contrast, categories like \textit{Security} and \textit{Performance \& Efficiency} have tighter, more specialized flows, particularly tied to cloud-native technologies. These seem to reflect more mature or stable architectural concerns rather than fast-moving trends.

Finally, on the right side, the \textbf{Context} dimension shows where and how these technologies are applied. Contexts like \textit{Kubernetes}, \textit{Cloud Native}, and \textit{Cloud} show up both as technology names and conceptual domains. Others, such as \textit{Applications}, \textit{Enterprise}, and \textit{Developers}, suggest a strong grounding in real-world implementation and organizational needs.

\begin{keyRQAnswer}[Purpose and Context of Adoption]
Practitioner' talks emphasize technologies primarily for orientation and overview, with Kubernetes, Cloud Native, and Cloud leading in frequency and relevance. These technologies are introduced not only for their technical merits but also for their strategic role across multiple contexts. While purposes like Security and Performance \& Efficiency exhibit focused and stable attention, categories such as Vision, Innovation, and Architecture are steadily gaining momentum. Practitioner' talks are often grounded in cloud-centric and enterprise-specific settings, reflecting practical and platform-driven concerns. Emerging AI-related contexts (e.g., LLMs, SaaS) signal a shift toward intelligent, adaptive architectures.
\end{keyRQAnswer}
\section{Discussion}
\label{sec:Discussion}
In this section, we reflect on the key findings from our empirical analysis by examining how technologies are adopted, discussed, and positioned in software architecture practice. 

\subsection{Core Technologies Shaping Modern Architecture}

Across five years of practitioners' conferences, a handful of technologies have shaped the structure of software architecture practice. Kubernetes, Cloud Native, Serverless, and Container technologies consistently emerge as frequently cited terms (Table~\ref{tab:top10_quartile_tech}) and are most structurally embedded within architectural ecosystems. These tools seem to form the technological backbone of modern cloud-native systems.

Indeed, focusing on the co-occurrence networks (Figure~\ref{fig:technologiesnetwork}), such three technologies connect to over 80\% of all other technologies, both directly and indirectly. Their prominence in centrality metrics, whether through weighted degree, closeness, or betweenness (Figure~\ref{fig:centrality_comparison}), highlights how such technologies act both as primary innovation force as well as bridge between otherwise unconnected ``technological communities'''. For instance, we observe how Kubernetes is widely used as a bridge across domains, bridging cloud infrastructure, deployment pipelines, monitoring frameworks, and even AI services.

Therefore, it is not surprising that the core three technologies serve different purposes in many contexts ( Figure~\ref{fig:RQ2Sankey}). They are employed \textbf{for the purpose of}  performance optimization and security enhancement to implementation, development, and architectural modernization. Moreover, they also repeatedly \textbf{appear in the context of}  Cloud, Kubernetes, Software Development, and Enterprise (Table~\ref{tab:top10_quartile_context}) spanning all DevOps phases.


\begin{implications}[Modern Architecture]
    A small group of technologies, Kubernetes, Cloud Native, Serverless, and Containers, consistently shape software architecture practice. Their central role goes beyond popularity: they connect and unify the broader tech ecosystem, enabling scalable, strategic, and interoperable architectures.
\end{implications}

\subsection{The Gaps in Modern Architecture}

We could identify many vast specialized ecosystems seemingly built around the core technologies. Such ecosystems provide a novel ``reading key'' to the current interaction of technology in the state of the art, hence highlighting latent interactions that are yet underexplored in research.

For instance, it is evident that while modern architecture thrives in \textbf{Build, Deploy, Opearte and Monitor} phases, there are interesting \textbf{blind spots} in early lifecycle stages (Table~\ref{tab:450Tech}) shows that technologies cluster overwhelmingly around the Build, Deploy, Operate, and Monitor phases, where automation, orchestration, and observability dominate. 

By contrast, Plan, Code, and Release phases show minimal technological support, each representing less than 10\% of total classified technologies.

Indeed, Figure~\ref{fig:RQ2purpose} shows that most presentations in recent years skew toward Introduction, Implementation, and Performance. These categories focus on showcasing tools or demonstrating technical integration, often in downstream contexts. Meanwhile, fewer talks cover Vision, Roadmapping, or foundational Architecture \& Infrastructure. Even fewer are centered on Migration or Community building. 

This is an interesting finding since it reflects a tendency of the industry to early adopt new technologies, showcasing their use, but limiting their publicly discussed usage to specific use cases. We can therefore assume that the practitioners' conference can provide researchers a snapshot of the current use cases for technologies, more like a clue than a fundamental shake-up of the state-of-the-art.

For instance, discarding the core technologies, we observed that several technologies such as Generative AI, Prometheus, and Argo appear prominently when considering the betweenness centrality (Figure~\ref{fig:betweenness}) but rank lower in weighted or closeness metrics. 
Therefore, we note that \textbf{the role of emerging technology trends} is mostly associated with bridging otherwise disconnected domains without becoming hubs. Furthermore the use of AI, though expected, is well represented by the sudden and predominant appearance in 2023 (Figure~\ref{fig:missingTredns}). 

\begin{implications}[Fast in Adoption but Weak in Direction]
    Modern software architecture heavily emphasizes the Build-to-Monitor phases, while early stages like Plan and Code receive little technological support. Practitioners' talks tend to highlight tool usage and performance rather than broader architectural vision or strategy. Emerging technologies such as Generative AI serve more as connectors between domains than as core architectural components.
\end{implications}

\subsection{The Practitioner's Lens}
Our findings highlight that while microservices and DevOps are increasingly adopted in cloud-native systems, many teams struggle to maintain architectural clarity and documentation over time. The study shows that architecture-related keywords (e.g., “design”, “architecture”, “pattern”) appear significantly less often than terms related to implementation and infrastructure (e.g., “Docker”, “Kubernetes”, “CI/CD”). These differences in frequency suggest that practitioners' conferences are mostly interested in feature-driven technology rather than documenting and promoting technologies for architectural design, documentation, and planning.

Moreover, the prevalence of keywords like “DevOps”, “Kubernetes”, and “Docker” across several years suggests that infrastructure technologies continue to dominate discussions in cloud-native application development. This highlights a risk: tooling may be mistaken for process maturity. DevOps is not just about using modern tools; it is about cultural and organizational transformation~\cite{bass2015devops}. Hence, teams should reassess whether their tooling is truly enabling faster feedback cycles and better cross-functional collaboration or simply introducing unnecessary complexity for the sake of ``being cutting-edge " ~\citep {esposito_generative_2025}.

Finally, the stagnation in discussion about quality attributes, especially reliability, testability, and modifiability, suggests that non-functional requirements (NFRs) are under-explored. This aligns with broader concerns in software engineering about architectural technical debt~\cite{esposito_correlation_2024}. 

Therefore, we can observe that while industry moves towards the latest technology that can grant an advantage edge over the competition, according to \citet{esposito_generative_2025}, the research field keeps the only holistic lens focusing on those phases that practitioners disregard in their conferences.

\begin{implications}[The Lack of an Holististic View]
    Practitioners' conferences are more keen towards impactful cutting-edge-feature-driven technologies, hence they tend to mostly present talks on specific technologies useful in a specific niche context rather than documenting the technologies they find interesting for the design and development phases, contrary to the current research state-of-the-art.
\end{implications}

\section{Related Work}
\label{sec:rw}
To position our study within the broader landscape of software architecture and technology trend analysis, we review both research-oriented and practitioner-oriented contributions. Prior academic efforts have largely relied on systematic literature reviews, surveys, and interviews to map architectural practices and emerging trends. In parallel, industry platforms such as the Gartner Hype Cycle, DORA Reports, and the Thoughtworks Technology Radar have provided timely insights into DevOps, cloud-native tooling, and developer experience.

However, as summarized in Table~\ref{tab:gap_analysis}, existing studies tend to focus either on academic literature or on tool-level industry foresight—leaving a clear methodological and empirical gap. Notably, no prior work systematically investigates \textit{practitioner conferences} as a data source for tracking technological adoption and discourse. Our study addresses this gap by triangulating online trend analyses with a large-scale review of practitioner events, offering a \textbf{unique practitioner's lens} into what technologies are gaining traction, how they are discussed, and where academic research may be lagging or leading.

\begin{table*}[ht]
\centering
\caption{Coverage of Analytical Dimensions Across Research-Oriented, Practitioner-Oriented Related Works and Our Study}
\label{tab:gap_analysis}
\resizebox{0.8\linewidth}{!}{
\begin{tabular}{ p{8cm} c c c }
\hline
\textbf{Analytical Dimension} & \textbf{Research} & \textbf{Practitioner Online Trends} & \textbf{\textit{Our Study}} \\
\hline
Empirical validation through surveys or interviews & $\checkmark$ &  & $\checkmark$ \\
Coverage of DevOps practices and tools & $\checkmark$ & $\checkmark$ & $\checkmark$ \\
Coverage of cloud-native application trends & $\checkmark$ & $\checkmark$ & $\checkmark$ \\
Forecasting technology maturity and hype cycles &  & $\checkmark$ &  \\
Toolchain-level insights and technology recommendation &  & $\checkmark$ & $\checkmark$ \\
Developer Experience and productivity focus &  & $\checkmark$ & $\checkmark$ \\
Architecture trends over time & $\checkmark$ & $\checkmark$ & $\checkmark$ \\
\textbf{Use of practitioner conferences as data source} &  &  & $\checkmark$ \\
Systematic cross-source analysis & Partial & Partial & $\checkmark$ \\
Generative AI and software development integration & Partial & $\checkmark$ & $\checkmark$ \\
Actionable insights tailored to practitioners and researchers & Partial & Partial & $\checkmark$ \\
\hline
\end{tabular}

}\end{table*}
\subsection{Research Oriented}

In this section, we review the related work on software architecture trends and practices from both academic and industry perspectives (Table~\ref{tab:SLR}). The first study was published in 2012~\citep{breivold2012systematic}, and the most recent in 2025~\citep{ca2025trends}.  

The concept of software architecture evolvability can provide a basic perspective to understand the trends in software architecture. ~\citep{breivold2012systematic} conducted a systematic literature review to explore the software architecture evolution. They identified the five core topics and highlighted software architecture's evolvability, multidimensionality, and interactive nature. This indicates the rationality of researching the trends in software architecture. However, the paper was published in 2012 and only focused on the literature before 2010, and it does not include recent trends and developments in software architecture.

Several works have recently focused on emerging trends in certain aspects of software architecture.~\citep{farshidi2020capturing} conducted a systematic literature review to investigate the use trends in software architecture patterns and analyzed the relationship between patterns and quality attributes. They aimed to build a decision model for the architectural pattern selection problem.
On the other hand,~\citep{batmetan2023future} conducted a systematic literature review to explore the future trends in enterprise architecture. This paper focused on the architectures and relevant models that apply to current enterprises, such as microservices, cloud-native, DevOps, etc. 
In addition, \cite{ca2025trends} conducted a systematic desktop review to explore the evolution and trends in software architecture design over the last ten years. The results highlighted key shifts, emerging paradigms, and architectural decision factors.

However, these studies have certain limitations. Agwenyi and Mbugua's research~\citep{ca2025trends} only has five pages, and the results were only described in bullet points, lacking data tables, detailed descriptions, literature analysis, structured discussions, etc. The included literature (only a dozen) is insufficient and systematic to support the recent ten-year evolution and trends. Most importantly, all results of these three papers are based on academic papers and lack information and data from an industry perspective. Although~\citep{farshidi2020capturing} conducted interviews and let practitioners evaluate the results, the data source is still secondary data rather than direct data from practitioners in the industry.

On the other hand, some research is from an industry perspective.~\citep{kassab2018software} conducted an empirical study to research the current usage status of software architecture patterns in practice. They used the industry questionnaire method to extract and analyze the practitioners’ behaviors, motivations, and challenges in using architectural patterns in real projects.~\citep{wan2023software} conducted qualitative research to investigate challenges that practitioners face in software architecture practice during software development and maintenance. The interview results revealed some issues, such as architectural erosion and refactoring, and reported the corresponding challenges and suggestions on how to address them. However, the surveys and interviews for these works targeted industry practitioners, and the sample was limited and unable to reflect an overall trend.

In summary, previous studies gave us an understanding of architectural evolution, pattern usage, and practice-driven challenges in the software architecture field. However, they either rely on the academic literature without the industry practitioners' perspective, or their surveys and interviews with industry practitioners are limited in sample and cannot reflect trends. Our study fills this gap by analyzing presentations from industry conferences over the last five years to explore the recent software architecture trends through the practitioner lens.

\begin{table*}
\centering
\scriptsize
\caption{Research-Oriented Related Work} 
\label{tab:SLR} 
\resizebox{\linewidth}{!}{%
\begin{tabular}
{m{2cm} c m{2.3cm} m{2cm} m{3.5cm} m{3.5cm} m{3.5cm}}
\hline
\textbf{Ref} & \textbf{Year} & \textbf{Title} & \textbf{Method} & \textbf{Findings} & \textbf{Contributions} & \textbf{Limitations} \\ \hline
\cite{breivold2012systematic} & 2012 & Software Architecture Evolution & SLR & Identified five core topics and emphasized evolvability & Knowledge map and maturity assessment & Outdated (2010 data), no industry validation \\ \hline
\cite{kassab2018software} & 2018 & Patterns in Practice & Empirical (survey) & Patterns are widely used; explored motivations and issues & Rich data on real-world use & Based solely on surveys; limited scope \\ \hline
\cite{farshidi2020capturing} & 2020 & Pattern-Driven Design & SLR + Interviews & Mapped patterns to quality attributes; decision model & Conceptual framework linking patterns and quality & Based on academic sources, lacks depth in trend analysis \\ \hline
\cite{batmetan2023future} & 2023 & Trends in Enterprise Architecture & SLR & Identified future trends to support IT adoption & Practical guidance for enterprise adaptation & No practitioner input; lacks implementation insight \\ \hline
\cite{wan2023software} & 2023 & Practice Challenges & Interviews & Challenges and solutions in architecture practice & Industry-driven view on erosion, refactoring, etc. & Biased regional sample; lacks trend analysis \\ \hline
\cite{ca2025trends} & 2025 & Trends in Software Architecture Designs: Evolution and Current State & Desktop Review & Key shifts, paradigms, and decision factors over 10 years & Industry + academic synthesis of trends & Superficial, short paper; weak literature base \\ \hline
\end{tabular}%
}
\end{table*}

\subsection{Practitioner Oriented}
In this section, we explore a spectrum of widely recognized online practitioner-oriented trend analyses, examining their relevance to our study of technology evolution in cloud-native applications, DevOps, and programming languages. Our review is structured into three thematic clusters. We begin by discussing three prominent platforms in the technology trend forecasting domain: the Gartner Hype Cycle, the InfoQ Trends Report, and the Thoughtworks Technology Radar. We then shift our attention to two DevOps-centric sources, the DORA Accelerate State of DevOps Report and the Periodic Table of DevOps Tools. Finally, we analyze programming language trends through the lenses of the TIOBE Index, RedMonk Developer Rankings, and IEEE Spectrum. Table~\ref{tab:trend_summary} summarizes the focus areas and comparative insights from each trend analysis, contextualized against our practitioner-driven study.

The \textbf{Gartner Hype Cycle (GHC)} remains a foundational tool in understanding how emerging technologies evolve over time, from the spark of innovation to the trough of disillusionment and, eventually, to productive use. Both the 2024 and 2025 editions provide a macro-level perspective on technology maturity and adoption, helping organizations anticipate the trajectory of innovations across multiple domains, including cloud-native development, DevOps tooling, and software architecture.

In 2024, Gartner organized its outlook around four core themes: \textit{Autonomous AI}, \textit{Developer Productivity}, \textit{Total Experience}, and \textit{Human-Centric Security and Privacy}~\citep{gartner2024}. Particularly relevant to our study are the emphases on developer productivity and security, where technologies like \textbf{GitOps}, \textbf{AI-Augmented Software Engineering}, and \textbf{Prompt Engineering} represent the intersection of cloud-native automation, generative tooling, and secure-by-design practices. The Hype Cycle estimates that such technologies will reach maturity in 2–5 years. Meanwhile, longer-term innovations such as humanoid robots and 6G are expected to take over a decade to materialize.

The 2025 edition continues in this trajectory but introduces a more grounded tone. It places a sharper focus on \textbf{Agentic AI}, a subdomain of autonomous systems involving self-directed task-solving agents. While promising, Gartner projects that over 40\% of such initiatives may be discontinued by 2027 due to governance challenges and uncertain ROI~\citep{gartner2025,reuters2025}. The enduring presence of \textbf{GitOps} and \textbf{AI-Augmented Development} reaffirms the maturity of DevOps-aligned automation. What distinguishes 2025’s perspective is its cautious realism: it shifts attention from unbounded opportunity to risk management and value realization. For our work, this evolution reinforces the need to supplement hype-based foresight with practitioner-grounded evidence, like deployment insights and community sentiment, as captured in our own study.

The \textbf{Thoughtworks Technology Radar (TTR)} complements Gartner with a more opinionated and practice-oriented analysis. TTR categorizes technologies across four quadrants, Techniques, Tools, Platforms, and Languages \& Frameworks, further stratified by adoption readiness: Adopt, Trial, Assess, and Hold. Volume 31 highlights the responsible integration of \textbf{Generative AI} and \textbf{Large Language Models (LLMs)} into development workflows~\citep{thoughtworks2024}. Techniques such as \textit{1\% canary releases}, \textit{component testing}, and \textit{continuous deployment} emerge as recommended practices for modern DevOps pipelines. Additionally, the radar notes the growing importance of \textbf{Small Language Models (SLMs)} in edge computing contexts, signaling an architectural shift toward decentralized intelligence.

The \textbf{InfoQ Trends Report} offers a complementary top-down perspective, curated by expert panels and grounded in software community insights~\citep{infoq2025}. This report organizes its analysis across strategic domains like Architecture \& Design, DevOps, and AI/ML. Within DevOps, it emphasizes \textbf{platform engineering}, \textbf{observability}, and \textbf{AI-assisted software development}. A key insight is the growing importance of \textbf{Developer Experience (DevEx)}, not just as a byproduct but as a central enabler of successful DevOps transformations. InfoQ also confirms the sustained relevance of \textbf{GitOps}, consistent with both Gartner and TTR.

Together, these three platforms offer a layered view: Gartner highlights macro-level innovation arcs, Thoughtworks delivers tactical insights for engineering teams, and InfoQ bridges strategy with community practice. This triangulation provides a robust lens for analyzing DevOps and cloud-native transformations.

We now transition to trend analyses with a more explicit DevOps focus. The \textbf{DORA Accelerate State of DevOps Report}, now under Google's stewardship, remains one of the most authoritative empirical sources in the field. The 2024 edition aggregates data from over 39,000 software professionals to uncover the organizational and technical factors that drive elite performance~\citep{dora2024}. Central to its findings are four performance metrics, deployment frequency, lead time, change failure rate, and mean time to recovery, that continue to predict software delivery success. What stands out in 2024 is the observed synergy between these metrics and \textbf{AI-augmented development workflows}, which are shown to enhance delivery throughput, job satisfaction, and team efficiency. DORA also echoes the rise of \textbf{platform engineering} as a structural necessity for enabling scalable, secure DevOps practices.

In contrast to DORA’s data-driven lens, the \textbf{Periodic Table of DevOps Tools}~\citep{periodic2025} provides a visually structured inventory of tooling ecosystems. Organized similarly to the chemical periodic table, it clusters tools into categories like source control, CI/CD, observability, testing, security, and governance. The 2025 edition introduces emerging areas such as \textit{DevSecOps}, \textit{AI-driven analytics}, and \textit{developer portals}, reflecting broader trends toward automation, compliance, and user-centric design. Unlike predictive frameworks, the periodic table captures the present state of tool adoption. Tools like \textbf{GitHub Actions}, \textbf{Argo CD}, and \textbf{Backstage} are highlighted as gaining traction across both community and enterprise use.

Together, DORA and the Periodic Table present a balanced picture: one emphasizes process and performance, the other charts tooling practice and maturity. For our analysis, these are invaluable references, grounding technological aspirations in organizational and practitioner realities.

Finally, we turn our attention to the trends in programming languages, a domain that underpins all software development. Three independent sources offer complementary viewpoints: the \textbf{TIOBE Index}, the \textbf{RedMonk Developer Rankings}, and the \textbf{IEEE Spectrum}.

The \textbf{TIOBE Index}~\citep{tiobe2025} quantifies language popularity based on search engine queries. As of June 2025, \textbf{Python} leads with a 25.87\% share, followed by \textbf{C}, \textbf{C++}, \textbf{Java}, and \textbf{JavaScript}. The rising adoption of \textbf{Go} and \textbf{Rust} further reflects developer interest in performance and concurrency. In contrast, the \textbf{RedMonk Rankings}~\citep{redmonk2025}, which analyze GitHub and Stack Overflow activity, offer a more community-driven view. Here, \textbf{JavaScript}, \textbf{Python}, \textbf{TypeScript}, and \textbf{Rust} emerge as dominant, indicating both usage and ecosystem vibrancy. The \textbf{IEEE Spectrum}~\citep{ieee2025} balances these two views by incorporating job market demand, research activity, and open-source engagement. Its rankings affirm the leading role of \textbf{Python}, \textbf{Java}, and \textbf{SQL}, with domain-specific languages like \textbf{R} and \textbf{MATLAB} maintaining strongholds in scientific computing.

Across all three sources, \textbf{Python} stands out as the de facto lingua franca of modern development, spanning AI, web, and data engineering. \textbf{JavaScript} retains its stronghold in the web ecosystem, while languages like \textbf{Rust}, \textbf{Go}, and \textbf{TypeScript} gain momentum due to their safety, concurrency, and developer ergonomics. Understanding these language dynamics helps position our trend analysis within broader shifts in tooling, education, and enterprise strategy.

\begin{table*}[htb]
\centering
\footnotesize
\caption{Practitioner-Oriented Related Work}
\label{tab:trend_summary}
\resizebox{\linewidth}{!}{
\begin{tabular}{m{2.2cm} m{0.9cm} m{3.3cm} m{2.2cm} m{3.2cm} m{3.5cm} m{3.5cm}}
\hline
\textbf{Ref} & \textbf{Year} & \textbf{Title / Source} & \textbf{Method} & \textbf{Findings} & \textbf{Contributions} & \textbf{Limitations} \\
\hline
\citet{gartner2024,gartner2025} & 2024–2025 & Gartner Hype Cycle for Emerging Technologies & Expert-curated trend forecast & Identifies GitOps, Prompt Engineering, Agentic AI; highlights maturity timelines & Strategic foresight for DevOps and AI tooling & Hype-prone; lacks empirical validation and practitioner interaction \\
\hline
\citet{thoughtworks2024} & 2024 & Thoughtworks Technology Radar Vol. 31 & Expert synthesis + community practices & Highlights canary releases, SLMs, responsible LLM adoption in DevOps & Offers tactical engineering guidance & Opinionated; lacks cross-domain triangulation \\
\hline
\citet{infoq2025} & 2025 & InfoQ Trends Report & Expert panel + industry interviews & Focuses on DevEx, GitOps, platform engineering, observability & Captures evolving practitioner discourse & Broad themes; lacks depth and empirical metrics \\
\hline
\citet{dora2024} & 2024 & DORA State of DevOps Report & Global survey (39,000+ practitioners) & Correlates DevOps practices with performance, validates AI-augmented DevOps & Data-driven insight into delivery, team culture & High-level view; limited tooling specificity \\
\hline
\citet{periodic2025} & 2025 & Periodic Table of DevOps Tools & Curated practitioner resource & Categorizes DevOps tools; emphasizes DevSecOps and developer portals & Operational visibility of current tool adoption & Snapshot-in-time; lacks methodological grounding \\
\hline
\citet{tiobe2025} & 2025 & TIOBE Programming Language Index & Search engine query analysis & Ranks Python, C++, Java, Rust, Go; reflects general interest trends & Industry awareness of PL evolution & Lacks developer sentiment and contextual depth \\
\hline
\citet{redmonk2025} & 2025 & RedMonk Developer Rankings & GitHub + Stack Overflow activity & Emphasizes JavaScript, Python, TypeScript; indicates community traction & Developer-centric adoption signals & Social signal bias; lacks formal trend modeling \\
\hline
\citet{ieee2025} & 2024 & IEEE Spectrum PL Rankings & Composite index (jobs, OSS, citations) & Highlights Python, SQL, Java, R; blends academic and industry signals & Balanced view of demand and research presence & Less frequent updates; not tool- or DevOps-specific \\
\hline
\end{tabular}
}
\end{table*}

\section{Threats to Validity}
\label{sec:Threats}
This section discusses the potential threats to validity using four common categories: construct, internal, external, and reliability validity \citep{wohlin_experimentation_2024}.

\subsection{Construct Validity}

Construct validity concerns whether the variables we studied accurately represent the concepts of interest \citep{wohlin_experimentation_2024}. In our case, we relied on presentation titles from practitioner' talks to extract technologies, purposes, and contexts. However, presentation titles are brief and may not fully capture the depth or nuance of the talk content. There is a risk that the stated purpose does not reflect the actual architectural focus, or that the context is inferred implicitly rather than explicitly stated. While we used large-scale extraction methods, followed by manual classification and validation, the mapping between textual fragments and conceptual categories may still be imperfect.

\subsection{Internal Validity}

Internal validity refers to the degree to which the results can be attributed to the phenomena under investigation rather than other factors \citep{wohlin_experimentation_2024}. Our co-occurrence and centrality analyses are data-driven, but correlation does not imply causation. A technology’s high frequency or central position in a network does not necessarily mean it is architecturally central in practice. For example, Kubernetes may appear frequently across talks for reasons unrelated to architecture (e.g., deployment culture, cloud platform popularity). We mitigated this threat by triangulating results from frequency, network analysis, and interconnection among the technology, the purpose of the talk, and its context, but latent confounding factors may remain.

\subsection{Conclusion Validity}

Conclusion validity concerns the extent to which the conclusions drawn from the analysis are supported by the data \citep{wohlin_experimentation_2024}. Our results are based on clear patterns in frequency counts, network properties, and categorical mappings, but some inferences, especially those involving practitioner intent or strategic positioning, rely on interpretation. For example, framing Kubernetes as a ``meta-construct'' is based on the richness of its connections and roles, but alternate interpretations could be drawn. We have grounded our interpretations in multiple sources of evidence, but caution that some conclusions are suggestive rather than definitive.

\subsection{External Validity}

External validity relates to how well the findings generalize beyond our sample \citep{wohlin_experimentation_2024}. Our data comes from a large set of practitioner' talks over a five-year span, covering a wide range of conferences and technologies. However, the results may be biased toward popular platforms (e.g., Kubernetes, AWS), English-speaking conferences, or Western industry trends. Smaller or non-English communities, academic venues, or specialized domains (e.g., safety-critical systems) may not be adequately represented. As such, the generalizability of our findings to all types of software architecture practice may be limited.

\subsection{Reliability}

Reliability reflects the reproducibility of our methods and results \citep{wohlin_experimentation_2024}. We used automated scripts for technology extraction, co-occurrence analysis, and centrality computation, and documented all parameters used. However, classification of purposes and contexts involved the use of large language models (LLMs) followed by human inspection, which introduces subjectivity. Although we applied consistency checks and manual reviews, some classification decisions may vary if repeated by a different team. We released our detailed replication package to support transparency and reproducibility (see Data Availability Statement).
\section{Conclusion}
\label{sec:Conclusion}
Our findings offer a grounded look at how software architecture technologies are actually used and talked about in industry. By analyzing thousands of practitioner talk titles over five years, we explored which technologies are in play, what they’re used for, and the settings in which they’re applied.

Standing out is the strong presence of a few key technologies, particularly Kubernetes, Cloud Native, Serverless, and Containers. These tools consistently show up across a variety of talks and serve as a kind of backbone for modern architectural practice, supporting a wide range of goals and adapting to different environments.

Futhermore, we noticed an imbalance in how technology supports different phases of the development lifecycle. Most tools and discussions focus on the later DevOps stages, like deployment, monitoring, and operations, while earlier phases, such as planning, coding, and release, receive far less attention. This suggests that while the industry has invested heavily in runtime performance and operational efficiency, the early, strategic side of architecture may still be lacking support.

While the industry focuses on short-term gains from adopting cutting-edge tools, \textbf{research continues to provide a more holistic lens} on architectural design, quality, and evolution~\citep{esposito_generative_2025}.

Finally, our study paints a data-driven picture of a fast-evolving field, one where a handful of technologies lead the way, but where practitioners emphasize on the one hand the stability of the adoption of such technologies and, in the other hand, fastly embracing the emerging trends hinting at new and, industry-oriented, possible research directions.

\section*{Data Availability Statement}
\label{Replicability}
We provide a replication package\footnote{\url{https://doi.org/10.5281/zenodo.15728939}} to allow researchers to replicate, read, or extend our research. This package presents complete raw data for our research process.
\section*{Acknowledgment}
This work has been funded by the Research Council of Finland (grants n. 359861 and 349488 - MuFAno), by Business Finland (grant 6GSoft \citep{akbar_6gsoft_2024}), and FAST, the Finnish Software Engineering Doctoral Research Network, funded by the Ministry of Education and Culture, Finland.

\bibliographystyle{cas-model2-names}
\bibliography{main.bib,references-gray.bib}

\begin{thebibliography}{51}
\expandafter\ifx\csname natexlab\endcsname\relax\def\natexlab#1{#1}\fi
\providecommand{\url}[1]{\texttt{#1}}
\providecommand{\href}[2]{#2}
\providecommand{\path}[1]{#1}
\providecommand{\DOIprefix}{doi:}
\providecommand{\ArXivprefix}{arXiv:}
\providecommand{\URLprefix}{URL: }
\providecommand{\Pubmedprefix}{pmid:}
\providecommand{\doi}[1]{\href{http://dx.doi.org/#1}{\path{#1}}}
\providecommand{\Pubmed}[1]{\href{pmid:#1}{\path{#1}}}
\providecommand{\bibinfo}[2]{#2}
\ifx\xfnm\relax \def\xfnm[#1]{\unskip,\space#1}\fi
\bibitem[{Agwenyi and Mbugua(2025)}]{ca2025trends}
\bibinfo{author}{Agwenyi}, \bibinfo{author}{Mbugua}, \bibinfo{year}{2025}.
\newblock \bibinfo{title}{Trends in software architecture designs: Evolution and current state}.
\newblock \bibinfo{journal}{International Journal of Innovative Science and Research Technology} \bibinfo{volume}{10}, \bibinfo{pages}{1725--1729}.
\bibitem[{Akbar et~al.(2024)Akbar, Esposito, Hyrynsalmi, Kumar, Lcnarduzzi, Li, Mehraj, Mikkonen, Moreschini, Makitalo, Oivo, Paavonen, Parveen, Smolander, Su, Systa, Taibi, Yang, Zhang and Zohaib}]{akbar_6gsoft_2024}
\bibinfo{author}{Akbar, M.A.}, \bibinfo{author}{Esposito, M.}, \bibinfo{author}{Hyrynsalmi, S.}, \bibinfo{author}{Kumar, K.D.}, \bibinfo{author}{Lcnarduzzi, V.}, \bibinfo{author}{Li, X.}, \bibinfo{author}{Mehraj, A.}, \bibinfo{author}{Mikkonen, T.}, \bibinfo{author}{Moreschini, S.}, \bibinfo{author}{Makitalo, N.}, \bibinfo{author}{Oivo, M.}, \bibinfo{author}{Paavonen, A.S.}, \bibinfo{author}{Parveen, R.}, \bibinfo{author}{Smolander, K.}, \bibinfo{author}{Su, R.}, \bibinfo{author}{Systa, K.}, \bibinfo{author}{Taibi, D.}, \bibinfo{author}{Yang, N.}, \bibinfo{author}{Zhang, Z.}, \bibinfo{author}{Zohaib, M.}, \bibinfo{year}{2024}.
\newblock \bibinfo{title}{{6GSoft}: {Software} for {Edge}-to-{Cloud} {Continuum}}, in: \bibinfo{booktitle}{2024 50th {Euromicro} {Conference} on {Software} {Engineering} and {Advanced} {Applications} ({SEAA})}, \bibinfo{publisher}{IEEE Computer Society}, \bibinfo{address}{Los Alamitos, CA, USA}. pp. \bibinfo{pages}{499--506}.
\newblock \DOIprefix\doi{10.1109/SEAA64295.2024.00082}.
\bibitem[{Andrikopoulos et~al.(2013)Andrikopoulos, Binz, Leymann and Strauch}]{andrikopoulos2013adapt}
\bibinfo{author}{Andrikopoulos, V.}, \bibinfo{author}{Binz, T.}, \bibinfo{author}{Leymann, F.}, \bibinfo{author}{Strauch, S.}, \bibinfo{year}{2013}.
\newblock \bibinfo{title}{How to adapt applications for the cloud environment: Challenges and solutions in migrating applications to the cloud}.
\newblock \bibinfo{journal}{Computing} \bibinfo{volume}{95}, \bibinfo{pages}{493--535}.
\bibitem[{Barrat et~al.(2004)Barrat, Barthelemy, Pastor-Satorras and Vespignani}]{barrat2004architecture}
\bibinfo{author}{Barrat, A.}, \bibinfo{author}{Barthelemy, M.}, \bibinfo{author}{Pastor-Satorras, R.}, \bibinfo{author}{Vespignani, A.}, \bibinfo{year}{2004}.
\newblock \bibinfo{title}{The architecture of complex weighted networks}.
\newblock \bibinfo{journal}{Proceedings of the national academy of sciences} \bibinfo{volume}{101}, \bibinfo{pages}{3747--3752}.
\bibitem[{Barroca et~al.(2018)Barroca, Sharp, Salah, Taylor and Gregory}]{barroca2018bridging}
\bibinfo{author}{Barroca, L.}, \bibinfo{author}{Sharp, H.}, \bibinfo{author}{Salah, D.}, \bibinfo{author}{Taylor, K.}, \bibinfo{author}{Gregory, P.}, \bibinfo{year}{2018}.
\newblock \bibinfo{title}{Bridging the gap between research and agile practice: an evolutionary model}.
\newblock \bibinfo{journal}{International Journal of System Assurance Engineering and Management} \bibinfo{volume}{9}, \bibinfo{pages}{323--334}.
\bibitem[{Bass et~al.(2021)Bass, Clements and Kazman}]{bass2021software}
\bibinfo{author}{Bass, L.}, \bibinfo{author}{Clements, P.}, \bibinfo{author}{Kazman, R.}, \bibinfo{year}{2021}.
\newblock \bibinfo{title}{Software architecture in practice, 4th Edition}.
\newblock \bibinfo{publisher}{Addison-Wesley Professional}.
\bibitem[{Bass et~al.(2015)Bass, Weber and Zhu}]{bass2015devops}
\bibinfo{author}{Bass, L.}, \bibinfo{author}{Weber, I.}, \bibinfo{author}{Zhu, L.}, \bibinfo{year}{2015}.
\newblock \bibinfo{title}{DevOps: A software architect's perspective}.
\newblock \bibinfo{publisher}{Addison-Wesley Professional}.
\bibitem[{Bastian et~al.(2009)Bastian, Heymann and Jacomy}]{bastian2009gephi}
\bibinfo{author}{Bastian, M.}, \bibinfo{author}{Heymann, S.}, \bibinfo{author}{Jacomy, M.}, \bibinfo{year}{2009}.
\newblock \bibinfo{title}{Gephi: an open source software for exploring and manipulating networks}, in: \bibinfo{booktitle}{Proceedings of the international AAAI conference on web and social media}, pp. \bibinfo{pages}{361--362}.
\bibitem[{Batmetan et~al.(2023)Batmetan, Rawis, Lengkong and Rotty}]{batmetan2023future}
\bibinfo{author}{Batmetan, J.R.}, \bibinfo{author}{Rawis, J.A.}, \bibinfo{author}{Lengkong, J.S.J.}, \bibinfo{author}{Rotty, V.N.J.}, \bibinfo{year}{2023}.
\newblock \bibinfo{title}{Future trends for direction in enterprise architecture: systematic literature review}.
\newblock \bibinfo{journal}{International Journal of Information Technology and Education} \bibinfo{volume}{2}, \bibinfo{pages}{1--20}.
\bibitem[{Blondel et~al.(2008)Blondel, Guillaume, Lambiotte and Lefebvre}]{blondel2008fast}
\bibinfo{author}{Blondel, V.D.}, \bibinfo{author}{Guillaume, J.L.}, \bibinfo{author}{Lambiotte, R.}, \bibinfo{author}{Lefebvre, E.}, \bibinfo{year}{2008}.
\newblock \bibinfo{title}{Fast unfolding of communities in large networks}.
\newblock \bibinfo{journal}{Journal of statistical mechanics: theory and experiment} \bibinfo{volume}{2008}, \bibinfo{pages}{P10008}.
\bibitem[{Bolscher and Daneva(2019)}]{bolscher2019designing}
\bibinfo{author}{Bolscher, R.}, \bibinfo{author}{Daneva, M.}, \bibinfo{year}{2019}.
\newblock \bibinfo{title}{Designing software architecture to support continuous delivery and devops: a systematic literature review}, in: \bibinfo{booktitle}{14th International Conference on Software Technologies, ICSOFT 2019}, \bibinfo{organization}{SCITEPRESS}. pp. \bibinfo{pages}{27--39}.
\bibitem[{Brandes et~al.(2016)Brandes, Borgatti and Freeman}]{brandes2016maintaining}
\bibinfo{author}{Brandes, U.}, \bibinfo{author}{Borgatti, S.P.}, \bibinfo{author}{Freeman, L.C.}, \bibinfo{year}{2016}.
\newblock \bibinfo{title}{Maintaining the duality of closeness and betweenness centrality}.
\newblock \bibinfo{journal}{Social networks} \bibinfo{volume}{44}, \bibinfo{pages}{153--159}.
\bibitem[{Breivold et~al.(2012)Breivold, Crnkovic and Larsson}]{breivold2012systematic}
\bibinfo{author}{Breivold, H.P.}, \bibinfo{author}{Crnkovic, I.}, \bibinfo{author}{Larsson, M.}, \bibinfo{year}{2012}.
\newblock \bibinfo{title}{A systematic review of software architecture evolution research}.
\newblock \bibinfo{journal}{Information and Software Technology} \bibinfo{volume}{54}, \bibinfo{pages}{16--40}.
\bibitem[{De~Meo et~al.(2011)De~Meo, Ferrara, Fiumara and Provetti}]{DeMeo2011}
\bibinfo{author}{De~Meo, P.}, \bibinfo{author}{Ferrara, E.}, \bibinfo{author}{Fiumara, G.}, \bibinfo{author}{Provetti, A.}, \bibinfo{year}{2011}.
\newblock \bibinfo{title}{Generalized louvain method for community detection in large networks}, in: \bibinfo{booktitle}{International Conference on Intelligent Systems Design and Applications}, pp. \bibinfo{pages}{88--93}.
\bibitem[{{Digital.ai}(2025)}]{periodic2025}
\bibinfo{author}{{Digital.ai}}, \bibinfo{year}{2025}.
\newblock \bibinfo{title}{Periodic table of devops tools}.
\newblock \bibinfo{howpublished}{\texttt{Online Resource}}.
\newblock \bibinfo{note}{\url{https://intellipaat.com/blog/devops-periodic-table/}}.
\bibitem[{Esposito et~al.(2025a)Esposito, Li, Moreschini, Ahmad, Cerny, Vaidhyanathan, Lenarduzzi and Taibi}]{esposito_generative_2025}
\bibinfo{author}{Esposito, M.}, \bibinfo{author}{Li, X.}, \bibinfo{author}{Moreschini, S.}, \bibinfo{author}{Ahmad, N.}, \bibinfo{author}{Cerny, T.}, \bibinfo{author}{Vaidhyanathan, K.}, \bibinfo{author}{Lenarduzzi, V.}, \bibinfo{author}{Taibi, D.}, \bibinfo{year}{2025}a.
\newblock \bibinfo{title}{Generative {AI} for {Software} {Architecture}. {Applications}, {Trends}, {Challenges}, and {Future} {Directions}}.
\newblock \bibinfo{journal}{arXiv preprint arXiv:2503.13310} .
\bibitem[{Esposito et~al.(2024a)Esposito, Palagiano, Lenarduzzi and Taibi}]{esposito_beyond_2024}
\bibinfo{author}{Esposito, M.}, \bibinfo{author}{Palagiano, F.}, \bibinfo{author}{Lenarduzzi, V.}, \bibinfo{author}{Taibi, D.}, \bibinfo{year}{2024}a.
\newblock \bibinfo{title}{Beyond {Words}: {On} {Large} {Language} {Models} {Actionability} in {Mission}-{Critical} {Risk} {Analysis}}, in: \bibinfo{booktitle}{{International} {Symposium} on {Empirical} {Software} {Engineering} and {Measurement}, {ESEM} 2024}, pp. \bibinfo{pages}{517--527}.
\bibitem[{Esposito et~al.(2024b)Esposito, Palagiano, Lenarduzzi and Taibi}]{esposito_large_2024}
\bibinfo{author}{Esposito, M.}, \bibinfo{author}{Palagiano, F.}, \bibinfo{author}{Lenarduzzi, V.}, \bibinfo{author}{Taibi, D.}, \bibinfo{year}{2024}b.
\newblock \bibinfo{title}{On {Large} {Language} {Models} in {Mission}-{Critical} {IT} {Governance}: {Are} {We} {Ready} {Yet}?}
\newblock \bibinfo{journal}{ICSE-SEIP '25} .
\bibitem[{Esposito et~al.(2024c)Esposito, Robredo, Fontana and Lenarduzzi}]{esposito_correlation_2024}
\bibinfo{author}{Esposito, M.}, \bibinfo{author}{Robredo, M.}, \bibinfo{author}{Fontana, F.A.}, \bibinfo{author}{Lenarduzzi, V.}, \bibinfo{year}{2024}c.
\newblock \bibinfo{title}{On the correlation between {Architectural} {Smells} and {Static} {Analysis} {Warnings}}.
\newblock \bibinfo{journal}{CoRR} \bibinfo{volume}{abs/2406.17354}.
\newblock \DOIprefix\doi{10.48550/ARXIV.2406.17354}. \bibinfo{note}{arXiv: 2406.17354}.
\bibitem[{Esposito et~al.(2025b)Esposito, Robredo, Sridharan, Travassos, Peñaloza and Lenarduzzi}]{esposito_call_2025}
\bibinfo{author}{Esposito, M.}, \bibinfo{author}{Robredo, M.}, \bibinfo{author}{Sridharan, M.}, \bibinfo{author}{Travassos, G.H.}, \bibinfo{author}{Peñaloza, R.}, \bibinfo{author}{Lenarduzzi, V.}, \bibinfo{year}{2025}b.
\newblock \bibinfo{title}{A {Call} for {Critically} {Rethinking} and {Reforming} {Data} {Analysis} in {Empirical} {Software} {Engineering}}.
\newblock \bibinfo{journal}{arXiv preprint arXiv:2501.12728} .
\bibitem[{Farshidi et~al.(2020)Farshidi, Jansen and van~der Werf}]{farshidi2020capturing}
\bibinfo{author}{Farshidi, S.}, \bibinfo{author}{Jansen, S.}, \bibinfo{author}{van~der Werf, J.M.}, \bibinfo{year}{2020}.
\newblock \bibinfo{title}{Capturing software architecture knowledge for pattern-driven design}.
\newblock \bibinfo{journal}{Journal of Systems and Software} \bibinfo{volume}{169}, \bibinfo{pages}{110714}.
\bibitem[{Fortunato and Barthelemy(2007)}]{fortunato2007resolution}
\bibinfo{author}{Fortunato, S.}, \bibinfo{author}{Barthelemy, M.}, \bibinfo{year}{2007}.
\newblock \bibinfo{title}{Resolution limit in community detection}.
\newblock \bibinfo{journal}{Proceedings of the national academy of sciences} \bibinfo{volume}{104}, \bibinfo{pages}{36--41}.
\bibitem[{Freeman et~al.(2002)}]{freeman2002centrality}
\bibinfo{author}{Freeman, L.C.}, et~al., \bibinfo{year}{2002}.
\newblock \bibinfo{title}{Centrality in social networks: Conceptual clarification}.
\newblock \bibinfo{journal}{Social network: critical concepts in sociology. Londres: Routledge} \bibinfo{volume}{1}, \bibinfo{pages}{238--263}.
\bibitem[{Garousi et~al.(2016)Garousi, Petersen and Ozkan}]{garousi2016challenges}
\bibinfo{author}{Garousi, V.}, \bibinfo{author}{Petersen, K.}, \bibinfo{author}{Ozkan, B.}, \bibinfo{year}{2016}.
\newblock \bibinfo{title}{Challenges and best practices in industry-academia collaborations in software engineering: A systematic literature review}.
\newblock \bibinfo{journal}{Information and Software Technology} \bibinfo{volume}{79}, \bibinfo{pages}{106--127}.
\bibitem[{{Gartner, Inc.}(2024)}]{gartner2024}
\bibinfo{author}{{Gartner, Inc.}}, \bibinfo{year}{2024}.
\newblock \bibinfo{title}{Gartner hype cycle for emerging technologies 2024}.
\newblock \bibinfo{howpublished}{\texttt{Online Report}}.
\newblock \bibinfo{note}{\url{https://www.gartner.com/en/articles/what-s-new-in-the-2024-gartner-hype-cycle-for-emerging-technologies}}.
\bibitem[{{Gartner, Inc.}(2025)}]{gartner2025}
\bibinfo{author}{{Gartner, Inc.}}, \bibinfo{year}{2025}.
\newblock \bibinfo{title}{Gartner hype cycle for emerging technologies 2025}.
\newblock \bibinfo{howpublished}{\texttt{Online Report}}.
\newblock \bibinfo{note}{\url{https://www.gartner.com/en/articles/hype-cycle-for-emerging-technologies}}.
\bibitem[{{Google Cloud DORA Team}(2024)}]{dora2024}
\bibinfo{author}{{Google Cloud DORA Team}}, \bibinfo{year}{2024}.
\newblock \bibinfo{title}{2024 accelerate state of devops report}.
\newblock \bibinfo{howpublished}{\texttt{Online Report}}.
\newblock \bibinfo{note}{\url{https://dora.dev/research/2024/dora-report/}}.
\bibitem[{{IEEE Spectrum}(2024)}]{ieee2025}
\bibinfo{author}{{IEEE Spectrum}}, \bibinfo{year}{2024}.
\newblock \bibinfo{title}{Top programming languages 2024}.
\newblock \bibinfo{howpublished}{\texttt{Online Report}}.
\newblock \bibinfo{note}{\url{https://spectrum.ieee.org/top-programming-languages-2024}}.
\bibitem[{{InfoQ}(2025)}]{infoq2025}
\bibinfo{author}{{InfoQ}}, \bibinfo{year}{2025}.
\newblock \bibinfo{title}{Infoq trends report}.
\newblock \bibinfo{howpublished}{\texttt{Online Report}}.
\newblock \bibinfo{note}{\url{https://www.infoq.com/infoq-trends-report/}}.
\bibitem[{Jamshidi et~al.(2013)Jamshidi, Ahmad and Pahl}]{jamshidi2013cloud}
\bibinfo{author}{Jamshidi, P.}, \bibinfo{author}{Ahmad, A.}, \bibinfo{author}{Pahl, C.}, \bibinfo{year}{2013}.
\newblock \bibinfo{title}{Cloud migration research: a systematic review}.
\newblock \bibinfo{journal}{IEEE transactions on cloud computing} \bibinfo{volume}{1}, \bibinfo{pages}{142--157}.
\bibitem[{Kassab et~al.(2018)Kassab, Mazzara, Lee and Succi}]{kassab2018software}
\bibinfo{author}{Kassab, M.}, \bibinfo{author}{Mazzara, M.}, \bibinfo{author}{Lee, J.}, \bibinfo{author}{Succi, G.}, \bibinfo{year}{2018}.
\newblock \bibinfo{title}{Software architectural patterns in practice: an empirical study}.
\newblock \bibinfo{journal}{Innovations in Systems and Software Engineering} \bibinfo{volume}{14}, \bibinfo{pages}{263--271}.
\bibitem[{Kitchenham and Brereton(2013)}]{kitchenham2013systematic}
\bibinfo{author}{Kitchenham, B.}, \bibinfo{author}{Brereton, P.}, \bibinfo{year}{2013}.
\newblock \bibinfo{title}{A systematic review of systematic review process research in software engineering}.
\newblock \bibinfo{journal}{Information and software technology} \bibinfo{volume}{55}, \bibinfo{pages}{2049--2075}.
\bibitem[{Li and Zhang(2020)}]{li2020preliminary}
\bibinfo{author}{Li, X.}, \bibinfo{author}{Zhang, B.}, \bibinfo{year}{2020}.
\newblock \bibinfo{title}{A preliminary network analysis on steam game tags: Another way of understanding game genres}, in: \bibinfo{booktitle}{Proceedings of the 23rd International Conference on Academic Mindtrek}, pp. \bibinfo{pages}{65--73}.
\bibitem[{Mens et~al.(2008)Mens, Demeyer, Barais, Le~Meur, Duchien and Lawall}]{mens2008software}
\bibinfo{author}{Mens, T.}, \bibinfo{author}{Demeyer, S.}, \bibinfo{author}{Barais, O.}, \bibinfo{author}{Le~Meur, A.F.}, \bibinfo{author}{Duchien, L.}, \bibinfo{author}{Lawall, J.}, \bibinfo{year}{2008}.
\newblock \bibinfo{title}{Software architecture evolution}.
\newblock \bibinfo{journal}{Software Evolution} , \bibinfo{pages}{233--262}.
\bibitem[{Newman(2004)}]{newman2004analysis}
\bibinfo{author}{Newman, M.E.}, \bibinfo{year}{2004}.
\newblock \bibinfo{title}{Analysis of weighted networks}.
\newblock \bibinfo{journal}{Physical Review E—Statistical, Nonlinear, and Soft Matter Physics} \bibinfo{volume}{70}, \bibinfo{pages}{056131}.
\bibitem[{Newman(2006)}]{newman2006modularity}
\bibinfo{author}{Newman, M.E.}, \bibinfo{year}{2006}.
\newblock \bibinfo{title}{Modularity and community structure in networks}.
\newblock \bibinfo{journal}{Proceedings of the national academy of sciences} \bibinfo{volume}{103}, \bibinfo{pages}{8577--8582}.
\bibitem[{Nivedhaa(2024)}]{nivedhaa2024software}
\bibinfo{author}{Nivedhaa, N.}, \bibinfo{year}{2024}.
\newblock \bibinfo{title}{Software architecture evolution: Patterns, trends, and best practices}.
\newblock \bibinfo{journal}{International Journal of Computer Sciences and Engineering (IJCSE)} \bibinfo{volume}{1}, \bibinfo{pages}{1--14}.
\bibitem[{O'Grady(2025)}]{redmonk2025}
\bibinfo{author}{O'Grady, S.}, \bibinfo{year}{2025}.
\newblock \bibinfo{title}{The redmonk programming language rankings: January 2025}.
\newblock \bibinfo{howpublished}{\texttt{Online Report}}.
\newblock \bibinfo{note}{\url{https://redmonk.com/sogrady/2025/06/18/language-rankings-1-25/}}.
\bibitem[{Opsahl et~al.(2010)Opsahl, Agneessens and Skvoretz}]{opsahl2010node}
\bibinfo{author}{Opsahl, T.}, \bibinfo{author}{Agneessens, F.}, \bibinfo{author}{Skvoretz, J.}, \bibinfo{year}{2010}.
\newblock \bibinfo{title}{Node centrality in weighted networks: Generalizing degree and shortest paths}.
\newblock \bibinfo{journal}{Social networks} \bibinfo{volume}{32}, \bibinfo{pages}{245--251}.
\bibitem[{Reuters(2025)}]{reuters2025}
\bibinfo{author}{Reuters}, \bibinfo{year}{2025}.
\newblock \bibinfo{title}{Over 40\% of agentic ai projects to be scrapped by 2027, gartner says}.
\newblock \bibinfo{howpublished}{\texttt{News Article}}.
\newblock \bibinfo{note}{\url{https://www.reuters.com/business/over-40-agentic-ai-projects-will-be-scrapped-by-2027-gartner-says-2025-06-25}}.
\bibitem[{Sabidussi(1966)}]{sabidussi1966centrality}
\bibinfo{author}{Sabidussi, G.}, \bibinfo{year}{1966}.
\newblock \bibinfo{title}{The centrality index of a graph}.
\newblock \bibinfo{journal}{Psychometrika} \bibinfo{volume}{31}, \bibinfo{pages}{581--603}.
\bibitem[{Sattar and Arifuzzaman(2022)}]{Sattar2022Scalable}
\bibinfo{author}{Sattar, N.S.}, \bibinfo{author}{Arifuzzaman, S.}, \bibinfo{year}{2022}.
\newblock \bibinfo{title}{Scalable distributed louvain algorithm for community detection in large graphs}.
\newblock \bibinfo{journal}{Journal of Supercomputing} \bibinfo{volume}{78}, \bibinfo{pages}{10275--10309}.
\bibitem[{Sen et~al.(2022)Sen, Falter and Mayer}]{sen2022using}
\bibinfo{author}{Sen, A.}, \bibinfo{author}{Falter, S.}, \bibinfo{author}{Mayer, N.}, \bibinfo{year}{2022}.
\newblock \bibinfo{title}{Using devops paradigm to deploy web applications.}
\newblock \bibinfo{journal}{Issues in Information Systems} \bibinfo{volume}{23}.
\bibitem[{Su et~al.(2024)Su, Li and Taibi}]{su2024microservice}
\bibinfo{author}{Su, R.}, \bibinfo{author}{Li, X.}, \bibinfo{author}{Taibi, D.}, \bibinfo{year}{2024}.
\newblock \bibinfo{title}{From microservice to monolith: A multivocal literature review}.
\newblock \bibinfo{journal}{Electronics} \bibinfo{volume}{13}, \bibinfo{pages}{1452}.
\bibitem[{Taibi et~al.(2017)Taibi, Lenarduzzi and Pahl}]{taibi2017processes}
\bibinfo{author}{Taibi, D.}, \bibinfo{author}{Lenarduzzi, V.}, \bibinfo{author}{Pahl, C.}, \bibinfo{year}{2017}.
\newblock \bibinfo{title}{Processes, motivations, and issues for migrating to microservices architectures: An empirical investigation}.
\newblock \bibinfo{journal}{IEEE Cloud Computing} \bibinfo{volume}{4}, \bibinfo{pages}{22--32}.
\bibitem[{{Thoughtworks}(2024)}]{thoughtworks2024}
\bibinfo{author}{{Thoughtworks}}, \bibinfo{year}{2024}.
\newblock \bibinfo{title}{Technology radar volume 31}.
\newblock \bibinfo{howpublished}{\texttt{Online Report}}.
\newblock \bibinfo{note}{\url{https://www.thoughtworks.com/radar}}.
\bibitem[{{TIOBE Software}(2025)}]{tiobe2025}
\bibinfo{author}{{TIOBE Software}}, \bibinfo{year}{2025}.
\newblock \bibinfo{title}{Tiobe index for june 2025}.
\newblock \bibinfo{howpublished}{\texttt{Online Report}}.
\newblock \bibinfo{note}{\url{https://www.tiobe.com/tiobe-index/}}.
\bibitem[{Wan et~al.(2023)Wan, Zhang, Xia, Jiang and Lo}]{wan2023software}
\bibinfo{author}{Wan, Z.}, \bibinfo{author}{Zhang, Y.}, \bibinfo{author}{Xia, X.}, \bibinfo{author}{Jiang, Y.}, \bibinfo{author}{Lo, D.}, \bibinfo{year}{2023}.
\newblock \bibinfo{title}{Software architecture in practice: Challenges and opportunities}, in: \bibinfo{booktitle}{Proceedings of the 31st ACM Joint European Software Engineering Conference and Symposium on the Foundations of Software Engineering}, pp. \bibinfo{pages}{1457--1469}.
\bibitem[{Wei et~al.(2022)Wei, Wang, Schuurmans, Bosma, Xia, Chi, Le, Zhou et~al.}]{wei2022chain}
\bibinfo{author}{Wei, J.}, \bibinfo{author}{Wang, X.}, \bibinfo{author}{Schuurmans, D.}, \bibinfo{author}{Bosma, M.}, \bibinfo{author}{Xia, F.}, \bibinfo{author}{Chi, E.}, \bibinfo{author}{Le, Q.V.}, \bibinfo{author}{Zhou, D.}, et~al., \bibinfo{year}{2022}.
\newblock \bibinfo{title}{Chain-of-thought prompting elicits reasoning in large language models}.
\newblock \bibinfo{journal}{Advances in neural information processing systems} \bibinfo{volume}{35}, \bibinfo{pages}{24824--24837}.
\bibitem[{Wen et~al.(2011)Wen, Leicht and D’Souza}]{wen2011improving}
\bibinfo{author}{Wen, H.}, \bibinfo{author}{Leicht, E.}, \bibinfo{author}{D’Souza, R.M.}, \bibinfo{year}{2011}.
\newblock \bibinfo{title}{Improving community detection in networks by targeted node removal}.
\newblock \bibinfo{journal}{Physical Review E—Statistical, Nonlinear, and Soft Matter Physics} \bibinfo{volume}{83}, \bibinfo{pages}{016114}.
\bibitem[{Wohlin et~al.(2024)Wohlin, Runeson, Höst, Ohlsson, Regnell and Wesslén}]{wohlin_experimentation_2024}
\bibinfo{author}{Wohlin, C.}, \bibinfo{author}{Runeson, P.}, \bibinfo{author}{Höst, M.}, \bibinfo{author}{Ohlsson, M.C.}, \bibinfo{author}{Regnell, B.}, \bibinfo{author}{Wesslén, A.}, \bibinfo{year}{2024}.
\newblock \bibinfo{title}{Experimentation in {Software} {Engineering}, {Second} {Edition}}.
\newblock \bibinfo{publisher}{Springer}.
\newblock \DOIprefix\doi{10.1007/978-3-662-69306-3}.

\end{thebibliography}
\appendix

\section{Appendix A: Background}
\label{sec:Background}
This section introduces the centrality metrics, the Gephi tool to measure the influence and connectivity of technologies within the network, and the Louvain method for detecting community structures to cluster related technologies.
\subsection{Centrality Metrics}
\label{centralitymeasure}

In the network analysis, we selected three common classic metrics to measure the centrality (importance, influence, and relationship) of the vertices in a network~\citep{opsahl2010node}: Weighted Degree~\citep{barrat2004architecture}, Closeness Centrality~\citep{freeman2002centrality}, and Betweenness Centrality~\citep{freeman2002centrality}.

\textbf{Weighted Degree} is the sum of the edge weights connected to the vertex when analyzing the weighted network or the strength of the labeled vertex~\citep{newman2004analysis}. It enables the assignment of weights to the edges connected to the vertex to accurately measure and reflect the strength of the connection of the vertex~\citep{opsahl2010node}. A higher Weighted Degree value indicates a higher connection strength of the vertex and may play an essential role in the network.
Let $v$ be a vertex in the network, and let $N(v)$ be the set of neighborhoods of the vertex $v$. The weight of the edge between $v$ and another vertex $u \in N(v)$ is denoted $w(v, u)$.
Thus, the Weighted Degree of $v$, $D_W(v)$, is calculated as follows:

\begin{equation*}
D_W(v) = \sum_{u \in N(v)} w(v, u) \tag{1}
\end{equation*}

\textbf{Closeness Centrality} is calculated as the inverse of the shortest path distance from a vertex to all other vertices of the network~\citep{brandes2016maintaining}. This metric reflects the degree of closeness between a vertex and other vertices. The larger the sum of associated distances, the farther the vertex is from the others~\citep{sabidussi1966centrality}. A high Closeness Centrality value means that the vertex can spread quickly and may be the center of the network.
Let $V$ be the set of all vertexes in the network and $v$ is one of them. The shortest path distance between $i$ and another vertex $u \in V$ is denoted $d(v, u)$. 
Thus, according to the above definition, the closeness centrality of $v$, $C_C(v)$, is calculated as follows:

\begin{equation*}
C_C(v) = \frac{1}{\sum_{u \in V} d(v, u)} \tag{2}
\end{equation*}

\textbf{Betweenness centrality} is the frequency with a vertex appearing on the shortest paths between all pairs of vertexes in the network~\citep{brandes2016maintaining}. This metric can identify the key vertex that plays a role as the ``bridge,'' and reflect its intermediary ability. A high Betweenness Centrality value means the vertex is the bridge for many shortest paths in the network and influences connections between more vertices.
Let $v, u, s \in V$ be different vertices in the network. The number of paths through $v$ in the shortest path from $s$ to $u$ is denoted as $\sigma_{su}(v)$, and the number of all shortest paths from $s$ to $u$ is denoted as $\sigma_{su}$. Thus, the fraction $\frac{\sigma_{su}(v)}{\sigma_{su}}$ represents the proportion of all the shortest paths from $s$ to $u$ that go through $v$. The higher this value, the more important the effect of $v$ as a 'bridge' between $s$ and $u$. The betweenness centrality of $v$, $C_B(v)$, is calculated as follows:

\begin{equation*}
\centering
C_B(v) = \sum_{s \neq v \neq u} \frac{\sigma_{su}(v)}{\sigma_{su}} \tag{3}
\end{equation*}

\subsection{Louvain Method for Community Detection}
\label{louvain}

The Louvain Method for Community Detection is a greedy optimization method~\citep{blondel2008fast} to find the community structure in complex social networks or systems. The goal is to use maximum modularity to optimize community division~\citep{li2020preliminary}. Network Modularity is the metric for modularity optimization and can measure the strength of the connection between the vertices of a community and the relationships between communities~\citep{newman2006modularity}.

The calculation equation for network modularity $Q$ is calculated as follows:

\begin{equation*}
\centering
Q = \sum_{c=1}^{m} \left[ \frac{E_C}{|E|} - \gamma \left( \frac{K_C}{2|E|} \right)^2 \right] \tag{4}
\end{equation*}

In equation (4), let $m$ be the number of communities in the network, and $C$ be one of them. The number of edges in the community $C$ is denoted as $E_C$, the sum of all the degrees of the vertices in the community $C$ is denoted as $K_C$, and the number of edges in the network is denoted as $|E|$. The parameter $\gamma$ is a resolution parameter to adjust the community detection scale. The higher the value ($Q$) between the actual percentage of occupied edges in the community and the expected percentage of occupied edges in a random network, the more pronounced the community structure of the network.

Based on the Louvain method above, when $Q$ is maximized, if a vertex $i$ moves from one community to a new community $C$, the modularity gained $\Delta Q$ is calculated as follows:

{
\footnotesize
\begin{equation}
\centering
\Delta Q = \left[ \frac{\sum_{C} + k_i^C}{2n} - \left(\frac{\sum_{\bar{C}} + k_i}{2n}\right)^2 \right] - \left[ \frac{\sum_{C}}{2n} - \left(\frac{\sum_{\bar{C}}}{2n}\right)^2 - \left(\frac{k_i}{2n}\right)^2 \right] \tag{5}
\end{equation}
}

In equation (5), let the degree of vertex $i$ be denoted as $k_i$, the weighted edges from $i$ to the vertices in the community $C$ be denoted as $k_i^C$, and $n$ be the sum of the weighted edges in the network. $\sum_{C}$ is the sum of weighted edges in community C, and $\sum_{\bar{C}}$ represents the sum of weighted edges connected to the vertices in community C. Thus, the difference between modularity that $i$ has moved to community $C$ and modularity that has not moved to community $C$ is the gain in modularity.
\section{Appendix B: Data Collection} 
\label{AppendixA}

\begin{table*}
\centering
    \caption{Technology Frequency and Classification List 1 - the Fourth Quartile Top 25\%}
    \label{tab:tech_frequencies_checkmark}
\resizebox{0.73\linewidth}{!}{%
    \begin{tabular}{r |l| r| r r r r r r r r| r| r}
        \hline 
        \textbf{Ranking} & \textbf{Technology} & \textbf{\#}& \textbf{Plan} & \textbf{Code} &\textbf{Build} &  \textbf{Test} & \textbf{Release} & \textbf{Deploy} & \textbf{Operate} & \textbf{Monitor}  & \textbf{C/O/B} & \textbf{CPs} \\
        \hline
1	&	Kubernetes	&	1054	&		&		&	\checkmark	&		&		&	\checkmark	&	\checkmark	&	\checkmark	&	B	&	Other	\\
2	&	Cloud Native	&	584	&		&		&	\checkmark	&		&		&	\checkmark	&	\checkmark	&	\checkmark	&	B	&	Other	\\
3	&	Serverless	&	325	&		&		&	\checkmark	&		&		&	\checkmark	&	\checkmark	&	\checkmark	&	B	&	Other	\\
4	&	Cloud	&	297	&		&		&	\checkmark	&		&		&	\checkmark	&	\checkmark	&	\checkmark	&	C	&	Other	\\
5	&	Container	&	203	&		&		&	\checkmark	&		&		&	\checkmark	&	\checkmark	&	\checkmark	&	B	&	Other	\\
6	&	AI	&	97	&		&		&	\checkmark	&	\checkmark	&		&	\checkmark	&	\checkmark	&	\checkmark	&	B	&	Other	\\
7	&	Google Cloud	&	95	&		&		&	\checkmark	&		&		&	\checkmark	&	\checkmark	&	\checkmark	&	C	&	GCP	\\
8	&	OpenTelemetry	&	82	&		&		&		&		&		&		&	\checkmark	&	\checkmark	&	B	&	Other	\\
9	&	Service Mesh	&	81	&		&		&	\checkmark	&		&		&	\checkmark	&	\checkmark	&	\checkmark	&	B	&	Other	\\
10	&	Microservices	&	80	&		&		&	\checkmark	&		&		&	\checkmark	&	\checkmark	&	\checkmark	&	B	&	Other	\\
11	&	WebAssembly	&	79	&		&		&	\checkmark	&		&		&	\checkmark	&	\checkmark	&	\checkmark	&	B	&	Other	\\
12	&	API	&	75	&		&		&	\checkmark	&	\checkmark	&		&	\checkmark	&	\checkmark	&	\checkmark	&	B	&	Other	\\
13	&	AWS S3	&	65	&		&		&	\checkmark	&		&		&	\checkmark	&	\checkmark	&	\checkmark	&	B	&	AWS	\\
14	&	AWS EKS	&	61	&		&		&	\checkmark	&		&		&	\checkmark	&	\checkmark	&	\checkmark	&	C	&	AWS	\\
15	&	AWS SageMaker	&	60	&		&		&	\checkmark	&		&		&	\checkmark	&	\checkmark	&	\checkmark	&	C	&	AWS	\\
16	&	Prometheus	&	59	&		&		&		&		&		&		&		&	\checkmark	&	B	&	Other	\\
17	&	Generative AI	&	55	&		&		&	\checkmark	&	\checkmark	&		&	\checkmark	&	\checkmark	&	\checkmark	&	B	&	Other	\\
18	&	GitOps	&	48	&	\checkmark	&		&	\checkmark	&		&		&	\checkmark	&	\checkmark	&	\checkmark	&	B	&	Other	\\
19	&	eBPF	&	46	&		&		&	\checkmark	&	\checkmark	&		&	\checkmark	&	\checkmark	&	\checkmark	&	B	&	Other	\\
20	&	AWS Lambda	&	46	&		&		&	\checkmark	&		&		&	\checkmark	&	\checkmark	&	\checkmark	&	C	&	AWS	\\
21	&	Alibaba Cloud	&	44	&		&		&	\checkmark	&		&		&	\checkmark	&	\checkmark	&	\checkmark	&	C	&	Other	\\
22	&	Google Kubernetes Engine	&	42	&		&		&	\checkmark	&		&		&	\checkmark	&	\checkmark	&	\checkmark	&	B	&	GCP	\\
23	&	AWS Aurora	&	38	&		&		&	\checkmark	&		&		&	\checkmark	&	\checkmark	&	\checkmark	&	C	&	AWS	\\
24	&	Linkerd	&	37	&		&		&		&		&		&	\checkmark	&	\checkmark	&	\checkmark	&	B	&	Other	\\
25	&	.Net	&	36	&		&		&	\checkmark	&	\checkmark	&		&	\checkmark	&	\checkmark	&	\checkmark	&	B	&	Other	\\
26	&	Observability	&	35	&		&		&		&		&		&		&	\checkmark	&	\checkmark	&	B	&	Other	\\
27	&	AWS ECS	&	35	&		&		&	\checkmark	&		&		&	\checkmark	&	\checkmark	&	\checkmark	&	C	&	AWS	\\
28	&	Argo	&	34	&		&		&	\checkmark	&		&		&	\checkmark	&	\checkmark	&	\checkmark	&	B	&	Other	\\
29	&	CI/CD	&	34	&		&		&	\checkmark	&	\checkmark	&	\checkmark	&	\checkmark	&	\checkmark	&	\checkmark	&	B	&	Other	\\
30	&	Machine Learning	&	33	&		&		&	\checkmark	&	\checkmark	&		&	\checkmark	&	\checkmark	&	\checkmark	&	B	&	Other	\\
31	&	SaaS	&	33	&		&		&	\checkmark	&		&		&	\checkmark	&	\checkmark	&	\checkmark	&	B	&	Other	\\
32	&	Hybrid-Cloud	&	32	&		&		&	\checkmark	&		&		&	\checkmark	&	\checkmark	&	\checkmark	&	B	&	Other	\\
33	&	AWS Redshift	&	32	&		&		&	\checkmark	&		&		&	\checkmark	&	\checkmark	&	\checkmark	&	C	&	AWS	\\
34	&	AWS EC2	&	32	&		&		&		&		&		&	\checkmark	&	\checkmark	&	\checkmark	&	C	&	AWS	\\
35	&	Anthos	&	32	&		&		&	\checkmark	&		&		&	\checkmark	&	\checkmark	&	\checkmark	&	B	&	GCP	\\
36	&	AWS Bedrock	&	30	&		&		&	\checkmark	&		&		&	\checkmark	&	\checkmark	&	\checkmark	&	C	&	AWS	\\
37	&	Containerd	&	30	&		&		&	\checkmark	&		&		&	\checkmark	&	\checkmark	&	\checkmark	&	B	&	Other	\\
38	&	Istio	&	29	&		&		&		&		&		&	\checkmark	&	\checkmark	&	\checkmark	&	B	&	Other	\\
39	&	Helm	&	28	&		&		&		&		&	\checkmark	&	\checkmark	&	\checkmark	&		&	B	&	Other	\\
40	&	LLMs	&	28	&		&		&	\checkmark	&		&		&	\checkmark	&	\checkmark	&	\checkmark	&	B	&	Other	\\
41	&	AWS Dynamodb	&	27	&		&		&	\checkmark	&		&		&	\checkmark	&	\checkmark	&	\checkmark	&	B	&	AWS	\\
42	&	Open Source	&	26	&	\checkmark	&	\checkmark	&	\checkmark	&	\checkmark	&	\checkmark	&	\checkmark	&	\checkmark	&	\checkmark	&	B	&	Other	\\
43	&	Apache Kafka	&	26	&		&		&	\checkmark	&		&		&	\checkmark	&	\checkmark	&	\checkmark	&	B	&	Other	\\
44	&	Cilium	&	25	&		&		&	\checkmark	&		&		&	\checkmark	&	\checkmark	&	\checkmark	&	B	&	Other	\\
45	&	Edge	&	25	&		&		&	\checkmark	&		&		&	\checkmark	&	\checkmark	&	\checkmark	&	B	&	Other	\\
46	&	Harbor	&	25	&		&		&	\checkmark	&		&	\checkmark	&	\checkmark	&	\checkmark	&	\checkmark	&	B	&	Other	\\
47	&	Kubeflow	&	25	&		&		&	\checkmark	&		&		&	\checkmark	&	\checkmark	&	\checkmark	&	B	&	Other	\\
48	&	Postgresql	&	25	&		&		&	\checkmark	&		&		&	\checkmark	&	\checkmark	&	\checkmark	&	B	&	Other	\\
49	&	Crossplane	&	25	&		&		&	\checkmark	&		&		&	\checkmark	&	\checkmark	&	\checkmark	&	B	&	Other	\\
50	&	Multi-Cloud	&	25	&		&		&	\checkmark	&		&		&	\checkmark	&	\checkmark	&	\checkmark	&	B	&	Other	\\
51	&	Edge Computing	&	24	&		&		&	\checkmark	&		&		&	\checkmark	&	\checkmark	&	\checkmark	&	B	&	Other	\\
52	&	Ingress	&	23	&		&		&		&		&		&	\checkmark	&	\checkmark	&	\checkmark	&	B	&	Other	\\
53	&	Bigquery	&	23	&		&		&	\checkmark	&		&		&	\checkmark	&	\checkmark	&	\checkmark	&	C	&	GCP	\\
54	&	AWS EBS	&	23	&		&		&	\checkmark	&		&		&	\checkmark	&	\checkmark	&	\checkmark	&	C	&	AWS	\\
55	&	Vmware	&	22	&		&		&	\checkmark	&		&		&	\checkmark	&	\checkmark	&	\checkmark	&	B	&	Other	\\
56	&	Open Policy Agent	&	22	&		&		&	\checkmark	&		&		&	\checkmark	&	\checkmark	&	\checkmark	&	B	&	Other	\\
57	&	Gpu	&	22	&		&		&	\checkmark	&		&		&	\checkmark	&	\checkmark	&	\checkmark	&	B	&	Other	\\
58	&	Nats	&	22	&		&		&	\checkmark	&		&		&	\checkmark	&	\checkmark	&	\checkmark	&	B	&	Other	\\
59	&	AWS Amplify	&	22	&		&		&	\checkmark	&		&		&	\checkmark	&	\checkmark	&	\checkmark	&	B	&	AWS	\\
60	&	IaC	&	21	&	\checkmark	&		&	\checkmark	&		&		&	\checkmark	&	\checkmark	&	\checkmark	&	B	&	Other	\\
        \hline 
\multicolumn{13}{l}{\textbf{\checkmark}: Yes,  \textbf{C}: Cloud Specific; \textbf{O}: On-premise; \textbf{B}: Both, \textbf{CPs}: Cloud Providers; \textbf{GCP}: Google Cloud; \textbf{AWS}: Amazon; \textbf{Azure}: Azure} \\

    \end{tabular}
    }
\end{table*}

\begin{table*}
\centering
    \caption{Technology Frequency and Classification List 2 - the Fourth Quartile Top 25\%}
    \label{tab:tech_frequencies_checkmark_2}
\resizebox{0.73\linewidth}{!}{%
    \begin{tabular}{r |l| r| r r r r r r r r| r| r}
        \hline 
        \textbf{Ranking} & \textbf{Technology} & \textbf{\#}& \textbf{Plan} & \textbf{Code} &\textbf{Build} &  \textbf{Test} & \textbf{Release} & \textbf{Deploy} & \textbf{Operate} & \textbf{Monitor}  & \textbf{C/O/B} & \textbf{CPs} \\
        \hline
61	&	Kyverno	&	21	&		&		&	\checkmark	&		&		&	\checkmark	&	\checkmark	&	\checkmark	&	B	&	Other	\\
62	&	KubeEdge	&	21	&		&		&	\checkmark	&		&		&	\checkmark	&	\checkmark	&	\checkmark	&	B	&	Other	\\
63	&	Flux	&	21	&		&		&		&		&	\checkmark	&	\checkmark	&	\checkmark	&	\checkmark	&	B	&	Other	\\
64	&	AWS RDS	&	21	&		&		&	\checkmark	&		&		&	\checkmark	&	\checkmark	&	\checkmark	&	C	&	AWS	\\
65	&	AWS EMR	&	19	&		&		&	\checkmark	&		&		&	\checkmark	&	\checkmark	&	\checkmark	&	B	&	AWS	\\
66	&	Big Data	&	18	&		&		&	\checkmark	&	\checkmark	&		&	\checkmark	&	\checkmark	&	\checkmark	&	B	&	Other	\\
67	&	Cloud Network	&	18	&		&		&	\checkmark	&		&		&	\checkmark	&	\checkmark	&	\checkmark	&	B	&	Other	\\
68	&	Java	&	18	&		&	\checkmark	&	\checkmark	&	\checkmark	&		&	\checkmark	&	\checkmark	&	\checkmark	&	B	&	Other	\\
69	&	gRPC	&	18	&		&		&	\checkmark	&		&		&	\checkmark	&	\checkmark	&	\checkmark	&	B	&	Other	\\
70	&	AWS Fargate	&	18	&		&		&		&		&		&	\checkmark	&	\checkmark	&	\checkmark	&	B	&	AWS	\\
71	&	Cri-o	&	17	&		&		&	\checkmark	&		&		&	\checkmark	&	\checkmark	&	\checkmark	&	B	&	Other	\\
72	&	Polardb	&	17	&		&		&	\checkmark	&		&		&	\checkmark	&	\checkmark	&	\checkmark	&	B	&	Other	\\
73	&	Azure OpenAI	&	17	&		&		&	\checkmark	&	\checkmark	&		&	\checkmark	&	\checkmark	&	\checkmark	&	C	&	Azure	\\
74	&	Cloud Computing	&	17	&		&		&	\checkmark	&		&		&	\checkmark	&	\checkmark	&	\checkmark	&	B	&	Other	\\
75	&	Knative	&	17	&		&		&	\checkmark	&		&		&	\checkmark	&	\checkmark	&	\checkmark	&	B	&	Other	\\
76	&	CloudEvents	&	17	&		&		&	\checkmark	&	\checkmark	&		&	\checkmark	&	\checkmark	&	\checkmark	&	B	&	Other	\\
77	&	Fluent Bit	&	16	&		&		&	\checkmark	&	\checkmark	&		&	\checkmark	&		&	\checkmark	&	B	&	Other	\\
78	&	Terraform	&	16	&		&		&	\checkmark	&		&		&	\checkmark	&	\checkmark	&	\checkmark	&	B	&	Other	\\
79	&	IoT	&	16	&		&		&	\checkmark	&		&		&	\checkmark	&	\checkmark	&	\checkmark	&	B	&	Other	\\
80	&	Cloud Run	&	16	&		&		&	\checkmark	&		&		&	\checkmark	&	\checkmark	&	\checkmark	&	C	&	GCP	\\
81	&	Event-Driven Architecture	&	16	&		&		&	\checkmark	&		&		&	\checkmark	&	\checkmark	&	\checkmark	&	B	&	Azure	\\
82	&	Apache Spark	&	16	&		&		&	\checkmark	&	\checkmark	&		&	\checkmark	&	\checkmark	&	\checkmark	&	B	&	Other	\\
83	&	AWS FSX	&	15	&		&		&	\checkmark	&		&		&	\checkmark	&	\checkmark	&	\checkmark	&	B	&	AWS	\\
84	&	Cloud Infrastructure	&	15	&		&		&	\checkmark	&		&		&	\checkmark	&	\checkmark	&	\checkmark	&	B	&	Other	\\
85	&	Azure AI	&	15	&		&		&	\checkmark	&		&		&	\checkmark	&	\checkmark	&	\checkmark	&	C	&	Azure	\\
86	&	AWS VPC	&	15	&		&		&	\checkmark	&		&		&	\checkmark	&	\checkmark	&	\checkmark	&	C	&	AWS	\\
87	&	Falco	&	15	&		&		&	\checkmark	&	\checkmark	&		&	\checkmark	&		&	\checkmark	&	B	&	Other	\\
88	&	Thanos	&	15	&		&		&	\checkmark	&		&		&	\checkmark	&	\checkmark	&	\checkmark	&	O	&	Other	\\
89	&	Jaeger	&	15	&		&		&		&		&		&		&		&	\checkmark	&	B	&	Other	\\
90	&	AWS CloudFront	&	15	&		&		&		&		&		&	\checkmark	&	\checkmark	&	\checkmark	&	C	&	AWS	\\
91	&	Etcd	&	14	&		&		&	\checkmark	&		&		&	\checkmark	&	\checkmark	&	\checkmark	&	B	&	Other	\\
92	&	Vitess	&	14	&		&		&		&		&		&	\checkmark	&	\checkmark	&	\checkmark	&	B	&	Other	\\
93	&	Dapr	&	14	&		&		&	\checkmark	&		&		&	\checkmark	&	\checkmark	&	\checkmark	&	B	&	Other	\\
94	&	Longhorn	&	14	&		&		&	\checkmark	&		&		&	\checkmark	&	\checkmark	&	\checkmark	&	B	&	Other	\\
95	&	Rust	&	14	&		&	\checkmark	&	\checkmark	&	\checkmark	&		&	\checkmark	&	\checkmark	&	\checkmark	&	B	&	Other	\\
96	&	AWS Outposts	&	14	&		&		&		&		&		&	\checkmark	&	\checkmark	&	\checkmark	&	B	&	AWS	\\
97	&	Keptn	&	14	&		&		&	\checkmark	&	\checkmark	&		&	\checkmark	&	\checkmark	&	\checkmark	&	B	&	Other	\\
98	&	AWS Graviton	&	14	&		&		&	\checkmark	&		&		&	\checkmark	&	\checkmark	&	\checkmark	&	C	&	AWS	\\
99	&	Data Warehouse	&	14	&		&		&	\checkmark	&		&		&	\checkmark	&	\checkmark	&	\checkmark	&	B	&	Other	\\
100	&	Strimzi	&	13	&		&		&	\checkmark	&		&		&	\checkmark	&	\checkmark	&	\checkmark	&	B	&	Other	\\
101	&	AWS Opensearch	&	13	&		&		&	\checkmark	&		&		&	\checkmark	&	\checkmark	&	\checkmark	&	C	&	AWS	\\
102	&	Kubectl	&	13	&		&		&		&		&		&	\checkmark	&	\checkmark	&	\checkmark	&	B	&	Other	\\
103	&	Looker	&	12	&		&		&	\checkmark	&		&		&	\checkmark	&	\checkmark	&	\checkmark	&	B	&	GCP	\\
104	&	Orchestration	&	12	&		&		&	\checkmark	&		&		&	\checkmark	&	\checkmark	&	\checkmark	&	B	&	Other	\\
105	&	Oci	&	12	&		&		&	\checkmark	&		&		&	\checkmark	&	\checkmark	&	\checkmark	&	B	&	Other	\\
106	&	Graphql	&	12	&		&		&	\checkmark	&	\checkmark	&		&	\checkmark	&	\checkmark	&	\checkmark	&	B	&	Other	\\
107	&	DevOps	&	12	&	\checkmark	&	\checkmark	&	\checkmark	&	\checkmark	&	\checkmark	&	\checkmark	&	\checkmark	&	\checkmark	&	B	&	Other	\\
108	&	Cloud Migration	&	12	&	\checkmark	&		&	\checkmark	&		&		&	\checkmark	&	\checkmark	&	\checkmark	&	B	&	Other	\\
109	&	Kubevirt	&	12	&		&		&	\checkmark	&		&		&	\checkmark	&	\checkmark	&	\checkmark	&	B	&	Other	\\
110	&	AWS Appsync	&	12	&		&		&	\checkmark	&		&		&	\checkmark	&	\checkmark	&	\checkmark	&	B	&	AWS	\\
111	&	Zero Trust	&	11	&		&		&		&		&		&		&	\checkmark	&	\checkmark	&	B	&	Other	\\
112	&	Openshift	&	11	&		&		&	\checkmark	&		&		&	\checkmark	&	\checkmark	&	\checkmark	&	B	&	Other	\\
113	&	AWS Cloudwatch	&	11	&		&		&		&		&		&		&	\checkmark	&	\checkmark	&	C	&	AWS	\\
114	&	AWS Well-architected	&	11	&		&		&	\checkmark	&		&		&	\checkmark	&	\checkmark	&	\checkmark	&	B	&	AWS	\\
115	&	AWS Quicksight	&	11	&		&		&		&		&		&	\checkmark	&		&	\checkmark	&	B	&	AWS	\\
116	&	Keda	&	11	&		&		&		&		&		&	\checkmark	&	\checkmark	&	\checkmark	&	B	&	Other	\\
117	&	Vmware Cloud	&	11	&		&		&	\checkmark	&		&		&	\checkmark	&	\checkmark	&	\checkmark	&	B	&	Other	\\
118	&	Infrastructure	&	11	&		&		&	\checkmark	&		&		&	\checkmark	&	\checkmark	&	\checkmark	&	B	&	Other	\\
119	&	AWS X-ray	&	11	&		&		&		&		&		&		&	\checkmark	&	\checkmark	&	B	&	AWS	\\
120	&	AWS EFS	&	11	&		&		&	\checkmark	&		&		&	\checkmark	&	\checkmark	&	\checkmark	&	B	&	AWS	\\
        \hline 
\multicolumn{13}{l}{\textbf{\checkmark}: Yes,  \textbf{C}: Cloud Specific; \textbf{O}: On-premise; \textbf{B}: Both, \textbf{CPs}: Cloud Providers; \textbf{GCP}: Google Cloud; \textbf{AWS}: Amazon; \textbf{Azure}: Azure} \\
    \end{tabular}
    }
\end{table*}

\end{document}